%% file: axFF_cls.tex
\documentclass[11pt,a4paper]{article}
\pdfoutput=1

\usepackage{bigfoot}
\usepackage{xpatch}
\makeatletter
\patchcmd\FN@allmarks{266}{256}{}{\fail}
\makeatother
\usepackage{jheppub}
\allowdisplaybreaks
\usepackage{afterpage}
\usepackage{tikz}
\usepackage{bm}
\usepackage{slashed}
\usepackage{mathtools}
\usepackage{mathrsfs}
\usepackage[squaren,Gray]{SIunits}
\usepackage{wasysym}
\usepackage{enumerate}
\usepackage{lscape}
\usepackage{longtable}
\usepackage{booktabs}
\usepackage{widetable}
\usepackage[figureposition=bottom,font=small,labelfont=bf,labelsep=period,skip=5pt,tableposition=top]{caption}
\usepackage{dcolumn}
\usepackage{multirow}
\usepackage{printlen}

\usepackage{soul}
\newcommand{\na}{\text{\st{$a$}}}
\newcommand{\tP}{{\tilde{P}}}

\newcommand{\SU}[1]{\ensuremath{\text{SU}(#1)}}

\newcommand{\PCACFF}{PCAC${}_{\mathrm{FF}}$}
\newcommand{\la}{\langle}
\newcommand{\ra}{\rangle}
\renewcommand{\vec}{\mathbf}

\definecolor{matplotgreen}{rgb}{0,0.5,0}%

\newcommand{\pcacON}{\textcolor{green}{$\checkmark$}}
\newcommand{\pcacOF}{\textcolor{red}{$\times$}}
\newcommand{\M}{$\mathllap{\text{-}}$}

\makeatletter
\newif\iftag@here
\newcommand*{\taghere}[1][0pt]
{\ifmeasuring@\else
  \global\tag@heretrue
  \tikz[remember picture,overlay]{\coordinate (taghere) at (0pt,#1);}%
\fi}
\def\place@tag{%
    \iftagsleft@
      \kern-\tagshift@
      \iftag@here
        \global\tag@herefalse
        \tikz[remember picture,overlay]%
          {\path (taghere) -| node[anchor=base]{\rlap{\boxz@}} (0pt,0pt);}%
      \else
        \if1\shift@tag\row@\relax
            \rlap{\vbox{%
                \normalbaselines
                \boxz@
                \vbox to\lineht@{}%
                \raise@tag
            }}%
        \else
            \rlap{\boxz@}%
        \fi
        \kern\displaywidth@
      \fi
    \else
      \kern-\tagshift@
      \iftag@here
        \global\tag@herefalse
        \tikz[remember picture,overlay]%
          {\path  (taghere) -|  node[anchor=base]{\llap{\boxz@}} (0pt,0pt);}%
      \else
        \if1\shift@tag\row@\relax
            \llap{\vtop{%
                \raise@tag
                \normalbaselines
                \setbox\@ne\null
                \dp\@ne\lineht@
                \box\@ne
                \boxz@
            }}%
        \else \llap{\boxz@}%
        \fi
      \fi
    \fi
}
\makeatother

\title{Nucleon axial structure from lattice QCD}

\collaborationImg{\includegraphics{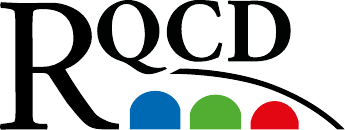}}

\author[a,b]{Gunnar~S.~Bali,}
\author[a]{Lorenzo~Barca,}
\author[a]{Sara~Collins,}
\author[a]{Michael~Gruber,}
\author[a]{Marius~L\"offler,}
\author[a]{Andreas~Sch{\"a}fer,}
\author[a]{Wolfgang~S{\"o}ldner,}
\author[a]{Philipp~Wein,}
\author[a]{Simon~Weish\"aupl,}
\author[a]{and Thomas~Wurm}

\emailAdd{gunnar.bali@ur.de}
\emailAdd{lorenzo1.barca@ur.de}
\emailAdd{sara.collins@ur.de}
\emailAdd{michael1.gruber@ur.de}
\emailAdd{marius.loeffler@ur.de}
\emailAdd{andreas.schaefer@ur.de}
\emailAdd{wolfgang.soeldner@ur.de}
\emailAdd{philipp.wein@ur.de}
\emailAdd{simon.weishaeupl@ur.de}
\emailAdd{thomas.wurm@ur.de}

\affiliation[a]{
	Institut f{\"u}r Theoretische Physik, Universit{\"a}t Regensburg,\\
	Universit{\"a}tsstra{\ss}e 31, D-93040 Regensburg, Germany
}
\affiliation[b]{
	Department of Theoretical Physics, Tata Institute of Fundamental Research,\\
	Homi Bhabha Road, Mumbai 400005, India
}

\abstract{We present a new analysis method that allows one to understand and model excited state contributions in observables that are dominated by a pion pole. We apply this method to extract axial and (induced) pseudoscalar nucleon isovector form factors, which satisfy the constraints due to the partial conservation of the axial current up to expected discretization effects. Effective field theory predicts that the leading contribution to the (induced) pseudoscalar form factor originates from an exchange of a virtual pion, and thus exhibits pion pole dominance. Using our new method, we can recover this behavior directly from lattice data. The numerical analysis is based on a large set of ensembles generated by the CLS effort, including physical pion masses, large volumes (with up to $96^3\times192$ sites and $Lm_\pi=6.4$), and lattice spacings down to~$\unit{0.039}{\femto\meter}$, which allows us to take all the relevant limits. We find that some observables are much more sensitive to the choice of parametrization of the form factors than others. On the one hand, the $z$-expansion leads to significantly smaller values for the axial dipole mass than the dipole ansatz ($M_A^{\text{$z$-exp}}=\unit{1.02(10)}{\giga\electronvolt}$ versus $M_A^{\text{dipole}}=\unit{1.31(8)}{\giga\electronvolt}$). On the other hand, we find that the result for the induced pseudoscalar coupling at the muon capture point is almost independent of the choice of parametrization ($g_P^{\star \ \text{$z$-exp}}=8.68(45)$ and $g_P^{\star \ \text{dipole}}=8.30(24)$), and is in good agreement with both, chiral perturbation theory predictions and experimental measurement via ordinary muon capture. We also determine the axial coupling constant~$g_A$.}

\keywords{Lattice QCD, Neutrino Physics, Nonperturbative Effects}

\arxivnumber{1911.13150}

\begin{document}
\maketitle
\flushbottom
\section{Introduction}
The axial structure of the nucleon is relevant for the description of experiments that involve weak interactions. The most precisely known quantity in this context is the axial coupling constant $g_A$, which corresponds to the axial form factor at vanishing momentum transfer and can be determined experimentally from $\beta$ decay (see refs.~\cite{Brown:2017mhw,Markisch:2018ndu,Gonzalez-Alonso:2018omy}; cf.\ also ref.~\cite{Hayen:2019nic}). At finite momentum transfer~$Q^2$, the axial and the induced pseudoscalar form factors are much less well known. They enter the description of exclusive pion electroproduction~\mbox{\cite{Choi:1993vt, Bernard:1994pk, Liesenfeld:1999mv, Fuchs:2003vw}} (e.g., $e^- p \to \pi^- p \nu$), (quasi-)elastic neutrino-nucleon scattering~\cite{Ahrens:1988rr, Kitagaki:1990vs, Bodek:2007vi, Meyer:2016oeg}, radiative muon capture~\cite{Hart:1977zz,Jonkmans:1996my,Wright:1998gi}, and ordinary muon capture~\cite{Gorringe:2002xx, Bardin:1980mi, Andreev:2012fj, Andreev:2015evt}. Via weak muon capture in muonic hydrogen a combination of the Dirac, Pauli, axial, and induced pseudoscalar form factors can be measured, constraining the latter at the muon capture point~\cite{Wright:1998gi, Winter:2011yp, Andreev:2012fj, Andreev:2015evt, Hill:2017wgb}. The direct determination of the induced pseudoscalar coupling in refs.~\cite{Andreev:2012fj,Andreev:2015evt} shows that, at small momentum transfer, the induced pseudoscalar form factor is indeed well approximated by a pion pole dominance (PPD) ansatz.\par%
From the theoretical side, one can gain insight into the form factors through various techniques. At small momentum transfer, chiral perturbation theory (ChPT) yields valuable low energy theorems~\cite{Bernard:1994wn, Fearing:1997dp, Bernard:2001rs, Fuchs:2003vw, Schindler:2006it} (motivating, e.g., the above mentioned PPD ansatz), while, at intermediate and large virtualities, the form factors can be determined (up to some systematic uncertainty of $\sim15\%$) using light-cone sum rules~\cite{Braun:2006hz, Anikin:2016teg}. Another interesting approach is the application of functional renormalization group methods~\cite{Eichmann:2011pv}.\par%
In this work, we will use lattice QCD, which enables a determination of hadronic observables from first principles. Once all systematic uncertainties are under control, this method provides the cleanest and most direct access to hadron form factors. Many studies of nucleon couplings and form factors have been carried out in the past using a wide variety of lattice actions and analysis methods (see, e.g., refs.~\mbox{\cite{Martinelli:1988rr, Lin:2008uz, Yamazaki:2008py, Yamazaki:2009zq, Bratt:2010jn, Alexandrou:2010hf, Capitani:2012gj, Green:2012ud, Horsley:2013ayv, Bhattacharya:2013ehc, Chambers:2014qaa, Bali:2014nma, vonHippel:2016wid, Bhattacharya:2016zcn, Meyer:2016kwb, Yoon:2016jzj, Liang:2016fgy, Bouchard:2016heu, Alexandrou:2017msl, Berkowitz:2017gql, Yao:2017fym, Chang:2018uxx, Alexandrou:2018lvq, Green:2017keo, Alexandrou:2017hac, Capitani:2017qpc, Rajan:2017lxk, Tsukamoto:2017fnm, Jang:2018lup, Ishikawa:2018rew, Liang:2018pis, Bali:2018qus, Shintani:2018ozy, Jang:2018djx, Green:2019zhh}}). Recent studies of form factors at finite virtualities with data close to physical pion masses have faced two problems: first of all, it is difficult to reconcile the data with the partial conservation of the axial current (PCAC). Even though PCAC is approximately fulfilled on the correlation function level, the corresponding relation between the ground state form factors, which are extracted using a spectral decomposition, is broken to a much larger extent. Secondly, the PPD ansatz for the induced pseudoscalar form factor fails to describe the data at small momentum transfer and small pion masses, which is the domain where one would expect this ansatz to give the best approximation. In both cases an explanation in terms of finite lattice spacing effects is unlikely, since the violation of PCAC is largest at small virtualities and masses. For nice presentations of these problems see, e.g., refs.~\cite{Rajan:2017lxk,Ishikawa:2018rew}. A prime suspect that may be responsible for both effects is a particularly large excited state contamination, albeit it was demonstrated in ref.~\cite{Jang:2018lup} that the problem persist if one uses a traditional fit ansatz with up to three free excited states. In ref.~\cite{Bali:2018qus} we have proposed a subtraction method that removes excited state contributions that violate the equations of motion for the nucleon. While this leads to a recovery of the PCAC relation on the form factor level, the PPD ansatz still remained strongly broken. While this is not impossible as such, the induced pseudoscalar charge at the muon capture point remained at variance with the experimental value~\cite{Wright:1998gi, Winter:2011yp, Andreev:2012fj, Andreev:2015evt, Hill:2017wgb}.\par%
A deeper insight into the excited state effects is possible using effective field theory (EFT). ChPT based analyses~\cite{Bar:2018akl, Bar:2018xyi,Bar:2019gfx} (along the lines of refs.~\cite{Bar:2017kxh, Bar:2017gqh,Tiburzi:2015tta, Tiburzi:2015sra} using interpolating currents from ref.~\cite{Wein:2011ix}) indicate that the subtraction method mentioned above does not remove all excited states and that the violation of the PPD ansatz is due to additional, large contributions from $N\pi$ exited states that predominantly affect the induced pseudoscalar and the pseudoscalar form factors. While an a posteriori subtraction of the effect performed in refs.~\cite{Bar:2018akl, Bar:2018xyi} leads to satisfying results, such a procedure appears inadequate from the lattice QCD perspective as it introduces a dependence on ChPT input parameters and cannot be consistently combined with standard excited state fits. Moreover, when truncating the ChPT expansion for the interpolating currents at leading order (as, e.g., in refs.~\cite{Bar:2018akl, Bar:2018xyi}) the $N$ and $N\pi$ overlap factors have the exact same dependence on the operator smearing, which may not be justified in lattice simulations where spatially extended (smeared) nucleon interpolators with radii that are not very small compared to the inverse pion mass are employed.\par%
The procedure advocated in this article makes use of the same effective field theory methods as refs.~\cite{Bar:2018akl, Bar:2018xyi, Bar:2019gfx, Bar:2019zkx} in order to calculate the leading excited state contribution to the correlation function explicitly for all axial and pseudoscalar channels with less assumptions. We will show that exploiting the EFT knowledge stabilizes excited state fits considerably and allows us to extract ground state form factors, which are found to obey both the PCAC relation on the form factor level and the PPD ansatz reasonably well. Recently an alternative analysis method has been proposed in ref.~\cite{Jang:2019vkm}, which also allows an extraction of nucleon form factors that satisfy PCAC. We will discuss similarities and differences between our approach and this method in sections~\ref{sec_excited_energies} and~\ref{sec_approx_pcac_ppd}.\par%
This article is structured as follows. In section~\ref{sec_correlation_functions} we will give a detailed description of the EFT calculation needed to determine the leading $N\pi$ contribution to the correlation function and how it can be combined with the usual excited state analysis. The lattice setup and the employed ratios of correlation functions are detailed in section~\ref{sec_data_analysis}. Section~\ref{sec_formfactors} contains the results for the form factors (using both, dipole fits and the $z$-expansion) and includes an analysis of the PCAC relation as well as of the PPD ansatz. We also explore parametrizations that are consistent with PCAC in the continuum. In the latter case the continuum limit is under much better control. We summarize our findings in section~\ref{sec_summary}.\par%
\section{Correlation functions\label{sec_correlation_functions}}%
\subsection{Definitions}
In order to study hadron structure using lattice QCD one has to calculate two- and three-point correlation functions, where hadron states with matching quantum numbers are created by a suitable interpolating current $\bar{\mathcal N}$ at the source time~$t_{\rm src}$, and are destroyed by $\mathcal N$ at the sink time~$t_{\rm snk}$ (here, we will always set $t_{\rm src}=0$ and $t_{\rm snk}=t$ without loss of generality). In the case of three-point correlation functions one inserts a local current $\mathcal O$ at some insertion time $\tau$ with $t>\tau>0$ and, usually, one is interested in the ground state matrix element of this current insertion. The momenta can be fixed by appropriate Fourier transforms, in our case at the sink and the insertion, such that the initial state and final state momenta are $\vec{p}$ and $\vec p^\prime$, respectively:
\begin{align}%
C_{{\rm 2pt},P_+}^{\vec{p}} (t) =  P_+^{\alpha\beta} C_{\rm 2pt,\beta\alpha}^{\vec{p}} (t) &= a^3 \sum_{\vec{x}} e^{-i\vec{p}\cdot\vec{x}} \, P_+^{\alpha\beta} \, \la \mathcal N^\beta(\vec{x}, t) \bar{\mathcal N}^{\alpha}(\vec{0}, 0) \ra \,, \label{eq_2ptdef} \\
C_{{\rm 3pt}, \Gamma}^{\vec{p}^{\mathrlap{\prime}}, \vec{p}, \mathcal{O}} (t, \tau) = \Gamma^{\alpha\beta} C_{{\rm 3pt}, \beta\alpha}^{\vec{p}^{\mathrlap{\prime}}, \vec{p}, \mathcal{O}} (t, \tau) &= a^6 \smashoperator{\sum_{\vec{x}, \vec{y}}} e^{-i\vec{p}^{\prime \!}\cdot\vec{x} + i(\vec{p}^{\mathrlap{\prime}} - \vec{p})\cdot\vec{y}} \, \Gamma^{\alpha\beta} \, \la \mathcal N^\beta(\vec{x}, t) \mathcal{O}(\vec{y}, \tau) \bar{\mathcal N}^{\alpha}(\vec{0}, 0) \ra \,, \label{eq_3ptdef}
\end{align}%
where $\smash{C_{\rm 2pt}^{\vec{p}}}$, $\smash{C_{{\rm 3pt}}^{\vec{p}^{\mathrlap{\prime}}, \vec{p}, \mathcal{O}}}$, $P_+$, and $\Gamma$ are matrices in Dirac space with the corresponding spin indices $\alpha$ and $\beta$. The three-quark nucleon interpolating current is defined via the usual quark-diquark structure with the charge conjugation matrix $C$,%
\begin{align}%
\mathcal N^\alpha(\vec x, t)=\bigl(u(\vec x, t)^T C\gamma_5 d(\vec x, t)\bigr) u^\alpha(\vec x, t) \,, 
\end{align}%
where each quark is smeared separately in the spatial directions using Wuppertal smearing~\cite{Gusken:1989qx} on spatially APE-smoothed links~\cite{Falcioni:1984ei}. Note that Minkowski scalar products and gamma matrix conventions are used throughout this work. At zero three-momentum $P_+=(1+\gamma_0)/2$ annihilates the leading negative parity contribution. For the analysis of the pseudoscalar and axialvector form factors we choose $\Gamma$ to be $P_+^i = P_+^{\phantom{i}} \gamma^i \gamma_5$, $i=1,2,3$. In order to relate the correlation functions to matrix elements, one inserts identity operators (corresponding to sums over all hadronic states) and uses the translational properties of the currents to carry out the Fourier transforms. When evaluating the result at large Euclidean times, $t$, $\tau$, and $t-\tau$, excited states are exponentially suppressed and the correlation functions can be approximated by the ground state contributions:%
\begin{align}
 C_{{\rm 2pt},P_+}^{\vec{p}} (t) &\approx \sum\limits_\sigma P_+^{\alpha\beta} \la 0 | \mathcal N^\beta | N_\sigma^{\vec{p}} \ra \la N_\sigma^{\vec{p}} | \bar{\mathcal N}^\alpha |  0 \ra \frac{e^{-E_{\vec{p}}t}}{2E_{\vec{p}}}  \,,
\label{eq_2ptgroundstate0} \\
C_{{\rm 3pt}, \Gamma}^{\vec{p}^{\mathrlap{\prime}}, \vec{p}, \mathcal{O}} (t, \tau) &\approx \sum\limits_{\sigma^\prime, \sigma} \Gamma^{\alpha\beta} \la 0 | \mathcal N^\beta | N_{\sigma^\prime}^{\vec{p^\prime}} \ra  \la N_{\sigma^\prime}^{\vec{p^\prime}} | \mathcal O |  N_\sigma^{\vec{p}} \ra \la N_\sigma^{\vec{p}} | \bar{\mathcal N}^\alpha |  0 \ra \frac{e^{-E_{\vec{p^\prime}}(t-\tau)} e^{-E_{\vec{p}}\tau}}{2E_{\vec{p^\prime}}2E_{\vec{p}}}  \,,
\label{eq_3ptgroundstate0}
\end{align}%
where all currents are located at the origin and $| N_\sigma^{\vec{p}} \ra$ corresponds to a nucleon state with three-momentum $\vec{p}$ and spin-projection $\sigma$. The parity projected overlap matrix elements can be parametrized as%
\begin{align}%
  P_{\pm}^{\alpha \beta}\la 0 | \mathcal N^\beta(\vec{0},0) | N_\sigma^{\vec{p}} \ra &= P_{\pm}^{\alpha \beta} \sqrt{Z_{\vec{p}}^\pm} u^\beta_{\vec{p}, \sigma} \,,
\end{align}%
where $u^\beta_{\vec{p}, \sigma}$ is a nucleon spinor and $\sqrt{\vphantom{Z}\smash{Z_{\vec{p}}^\pm}}$ are  momentum- and smearing-dependent overlap factors. For smeared currents $\sqrt{\vphantom{Z}\smash{Z_{\vec{p}}^+}}$ and $\sqrt{\vphantom{Z}\smash{Z_{\vec{p}}^-}}$ can differ from each other due to the explicit breaking of Lorentz invariance by the operator smearing, cf.\ refs.~\cite{Bowler:1997ej} and~\cite{Stokes:2018emx} for more details. Since in our analysis only positive parity projected overlap matrix elements occur, we define $\sqrt{Z_{\vec{p}}} \equiv \sqrt{\vphantom{Z}\smash{Z_{\vec{p}}^+}}$. The form factor decomposition for the nucleon-nucleon matrix element of a generic current can be written as
\begin{align}%
  \la  N_{\sigma^\prime}^{\vec{p^\prime}} | \mathcal O(\vec{0},0)  | N_\sigma^{\vec{p}} \ra &= \bar u_{\vec{p^\prime}, \sigma^\prime} J[\mathcal O] u_{\vec{p}, \sigma} \,, \label{eq_FFdecomp}
\end{align}%
where $J[\mathcal O]$ is matrix valued and can be parametrized in terms of form factors, cf.\ eqs.~\eqref{eq_PseudoscalarFF} and~\eqref{eq_AxialvectorFF} below. Using the spinor identity $\sum_\sigma u_{\vec{p}, \sigma} \bar u_{\vec{p}, \sigma}=\slashed p + m$, where $m$ is the nucleon mass, one arrives at the ground state contribution
\begin{align}
 C_{{\rm 2pt}}^{\vec{p}} (t) &\approx \frac{Z_{\vec{p}}}{2 E_{\vec{p}}} e^{-E_{\vec{p}}t} \, (\slashed p + m)  \,,
\label{eq_2ptgroundstate} \\
 C_{{\rm 3pt}}^{\vec{p}^{\mathrlap{\prime}}, \vec{p}, \mathcal{O}} (t, \tau) &\approx \frac{\sqrt{Z_{\vec{p^\prime}}}\sqrt{Z_{\vec{p}}}}{2E_{\vec{p^\prime}}2E_{\vec{p}}}e^{-E_{\vec{p^\prime}}(t-\tau)} e^{-E_{\vec{p}}\tau} \, (\slashed p^\prime + m) J[\mathcal O] (\slashed p + m) \,.
\label{eq_3ptgroundstate}
\end{align}%
For the two-point function we can explicitly evaluate the trace with $P_+$ to find
\begin{align}
 C_{{\rm 2pt},P_+}^{\vec{p}} (t) &\approx \frac{Z_{\vec{p}}}{2 E_{\vec{p}}} e^{-E_{\vec{p}}t}\operatorname{tr} \bigl\{ P_+ (\slashed p + m) \bigr\}  = Z_{\vec{p}}  \frac{E_{\vec{p}} + m}{E_{\vec{p}}} e^{-E_{\vec{p}}t}  \,.
\label{eq_2ptgroundstate_tr}
\end{align}%
For the three-point functions the trace with $\Gamma$ depends on the current-specific decomposition~\eqref{eq_FFdecomp}. In practice it turns out (in particular in case of the three-point functions) that, at Euclidean time distances $t$ and $\tau$ with acceptable signal-to-noise ratio, not only the ground state contributes. In most cases this problem can be treated by taking into account generic excited state contributions in the fit functions (see section~\ref{sec_final_param}). However, there are situations in which the excited states constitute (at the available temporal distances) the dominant contribution. In the latter case the generic excited state parametrizations fail to describe the data appropriately and further physical insight into the excited state structure is needed, cf.\ section~\ref{sec_eft}.\par%
For the isovector pseudoscalar and axialvector currents used in this work,%
\begin{align}%
\mathcal P &= \bar u \gamma_5 u - \bar d \gamma_5 d \,, &
\mathcal A_\mu &= \bar u \gamma_\mu \gamma_5 u -  \bar d \gamma_\mu \gamma_5 d \,, \label{eq_currents}
\end{align}%
the explicit decompositions in terms of the pseudoscalar form factor, $G_P(Q^2)$, as well as the axial and induced pseudoscalar form factors, $G_A(Q^2)$ and $G_\tP(Q^2)$, are%
\begin{align}%
J[\mathcal P] &= \gamma_5 G_P(Q^2) \,, \label{eq_PseudoscalarFF} \\
J[\mathcal A_\mu] &= \gamma_\mu \gamma_5 G_A(Q^2) + \frac{q_\mu}{2m} \gamma_5 G_\tP(Q^2) \,, \label{eq_AxialvectorFF}
\end{align}%
where $q=p^\prime-p$ is the momentum transfer and $Q^2 = -q^2$ is the virtuality. The three form factors used above are not independent in the continuum theory, since the axial Ward identity yields \mbox{$\partial^\mu \mathcal A_\mu = 2i \, m_\ell \mathcal P$} known as partial conservation of the axialvector current (PCAC). Here $m_\ell$ is the light quark mass. On the lattice this relation can be broken by discretization effects. For the nucleon matrix elements it implies that%
\begin{align}
2i \, m_\ell \la N^{\smash{\vec p^\prime}}_{\smash{\sigma^\prime}} | \mathcal P | N_\sigma^{\vec{p}} \ra &= \la N^{\smash{\vec p^\prime}}_{\smash{\sigma^\prime}} | \partial^\mu\!\mathcal A_\mu | N_\sigma^{\vec{p}} \ra +\mathcal{O}(a^2) \,,\label{eq_NucleonMatrixPCAC}
\end{align}%
where we can safely ignore discretization effects linear in the lattice spacing~$a$, since our analysis is fully order~$a$ improved, cf.\ section~\ref{sec_lattice_setup}. Using the definitions~\eqref{eq_PseudoscalarFF} and~\eqref{eq_AxialvectorFF} together with the equations of motion one can deduce the corresponding relation for the form factors (called~\PCACFF\ in ref.~\cite{Bali:2018qus}):%
\begin{align}%
\frac{m_\ell}{m} G_P(Q^2) &=G_A(Q^2) - \frac{Q^2}{4m^2} G_\tP(Q^2) + \mathcal{O}(a^2) \,. \label{eq_NucleonFFPCAC}
\end{align}%
Eqs.~\eqref{eq_NucleonMatrixPCAC} and~\eqref{eq_NucleonFFPCAC} should both be satisfied, once the ground state matrix elements have been extracted reliably.%
\subsection{EFT-based analysis\label{sec_eft}}%
Employing a theory where hadrons are the effective degrees of freedom (like baryon chiral perturbation theory) in order to elucidate the excited state structure in correlation functions is appealing, in particular if multi-hadron states with additional pions are the relevant excitations, see refs.~\cite{Bar:2015zwa,Tiburzi:2015tta, Tiburzi:2015sra, Bar:2017kxh, Bar:2017gqh}. In many cases, however, these contributions are relatively small and one can deal with them using standard methods like, e.g., source/sink-smearing and multi-exponential fits that allow for generic excited state contributions. As will be explained in detail in this section, the situation is different in the context of isovector axial and pseudoscalar form factors, where $N\pi$ contributions can actually be a leading term from the EFT point of view due to pion pole dominance (PPD). Especially for small pion masses this effect outweighs the exponential suppression at the currently available source-sink distances due to the small energy gap.\footnote{Note that, due to the exponential deterioration of the signal, one cannot expect the source-sink distances to become dramatically larger in future simulations.} In this situation multi-exponential fits with generic excited states become very unstable and usually fail to isolate the ground state contribution (see the discussions in refs.~\cite{Rajan:2017lxk,Ishikawa:2018rew,Jang:2018lup,Bali:2018qus}).\footnote{An alternative method has been proposed in ref.~\cite{Jang:2019vkm}, which appears to resolve the ground state contribution in this situation. We will comment on this method in some detail in sections~\ref{sec_excited_energies} and~\ref{sec_approx_pcac_ppd}.}\par%
In refs.~\cite{Bar:2018xyi, Bar:2019gfx} nucleon three-point functions with axialvector and pseudoscalar current insertions have been analyzed using ChPT and compelling qualitative evidence has been presented that the violations of the PCAC and PPD relations are indeed caused by $N\pi$ excited states. This is done as follows: first, one calculates the excited state contribution to the form factor using ChPT. The predicted, excited state contaminated form factor is found to agree quite well with recent data from the PACS collaboration~\cite{Ishikawa:2018rew}, cf.\ refs.~\cite{Bar:2019gfx, Bar:2019zkx}. In a second step, one may attempt to correct the error by subtracting the calculated excited state contaminations a posteriori (see, e.g., refs.~\cite{Bar:2018akl, Bar:2018xyi}, where such a subtraction has been performed for the induced pseudoscalar form factor). While this method yields convincing qualitative results, there are some open questions and limitations that need to be addressed:
\begin{enumerate}
\item In general, the operator smearing can have a different effect on $N$ and $N\pi$ overlap factors, which a leading order ChPT calculation does not allow for. There are heuristic arguments that this effect of the smearing should be negligible as long as the smearing radii $r_{\rm sm}$ are much smaller than the Compton wavelength of the pion $\lambda_\pi\approx\unit{1.41}{\femto\meter}$, cf.\ refs.~\cite{Bar:2015zwa, Tiburzi:2015sra, Bar:2017kxh, Bar:2017gqh}. This seems to contradict the observation that the operator smearing used in actual simulations has a strong impact on the signal of excited states. In refs.~\cite{Bhattacharya:2013ehc,Bali:2014nma} it has been found that smearing radii of roughly $r_{\rm sm}\sim\unit{0.5}{\femto\meter}$ maximize the ground state overlap. In the lattice analysis performed in this article, the optimized smearing radii are on some ensembles even larger (up to $\unit{0.8}{\femto\meter}$, cf.\ table~\ref{tab_ensembles}), and it is questionable whether a dependence on the smearing can be completely excluded for such smearing radii.\footnote{Note that our analysis in section~\ref{sec_data_analysis} suggests that there is no strong suppression of the $N\pi$ states due to the smearing and that the leading order ChPT approximation for the interpolating currents is actually quite good.}
\item So far, an a posteriori subtraction of the excited states has only been performed in combination with the ratio method on the lattice. It is unclear how one would avoid double counting, if one combines it with a standard excited state analysis, e.g., by using multi-exponential fits.
\item Estimating the systematic error tied to the ChPT based subtraction is challenging.
\end{enumerate}
From a lattice QCD perspective the situation is in our opinion quite clear concerning point~$2$. If there is a large $N\pi$ excited state contribution, then it should be taken into account explicitly in the multi-state fits to the correlation functions.\footnote{One can also try to circumvent the problem entirely by either suppressing or subtracting the unwanted excited state contributions. In ref.~\cite{Meyer:2018twz} the pion pole contribution is suppressed by analyzing the matrix elements of currents with a Gaussian profile instead of local currents. Ref.~\cite{Bali:2018qus} presents a method to subtract some of the excited state contributions.} In this approach point $1$ can be addressed simultaneously by allowing for a smearing dependence of the $N\pi$ coupling to the interpolating currents. Furthermore, we can avoid systematic uncertainties (point~$3$) by relaxing ChPT constraints. In the following, we will describe in detail how this can be achieved.\par%
\begin{figure}
\centering
\includegraphics[width=0.3\textwidth]{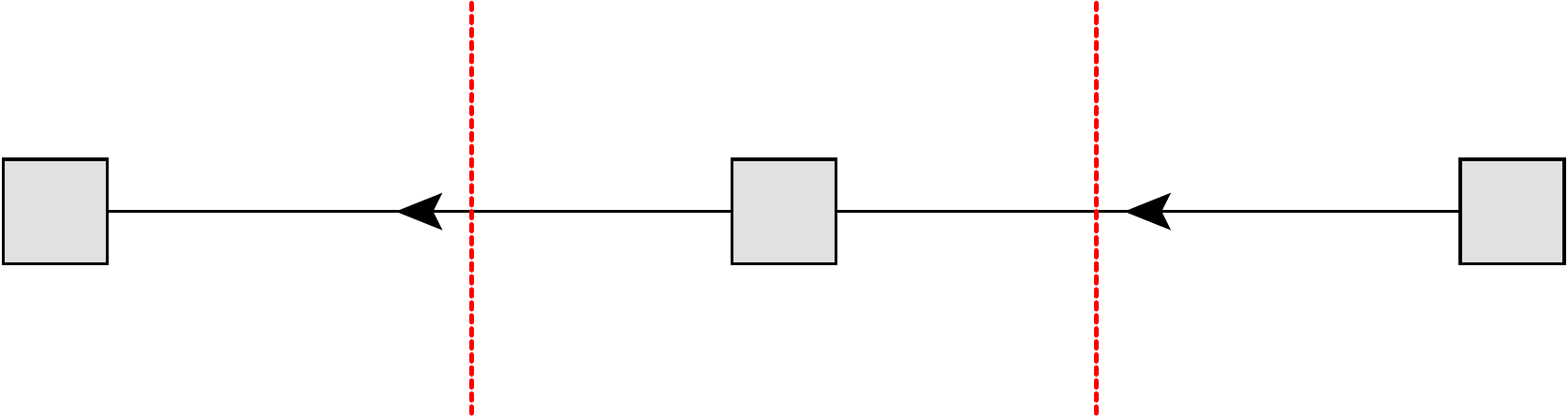}\\[1cm]
\includegraphics[width=0.3\textwidth]{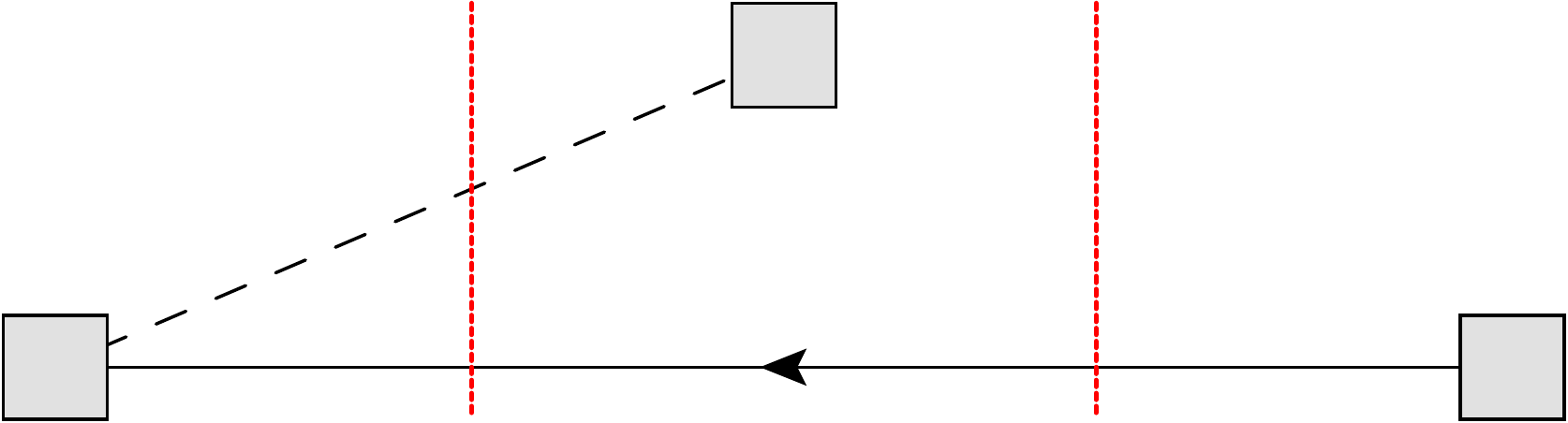}\hfill%
\includegraphics[width=0.3\textwidth]{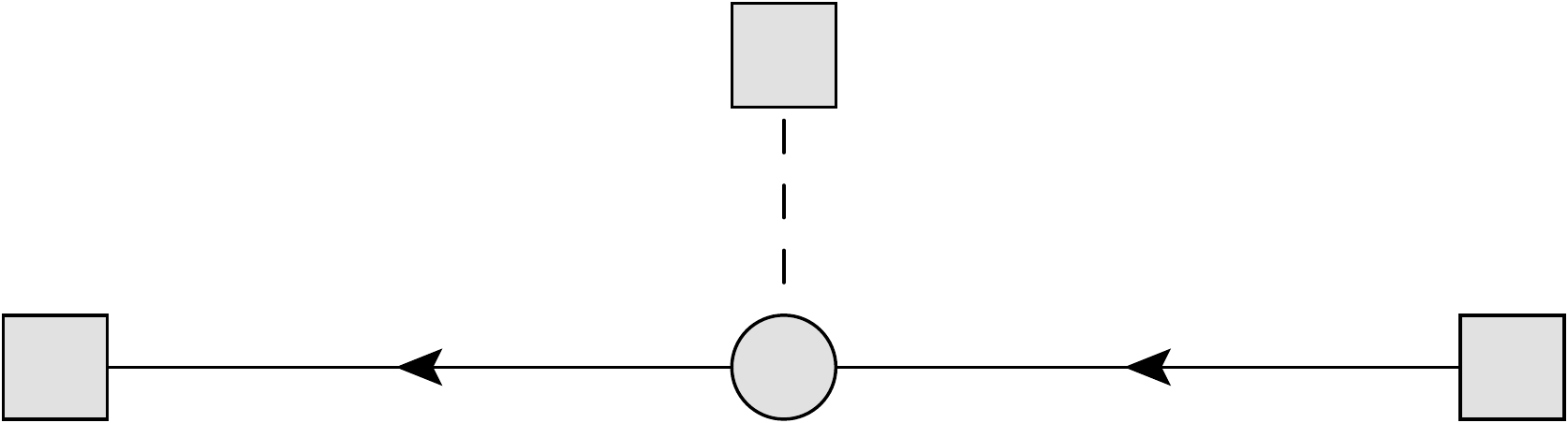}\hfill%
\includegraphics[width=0.3\textwidth]{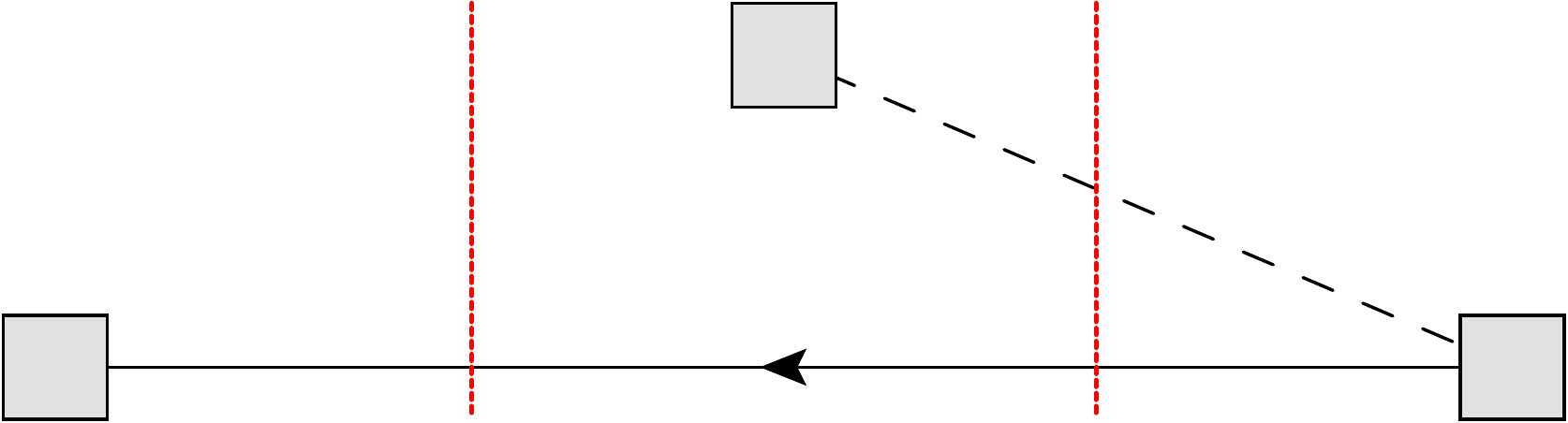}\\[0.3cm]
\includegraphics[width=0.5\textwidth]{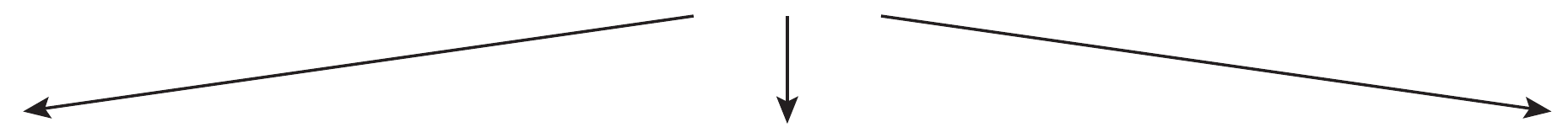}\\[0.3cm]
\includegraphics[width=0.3\textwidth]{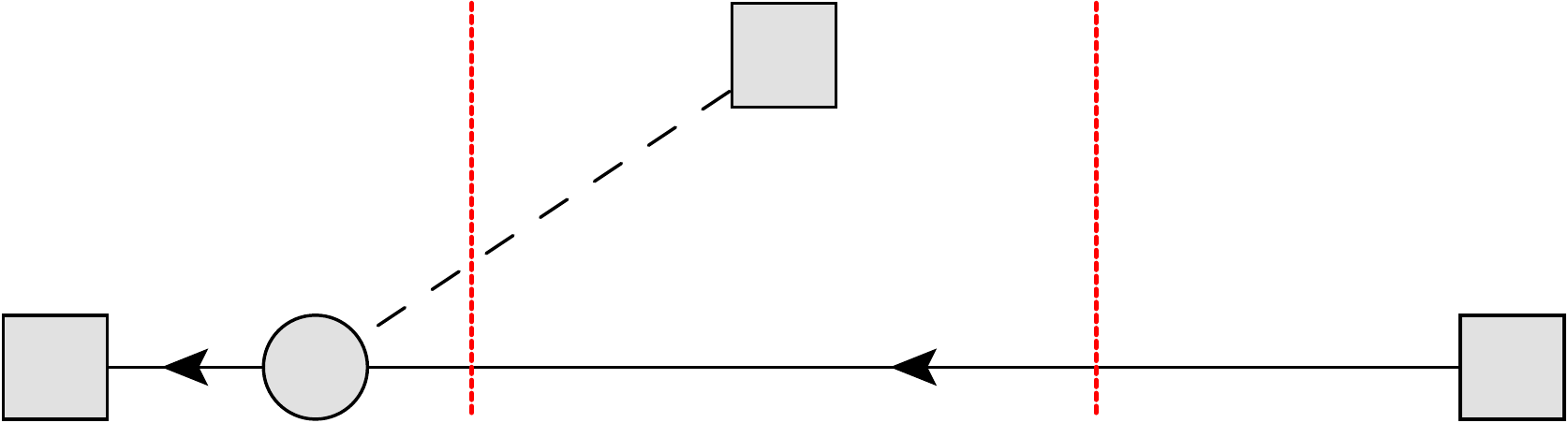}\hfill%
\includegraphics[width=0.3\textwidth]{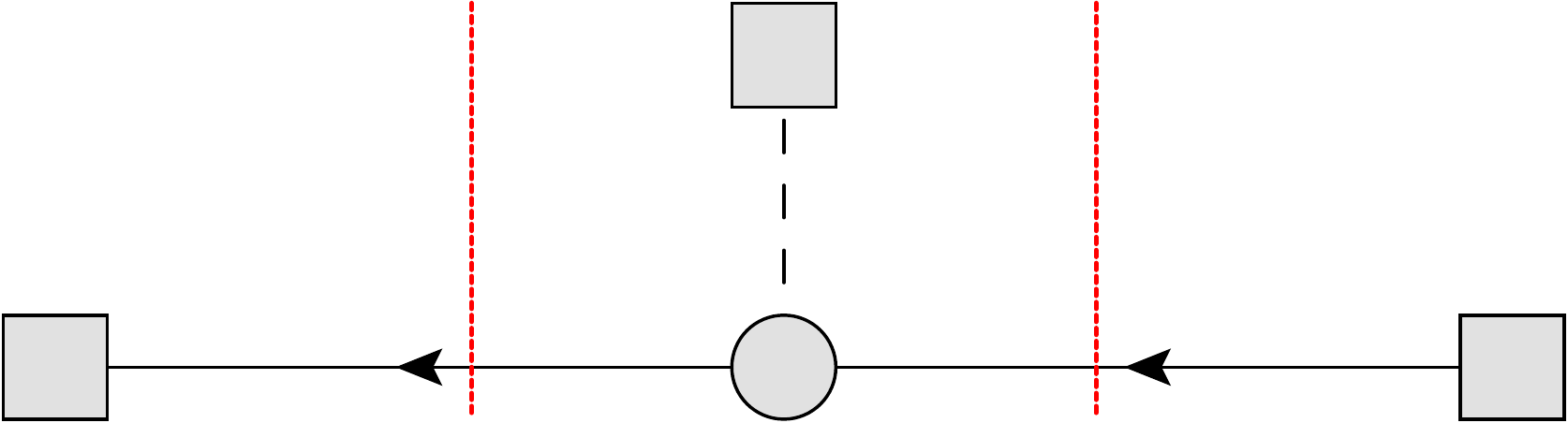}\hfill%
\includegraphics[width=0.3\textwidth]{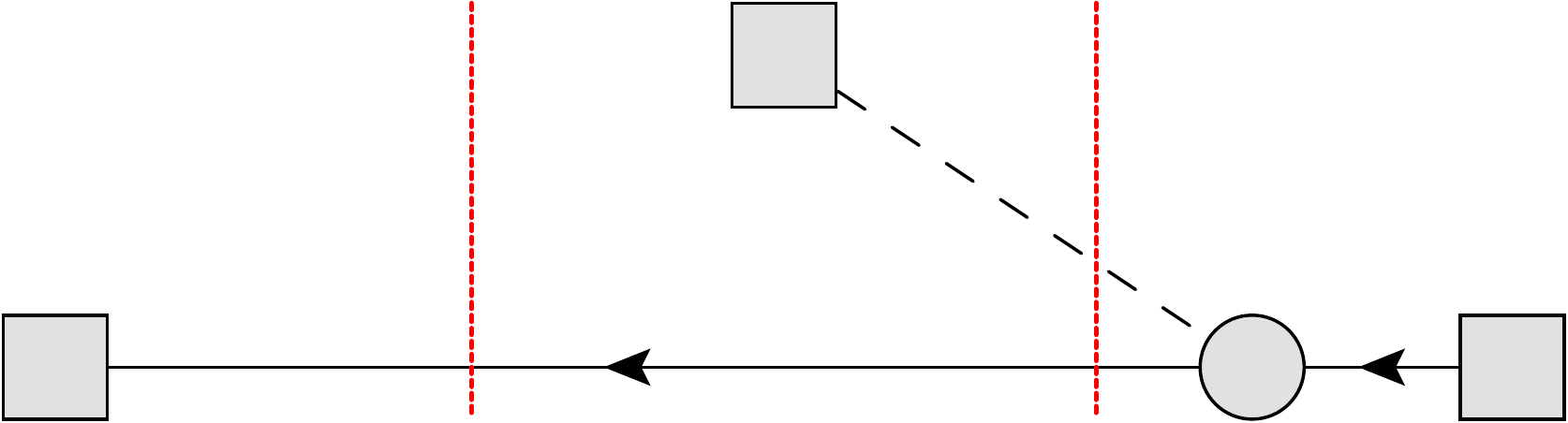}\\[\baselineskip]
\caption{\label{fig_diagrams}Feynman diagrams showing the most important (tree-level) contributions to the axial and pseudoscalar three-point functions. The squares correspond to explicitly inserted operators: the right and left ones correspond to smeared three-quark baryon interpolating currents at the source (at time~$0$) and the sink (at time~$t$), respectively, while the ones in the middle depict a pseudoscalar or an axialvector operator insertion (at time~$\tau$). The circles correspond to pion-nucleon interaction vertices, while the dashed and solid lines represent pion and nucleon propagators, respectively. The dotted red vertical lines indicate the sums over hadronic states one usually introduces to interpret correlation functions.}
\end{figure}
The first and second rows of figure~\ref{fig_diagrams} show the tree-level Feynman diagrams that contribute to the correlation functions. As discussed in ref.~\cite{Bar:2018xyi}, these yield the most important contribution to the correlation function. The squares on the right and left depict the smeared source and sink currents, while the one in the middle corresponds to the inserted local quark bilinears (axialvector or pseudoscalar currents in our case). The dashed and solid lines depict pion and nucleon propagators, while the circle stands for a pion-nucleon interaction vertex. The dotted red lines are for illustration only and indicate the identity operators (i.e., the sum over all hadronic states) that are usually inserted between source and current as well as between current and sink, cf.\ eq.~\eqref{eq_3ptgroundstate0}. This elucidates that the diagram in the first row yields a contribution to the ground state, while the diagrams on the left- and right-hand sides in the second row give rise to a nucleon-pion excitation in the final and initial state, respectively. For the diagram in the middle of the second row, however, the situation is not that simple, since the nucleon-pion interaction is not restricted to a specific time-slice. As a consequence, the diagram contributes to both the ground state and the excited states, as shown in the bottom row of figure~\ref{fig_diagrams}. This follows from an explicit calculation of the diagrams (see below). We emphasize that there is no one-to-one correspondence between the individual contributions in the spectral decomposition and the diagrams. For example, both the diagram in the first row and the diagram in the middle of the second row contribute to the ground state and, actually, an infinite number of diagrams will contribute to each state if one takes into account higher orders in ChPT (see ref.~\cite{Bar:2018xyi} for a list of one-loop diagrams). Finally, a single diagram can contribute to multiple states in the spectral decomposition, cf.\ the bottom row of figure~\ref{fig_diagrams}. We will exploit the fact that the pion pole contribution to the ground state automatically gives rise to an associated excited state.\par%
Before addressing the details, let us note that the following calculation is in large parts already contained in refs.~\cite{Bar:2018xyi, Bar:2019gfx}, where also one-loop diagrams are taken into account. Also the presentation in ref.~\cite{Meyer:2018twz} is based on similar considerations (cf.\ also ref.~\cite{Hansen:2016qoz}). However, we will present the result in a more general way (without using a particular spin projection or fixing initial and final state momenta to a predefined configuration) such that it can be used in a variety of simulation setups. The first ingredient we need in order to evaluate the diagrams in figure~\ref{fig_diagrams} are the corresponding Feynman rules. Here we follow the conventions of ref.~\cite{Scherer:2012xha}, but adapt them to our choices for the currents (see eq.~\eqref{eq_currents}) and convert them to position space. We work in two-flavor baryon ChPT here. However, since we only consider the nucleon sector and are only working at tree-level accuracy, a three-flavor calculation would give exactly the same result. Note that in this section all time variables are in Minkowski time and will be rotated to imaginary times only at the very end. The pion and nucleon propagators read%
\begin{align}%
  S_N(x)   &= i\int\!\!\frac{d^4q}{(2\pi)^4}e^{-iq \cdot x} \frac{\slashed{q}+m}{q^2-m^2+i\epsilon} \,,\\
  \mathllap{S^{ab}_\pi(x) = \delta^{ab}}  S_\pi(x) &= i\int\!\!\frac{d^4q}{(2\pi)^4}e^{-iq \cdot x} \frac{ \delta^{ab}}{q^2-m_\pi^2+i\epsilon} \,.
\end{align}%
For the vertices of the current insertions we have
\begin{align}%
\begin{tikzpicture}[baseline={([yshift=-0.6ex]current bounding box.center)}]
        \draw[shift={(-0.26, -0.26)}, thick] (0, 0) rectangle (0.52, 0.52) node[pos=.5] {$\mathcal{A}^\mu$};
        \draw[color=black, thick] (0.26, 0) to (1.3, 0);
        \draw[color=black, thick] (-0.26, 0) to (-1.3, 0);
\end{tikzpicture}
&= g_A \gamma^\mu \gamma_5 \sigma^3\,, \\[.2\baselineskip]
\begin{tikzpicture}[baseline={([yshift=-0.6ex]current bounding box.center)}]
        \draw[shift={(-0.26, -0.26)}, thick] (0, 0) rectangle (0.52, 0.52) node[pos=.5] {$\mathcal{P}$};
        \draw[color=black, thick] (0.26, 0) to (1.3, 0);
        \draw[color=black, thick] (-0.26, 0) to (-1.3, 0);
\end{tikzpicture}
&= 0 \,, \\[.2\baselineskip]
\begin{tikzpicture}[baseline={([yshift=-0.6ex]current bounding box.center)}]
        \draw[shift={(-0.26, -0.26)}, thick] (0, 0) rectangle (0.52, 0.52) node[pos=.5] {$\mathcal{A}^\mu$};
        \draw[color=black, thick, dashed] (0.26, 0) to (1.3, 0);
\end{tikzpicture}
&= -2F_\pi \partial^\mu \delta^{a3} \,, \\[.2\baselineskip]
\begin{tikzpicture}[baseline={([yshift=-0.6ex]current bounding box.center)}]
        \draw[shift={(-0.26, -0.26)}, thick] (0, 0) rectangle (0.52, 0.52) node[pos=.5] {$\mathcal{P}$};
        \draw[color=black, thick, dashed] (0.26, 0) to (1.3, 0);
\end{tikzpicture}
&= -2iF_\pi B \delta^{a3} \,,\qquad\smash[t]{B\equiv\frac{m_\pi^2}{2m_\ell}} \,,
\end{align}%
where we only take into account the leading contribution in the chiral counting\footnote{Note that $\gamma_5$ is counted as first order in baryon ChPT, while other elements of the Clifford algebra are counted as zeroth order, see, e.g., ref.~\cite{Oller:2006yh}. This explains why the $NN$ vertex of the pseudoscalar current vanishes at leading order.} and all derivatives are understood to act on the pion propagator. Here, $F_\pi$ and $g_A$ correspond to the pion decay constant and the axial coupling in the chiral limit, respectively, while $B$ is the condensate parameter and $\sigma^a$ are Pauli matrices. For the leading $N\pi$ interaction vertex we have%
\begin{align}%
\begin{tikzpicture}[baseline={([yshift=-3ex]current bounding box.center)}]
        \draw[thick] (0, 0) circle (0.2);
        \draw[color=black, thick] (0.2, 0) to (1.5, 0);
        \draw[color=black, thick] (-0.2, 0) to (-1.5, 0);
        \draw[color=black, thick, dashed] (0.0, 0.2) to (0.0, 1.0);
\end{tikzpicture}
&= -i\frac{g_A}{2 F_\pi}\slashed{\partial}\gamma_5 \sigma^a \,.
\end{align}%
The vertices for local three-quark currents have been derived in ref.~\cite{Wein:2011ix}. We adapt these to the smeared interpolating currents used here by allowing for momentum- and smearing-dependent couplings. With the nucleon isospinor $\Psi_N$, where $\Psi_p = (1,0)^T$ and $\Psi_n = (0,1)^T$, the leading order vertices read%
\def\sqrtstrut{\rule[-1pt]{0sp}{10pt}}
\begin{align}%
\begin{tikzpicture}[baseline={([yshift=-0.6ex]current bounding box.center)}]
        \draw[shift={(-0.2, -0.2)}, thick] (0, 0) rectangle (0.4, 0.4);
        \draw[color=black, thick] (0.2, 0) to (2.5, 0);
\end{tikzpicture}
&= \sqrt{\smash{Z_\vec{p^\prime}}\sqrtstrut}\bar\Psi_N\,, &
\begin{tikzpicture}[baseline={([yshift=-0.6ex]current bounding box.center)}]
        \draw[shift={(-0.2, -0.2)}, thick] (0, 0) rectangle (0.4, 0.4);
        \draw[color=black, thick] (-0.2, 0) to (-2.5, 0);
\end{tikzpicture}
&= \sqrt{\smash{Z_\vec{p}}\sqrtstrut}\Psi_N\,, \\
\begin{tikzpicture}[baseline={([yshift=-2ex]current bounding box.center)}]
        \draw[shift={(-0.2, -0.2)}, thick] (0, 0) rectangle (0.4, 0.4);
        \draw[color=black, thick] (0.2, 0) to (2.5, 0);
        \draw[color=black, thick, dashed] (0.2, 0.15) to (1.2, 0.7);
\end{tikzpicture}
&= \sqrt{\smash{\tilde{Z}_{\vec p,\vec q}}\sqrtstrut}\frac{i}{2F_\pi} \bar\Psi_N\gamma_5 \sigma^a\,,&
\begin{tikzpicture}[baseline={([yshift=-2ex]current bounding box.center)}]
        \draw[shift={(-0.2, -0.2)}, thick] (0, 0) rectangle (0.4, 0.4);
        \draw[color=black, thick] (-0.2, 0) to (-2.5, 0);
        \draw[color=black, thick, dashed] (-0.2, 0.15) to (-1.2, 0.7);
\end{tikzpicture}
&= \sqrt{\smash{\tilde{Z}_{\vec{p^\prime},\vec q}}\sqrtstrut}\frac{i}{2F_\pi}\gamma_5 \sigma^a \Psi_N \,,
\end{align}%
where one can actually assume $Z_\vec{p}=Z_\vec{p}(\vec p^2)$ and $Z_{\vec p,\vec q}=Z_{\vec p,\vec q}(\vec p^2,\vec p \cdot \vec q, \vec q^2)$ up to lattice artifacts (obviously, the couplings will also depend on the masses, the smearing method and the smearing radii). We will use $Z=Z_\vec{p}$, $Z^\prime=Z_\vec{p^\prime}$, $\tilde Z=\tilde Z_{\vec{p^\prime},\vec{q}}$, and $\tilde Z^\prime=\tilde Z_{\vec{p},\vec{q}}$ as shorthand notations. In the following we always consider protons, i.e., $\bar{\Psi}_p\sigma^3\Psi_p=1$. We will \emph{not} assume
\begin{align}%
  \sqrt{\smash{\tilde{Z}_{\vec p,\vec q}}\sqrtstrut}           &= \sqrt{\smash{Z_{\vec{p^\prime}}}\sqrtstrut} + \text{higher order}\,, &
  \sqrt{\smash{\tilde{Z}_{\vec {p^\prime},\vec  q}}\sqrtstrut} &= \sqrt{\smash{Z_\vec{p}}\sqrtstrut} + \text{higher order}\,,
\end{align}%
which should hold at least approximately for small smearing radii, as discussed above. Instead, we will test the validity of this assumption by comparing it to our data, cf.\ figure~\ref{fig_eft_prediction} in section~\ref{sec_correlation_function_fits}. We complete the setup with the definition of the following energies and four-momenta%
\begin{align}%
  E        &= \sqrt{\vec{p}^2+m^2} \,,&
  E^\prime &= \sqrt{\cramped{\vec{p}^\prime}^2+m^2}\,,&&&
  E_\pi    &= \sqrt{(\vec{p}^\prime-\vec{p})^2+m_\pi^2}\,,\\
  p        &= \begin{pmatrix}E\\\vec{p}\end{pmatrix},&
  p^\prime &= \begin{pmatrix}E^\prime\\\vec{p}^\prime\end{pmatrix},&
  q        &= \begin{pmatrix}E^\prime-E\\\vec{p}^\prime-\vec{p}\end{pmatrix},&
  r_\pm    &= \begin{pmatrix}E_\pi\\\pm(\vec{p}^\prime-\vec{p})\end{pmatrix}. \label{eq_vectors}
\end{align}%
\par
We will now consider one example for each type of diagram in figure~\ref{fig_diagrams} with an axialvector current insertion, starting with the purely nucleonic diagram (in the first row of figure~\ref{fig_diagrams}). Defining the four-vectors $x=(t,\vec x)$, $y=(\tau,\vec y)$ and the energies $E_i=q_i^0$, we obtain
\begin{flalign}%
\begin{split}\MoveEqLeft[1]
  \sqrt{Z^\prime}\sqrt{Z} \!\!\int\!\!d^3x\,e^{-i\mathbf{p}^\prime\cdot\mathbf{x}} \!\!\int\!\!d^3y\,e^{-i(\mathbf{p}-\mathbf{p}^\prime)\cdot\mathbf{y}} S_N(x-y)g_A\gamma^\mu\gamma_5 S_N(y)=\\
  &=-\sqrt{Z^\prime}\sqrt{Z} \!\!\int\!\!\frac{dE_2}{2\pi}e^{-iE_2(t-\tau)} \!\!\int\!\!\frac{dE_1}{2\pi}e^{-iE_1\tau} \frac{(\gamma_0E_2-\bm{\gamma}\cdot\mathbf{p}^\prime+m)g_A\gamma^\mu\gamma_5(\gamma_0E_1-\bm{\gamma}\cdot\mathbf{p}+m)}{(E_2^2-\mathbf{p}^{\prime2}-m^2+i\epsilon)(E_1^2-\mathbf{p}^2-m^2+i\epsilon)}\\
  &=\frac{\sqrt{Z^\prime}\sqrt{Z}}{2E^\prime 2E} e^{-iE^\prime(t-\tau)}e^{-iE\tau} (\slashed{p}^\prime+m)g_A\gamma^\mu\gamma_5(\slashed{p}+m) \,.
\end{split}\raisetag{.5cm}
\end{flalign}%
In the first step, one integrates over the positions which gives delta distributions in momentum space, which in turn eliminate the integrals over the three-momenta from the propagators. Then, we close both integration contours in the lower half of the complex plane and use Cauchy's residue theorem twice. Rotating to imaginary times ($t\rightarrow -it$ and $\tau\rightarrow-i\tau$) one obtains the axial part of eq.~\eqref{eq_3ptgroundstate} to zeroth order accuracy in ChPT, exactly as expected.\par%
Next, we consider the left diagram in the second row of figure~\ref{fig_diagrams}, where the current insertion couples to a pion that directly connects to the sink, while the nucleon propagates directly from source to sink. We find%
\begin{flalign}%
\begin{split}\MoveEqLeft[1]
  \sqrt{\tilde{Z}^\prime}\sqrt{Z} \!\!\int\!\!d^3x\,e^{-i\mathbf{p}^\prime\cdot\mathbf{x}} \!\!\int\!\!d^3y\,e^{-i(\mathbf{p}-\mathbf{p}^\prime)\cdot\mathbf{y}} \biggl(\frac{i}{2F_\pi}\gamma_5\biggr)\biggl(-2F_\pi\frac{\partial}{\partial y_\mu}\biggr)S_\pi(x-y)S_N(x)=\\
  &=- \sqrt{\tilde{Z}^\prime}\sqrt{Z}  \!\!\int\!\!\frac{dE_2}{2\pi}e^{-iE_2(t-\tau)}  \!\!\int\!\!\frac{dE_1}{2\pi}e^{-iE_1t} \frac{\begin{psmallmatrix}E_2\\\vec{q}\end{psmallmatrix}^\mu}{E_2^2-\mathbf{q}^2-m_\pi^2+i\epsilon} \frac{\gamma_5(\gamma_0E_1-\bm{\gamma}\cdot\mathbf{p}+m)}{E_1^2-\mathbf{p}^2-m^2+i\epsilon}\\
  &=+\frac{\sqrt{\tilde{Z}^\prime}\sqrt{Z}}{2E\,2E_\pi}  e^{-iE_\pi(t-\tau)}  e^{-iEt} r_+^\mu \gamma_5(\slashed{p}+m) \,, \label{eq_row2left}
\end{split}& \raisetag{.5cm}
\end{flalign}%
where we have introduced the notation $\begin{psmallmatrix}E_2\\\vec{q}\end{psmallmatrix}^\mu$, etc., to list the components of a $4$-vector. The pion carries the three-momentum~$\vec q$, while the nucleon propagates with momentum~$\vec p$. As in the first diagram, the integrals over the energies can be calculated independently. The diagram yields an $N\pi$ excitation in the final state with the energy $E+E_\pi$. In general this will not be the excited state with the smallest possible energy. For the diagram where the pion propagates from the source to the insertion (cf.\ the right diagram in the second row of figure~\ref{fig_diagrams}) one obtains, carrying out an analogous calculation,
\begin{align}%
-\frac{\sqrt{Z^\prime} \sqrt{\tilde{Z}}}{2E^\prime\,2E_\pi} e^{-iE^\prime t} e^{-iE_\pi\tau} r_-^\mu (\slashed{p}^\prime+m)\gamma_5 \,, \label{eq_row2right}
\end{align}%
which yields an $N\pi$ excitation in the initial state.\par%
Finally, the diagram where the nucleon-pion interaction happens dynamically (the middle diagram in the second row of figure~\ref{fig_diagrams}) gives%
\begin{align}%
\begin{split}\MoveEqLeft[1]
  \sqrt{Z^\prime}\sqrt{Z} \!\!\int\!\!d^3x\,e^{-i\mathbf{p}^\prime\cdot\mathbf{x}} \!\!\int\!\!d^3y\,e^{-i(\mathbf{p}-\mathbf{p}^\prime)\cdot\mathbf{y}} \!\!\int\!\!d^4z \\
 &\quad\times S_N(x-z)\biggl[\biggl(-i\frac{g_A}{2F_\pi}\gamma_\nu\gamma_5\frac{\partial}{\partial z_\nu}\biggr)\biggl(-2F_\pi\frac{\partial}{\partial y_\mu}\biggr)S_\pi(z-y) \biggr]S_N(z)=\\
  &=g_A\sqrt{Z^\prime}\sqrt{Z}  \!\!\int\!\!\frac{dE_2}{2\pi}e^{-iE_2(t-\tau)} \!\!\int\!\!\frac{dE_1}{2\pi}e^{-iE_1\tau}\\
 &\quad\times \frac{\begin{psmallmatrix}E_2-E_1\\\vec{q}\end{psmallmatrix}^\mu \begin{psmallmatrix}E_2-E_1\\\vec{q}\end{psmallmatrix}^\nu}{(E_2-E_1)^2-\mathbf{q}^2-m_\pi^2+i\epsilon} \frac{(\gamma_0E_2-\bm{\gamma}\cdot\mathbf{p}^\prime+m)\gamma_\nu\gamma_5(\gamma_0E_1-\bm{\gamma}\cdot\mathbf{p}+m)}{(E_2^2-\mathbf{p}^{\prime2}-m^2+i\epsilon)(E_1^2-\mathbf{p}^2-m^2+i\epsilon)} \,.
\end{split}&\raisetag{.5cm}
\end{align}%
In this case, where the virtual pion has the three-momentum~$\vec q$ and the energy~$E_2-E_1$, the remaining integrations over $E_1$ and $E_2$ are not independent of each other. We will perform them consecutively starting with $E_1$. Similarly to the procedure for the other diagrams, both integration contours can be closed in the lower half of the complex plane. There, the integrand has two single poles, which collapse to a double pole, if $E_2=E-E_\pi$. The latter case has to be treated separately. The result after the first integration is%
\begin{align}%
 g_A \sqrt{Z^\prime}\sqrt{Z} i \!\!\int\!\!\frac{dE_2}{2\pi} f(E_2) \,,
\end{align}%
where, for $E_2\neq E-E_\pi$,%
\begin{align}%
 f(E_2) &= e^{-iE_2(t-\tau)} e^{-iE\tau}  \begin{psmallmatrix}E_2-E\\\vec{q}\end{psmallmatrix}^\mu \begin{psmallmatrix}E_2-E\\\vec{q}\end{psmallmatrix}^\nu \frac{(\gamma_0E_2-\bm{\gamma}\cdot\mathbf{p}^\prime+m)\gamma_\nu\gamma_5(\slashed{p}+m)}{2E((E_2-E)^2-E_\pi^2+i\epsilon)(E_2^2-E^{\prime2}+i\epsilon)} \notag\\
  &+ e^{-iE_2t} e^{-iE_\pi\tau} \begin{psmallmatrix}-E_\pi\\\vec{q}\end{psmallmatrix}^\mu \begin{psmallmatrix}-E_\pi\\\vec{q}\end{psmallmatrix}^\nu \frac{(\gamma_0E_2-\bm{\gamma}\cdot\mathbf{p}^\prime+m)\gamma_\nu\gamma_5(\gamma_0(E_2+E_\pi)-\bm{\gamma}\cdot\mathbf{p}+m)}{2E_\pi(E_2^2-E^{\prime2}+i\epsilon)((E_2+E_\pi)^2-E^2+i\epsilon)} \,. \label{eq_poleterms}
\end{align}%
For $E_2 = E-E_\pi$, one can check that $f(E_2)$ is finite, which is the only relevant information since it means that there is no pole at this point when using the residue theorem for $E_2$ later on. Thus, one finds that $f(E_2)$ has three poles in the lower half of the complex plane. The first term in eq.~\eqref{eq_poleterms} has two single poles, while the second term in eq.~\eqref{eq_poleterms} has only one single pole. Its second, seeming pole is at $E_2 = E-E_\pi$, where eq.~\eqref{eq_poleterms} is not evaluated. One obtains three contributions that correspond to the diagrams in the bottom row of figure~\ref{fig_diagrams}:%
\begin{align}%
\begin{split}
  &-\frac{g_A\sqrt{Z^\prime}\sqrt{Z}}{2E^\prime\,2E} e^{-iE^\prime(t-\tau)} e^{-iE\tau} q^\mu q^\nu \frac{(\slashed{p}^\prime+m)\gamma_\nu\gamma_5(\slashed{p}+m)}{q^2-m_\pi^2}\\
  &-\frac{g_A\sqrt{Z^\prime}\sqrt{Z}}{2E\,2E_\pi}  e^{-iE_\pi(t-\tau)} e^{-iEt} r_+^\mu r_+^\nu \frac{(\slashed{p}+\slashed{r}_++m)\gamma_\nu\gamma_5(\slashed{p}+m)}{(p+r_+)^2-m^2}\\
  &-\frac{g_A\sqrt{Z^\prime}\sqrt{Z}}{2E^\prime\,2E_\pi} e^{-iE^\prime t} e^{-iE_\pi\tau} r_-^\mu r_-^\nu \frac{(\slashed{p}^\prime+m)\gamma_\nu\gamma_5(\slashed{p}^\prime+\slashed{r}_-+m)}{(p^\prime+r_-)^2-m^2} \,,
\end{split}%
\end{align}%
where we have written the result in terms of the four-vectors defined in eqs.~\eqref{eq_vectors}. The first term yields a contribution to the ground state. It is responsible for the leading, pole dominant contribution to the induced pseudoscalar form factor. The second and the third term contribute to the same $N\pi$ excitations in the final and initial states as those in eqs.~\eqref{eq_row2left} and~\eqref{eq_row2right}, respectively.\par%
This concludes our calculation of the tree-level diagrams shown in figure~\ref{fig_diagrams} for the axialvector current insertion. For the pseudoscalar current the calculation is analogous and we will not repeat it here. By matching the result obtained for the ground state with the usual form factor decompositions (using eq.~\eqref{eq_3ptgroundstate} in combination with eqs.~\eqref{eq_PseudoscalarFF} and~\eqref{eq_AxialvectorFF} after rotating to Euclidean times) one finds%
\begin{align}%
  G_A &= g_A + \text{higher order}\,, \label{eq_GA} \\
  G_\tP &= g_A\frac{4m^2}{Q^2+m_\pi^2} + \text{higher order}\,, \label{eq_GPtilde} \\
  G_P &= g_A\frac{m}{m_\ell}\frac{m_\pi^2}{Q^2+m_\pi^2} + \text{higher order} \,. \label{eq_GP}
\end{align}%
We emphasize that we will not enforce these results for the ground state contribution. In eq.~\eqref{eq_GA} this corresponds to augmenting the axial coupling in the chiral limit to the full axial form factor, which is justified at leading order accuracy. In the same spirit, we have already tacitly used the actual nucleon mass in the propagator instead of its chiral limit value, which is also correct to leading order accuracy in ChPT. It is consistent to perform the same replacement $g_A \mapsto G_A$ in the complete calculation. (We will show that this choice is in much better agreement with the data at nonzero $Q^2$, cf.\ section~\ref{sec_correlation_function_fits} and, in particular, figure~\ref{fig_eft_prediction}.) After doing so, eqs.~\eqref{eq_GPtilde} and~\eqref{eq_GP} yield the PPD assumptions~\cite{Nambu:1960xd,Adler:1965ga} for the (induced) pseudoscalar form factors, as expected.\par%
It turns out to be convenient to define the ratios
\begin{align}
 a &= \frac{\sqrt{\tilde Z}}{\sqrt{Z}} \,, &
 a^\prime &= \frac{\sqrt{\tilde Z^\prime}}{\sqrt{Z^\prime}} \,, \label{eq_NPi_coupling}
\end{align}
where $a=a^\prime=1$ would correspond to the assumption that the smearing does not affect the overlap of the interpolating currents with the $N\pi$ excited states (compared to the ground state). Note that in general $a$ and $a^\prime$ are functions of the momenta. Putting everything together and rotating to Euclidean time ($t\rightarrow -it$ and $\tau\rightarrow-i\tau$) we find
\begin{align}
C_{{\rm 3pt}}^{\vec{p}^{\mathrlap{\prime}}, \vec{p}, \mathcal A^\mu} &=+\frac{\sqrt{Z^\prime}\sqrt{Z}}{2E^\prime\,2E} e^{-E^\prime(t-\tau)} e^{-E\tau} (\slashed{p}^\prime+m)\biggl[G_A\gamma^\mu\gamma_5 + G_\tP\frac{q^\mu}{2m}\gamma_5\biggr](\slashed{p}+m) \notag\\
  &\quad-\frac{\sqrt{Z^\prime}\sqrt{Z}}{2E\,2E_\pi} e^{-(E+E_\pi)(t-\tau)} e^{-E\tau}  r_+^\mu \biggl(b^\prime \gamma_5(\slashed{p}+m) + G_A  \frac{(\slashed{p}+m)\slashed{r}_+\gamma_5(\slashed{p}+m)}{(p+r_+)^2-m^2}\biggr) \notag\\
  &\quad+\frac{\sqrt{Z^\prime}\sqrt{Z}}{2E^\prime\,2E_\pi} e^{-E^\prime (t-\tau)} e^{-(E^\prime + E_\pi)\tau} r_-^\mu \biggl(b \, (\slashed{p}^\prime+m)\gamma_5-G_A \frac{(\slashed{p}^\prime+m)\slashed{r}_-\gamma_5(\slashed{p}^\prime+m)}{(p^\prime+r_-)^2-m^2}\biggr) \notag\\&\quad+   \dots \,, \label{eq_res_A_notrace}\\
C_{{\rm 3pt}}^{\vec{p}^{\mathrlap{\prime}}, \vec{p}, \mathcal P} &=+\frac{\sqrt{Z^\prime}\sqrt{Z}}{2E^\prime\,2E} e^{-E^\prime(t-\tau)} e^{-E\tau}  (\slashed{p}^\prime+m) G_P\gamma_5(\slashed{p}+m) \notag\\
  &\quad-\frac{\sqrt{Z^\prime}\sqrt{Z}}{2E\,2E_\pi} e^{-(E+E_\pi)(t-\tau)} e^{-E\tau} B \biggl(b^\prime \gamma_5(\slashed{p}+m) + G_A \frac{(\slashed{p}+m)\slashed{r}_+\gamma_5(\slashed{p}+m)}{(p+r_+)^2-m^2}\biggr) \notag\\
  &\quad-\frac{\sqrt{Z^\prime}\sqrt{Z}}{2E^\prime\,2E_\pi} e^{-E^\prime (t-\tau)} e^{-(E^\prime + E_\pi)\tau}B \biggl(b \, (\slashed{p}^\prime+m)\gamma_5 - G_A \frac{(\slashed{p}^\prime+m)\slashed{r}_-\gamma_5(\slashed{p}^\prime+m)}{(p^\prime+r_-)^2-m^2}\biggr) \notag\\&\quad+ \dots \,, \label{eq_res_P_notrace}
\end{align}
where
\begin{align}
 b        &= -a       +G_A\frac{m_\pi^2}{(p^\prime+r_-)^2-m^2} \,, &
 b^\prime &= -a^\prime+G_A\frac{m_\pi^2}{(p+r_+)^2-m^2} \,,
\end{align}
and the dots represent additional excited state contributions. These results can be used for all momentum configurations and with arbitrary spin projections. After taking the trace with the specific matrices $P_+^i$ that we use here, the result can be further simplified, see below. We emphasize that the leading, pole enhanced $N\pi$ excited state contribution calculated here occurs either in the initial state or in the final state, but not in both simultaneously.
\subsection{Spectral decomposition\label{sec_final_param}}
In this section we will provide the explicit expressions for the correlation functions that are used in our analysis, including our parametrization of additional generic excited states. For the latter we will assume that they occur with the same energies in both, two- and three-point functions. State-of-the-art lattice analyses of form factors take into account up to three excited states in the two-point and up to two excited states in the three-point functions, see, e.g., ref.~\cite{Jang:2019jkn}. Whether this is necessary depends on the available statistics and on the applied source/sink smearing. In our simulation a relatively large number of smearing steps was performed, leading to large smearing radii, cf.\ table~\ref{tab_ensembles}. In this situation, we find it sufficient to add only one generic excited state to the two- and three-point correlators on top of the pion pole enhanced state that we have calculated in the last section. Including the additional generic excited state term, we obtain for the two-point function
\begin{align}
 C_{{\rm 2pt},P_+}^{\vec{p}} (t) &= Z_{\vec{p}}  \frac{E_{\vec{p}} + m}{E_{\vec{p}}} e^{-E_{\vec{p}}t} \bigl( 1 + A_{\vec p} e^{-\Delta E_{\vec{p}} t} \bigr)  \,. \label{eq_2pt_spectral_decomposition}
\end{align}
In the following we will abbreviate $\Delta E = \Delta E_{\vec{p}}$ and $\Delta E^\prime = \Delta E_{\vec{p^\prime}}$. Note that we do not assume any dispersion relation for the excited state energies, nor do we assume that these are single hadron states. Instead, we treat them as free fit parameters. We define the trace occurring in the ground state contribution to the three-point function as
\begin{align} \label{eq_gs_trace}
 B_{\Gamma, \mathcal{O}}^{\vec{p}^{\mathrlap{\prime}}, \vec{p}} &= \operatorname{Tr}\bigl\{\Gamma(\slashed{p}^\prime+m)J[\mathcal O](\slashed{p}+m)\bigr\} \,.
\end{align}
The explicit results can be found in appendix~\ref{app_traces}, together with the remaining traces needed to evaluate eqs.~\eqref{eq_res_A_notrace} and~\eqref{eq_res_P_notrace}. For the three-point functions we obtain the parametrization
\begin{align} 
\begin{split}
C_{{\rm 3pt},P_+^i}^{\vec{p}^{\mathrlap{\prime}}, \vec{p}, \mathcal A^\mu} &=\frac{\sqrt{Z^\prime}\sqrt{Z}}{2E^\prime\,2E} e^{-E^\prime(t-\tau)} e^{-E\tau} \\
 &\quad\times \biggl[ \begin{aligned}[t] & B_{P_+^i,\mathcal A^\mu}^{\vec{p}^{\mathrlap{\prime}}, \vec{p}} \biggl( 1 + B_{10} e^{-\Delta E^\prime (t-\tau)} +  B_{01} e^{-\Delta E \tau} + B_{11} e^{-\Delta E^\prime (t-\tau)}  e^{-\Delta E \tau} \biggr) \\
  &+e^{-\Delta E_{N\pi}^\prime (t-\tau)}\frac{E^\prime}{E_\pi} r_+^\mu \biggl(c^\prime p^i + d^\prime q^i\biggr)+e^{-\Delta E_{N\pi} \tau}\frac{E}{E_\pi} r_-^\mu \biggl(c \, p^{\prime i} + d \, q^i\biggr) \biggr] \,, \taghere \end{aligned}  \label{eq_res_A_noChPT} \end{split} \\
\begin{split}
C_{{\rm 3pt},P_+^i}^{\vec{p}^{\mathrlap{\prime}}, \vec{p}, \mathcal P} &=\frac{\sqrt{Z^\prime}\sqrt{Z}}{2E^\prime\,2E} e^{-E^\prime(t-\tau)} e^{-E\tau} \\
 &\quad\times \biggl[ \begin{aligned}[t] & B_{P_+^i,\mathcal P}^{\vec{p}^{\mathrlap{\prime}}, \vec{p}} \biggl( 1 + B_{10} e^{-\Delta E^\prime (t-\tau)} +  B_{01} e^{-\Delta E \tau} + B_{11} e^{-\Delta E^\prime (t-\tau)}  e^{-\Delta E \tau} \biggr) \\
&+e^{-\Delta E_{N\pi}^\prime (t-\tau)}\frac{E^\prime}{E_\pi} \frac{m_\pi^2}{2m_\ell} \biggl(c^\prime p^i + d^\prime q^i\biggr)-e^{-\Delta E_{N\pi} \tau}\frac{E}{E_\pi}  \frac{m_\pi^2}{2m_\ell} \biggl(c \, p^{\prime i} + d \, q^i\biggr) \biggr] \,,  \end{aligned}  \label{eq_res_P_noChPT} \end{split}
\end{align}
where we have suppressed the dependence of the excited state parameters on the momenta, the spin-projection and the current insertion: $B_{ij}=B_{ij}(\vec{p^\prime},\vec{p},\Gamma,\mathcal O)$. We have defined $\Delta E_{N\pi} = E_\pi + (E^\prime - E) $, $\Delta E_{N\pi}^\prime = E_\pi - (E^\prime - E)$ and%
\begin{align}
  c        &= -2b-4G_A\frac{mE_\pi+p^\prime \cdot r_-}{(p^\prime+r_-)^2-m^2} \,, &
  c^\prime &= -2b^\prime-4G_A\frac{mE_\pi+p \cdot r_+}{(p+r_+)^2-m^2} \,, \label{eq_c}\\
  d &= -G_A \frac{4m(m+E^\prime)}{(p^\prime+r_-)^2-m^2}\,, & 
  d^\prime &= G_A \frac{4m(m+E)}{(p+r_+)^2-m^2} \,. \label{eq_d}
\end{align}\par%
Equations~\eqref{eq_c} and~\eqref{eq_d} are only valid up to higher order corrections in ChPT. For instance, one could replace $G_A$ by \mbox{$(Q^2+m_\pi^2)G_\tP/(4m^2)$} or by \mbox{$(Q^2+m_\pi^2) m_\ell G_P/(m m_\pi^2)$} in the $N\pi$ excited state contributions (cf.\ eqs.~\eqref{eq_GA}, \eqref{eq_GPtilde} and~\eqref{eq_GP}) and the result would still be valid at leading order. From a plain vanilla ChPT power-counting point of view one could even replace $G_A$ by $g_A$. Therefore, in anticipation of possible higher order corrections, we may relax the assumptions even further by using $c$, $c^\prime$, $d$, and $d^\prime$ as free fit parameters, which reduces the ChPT input. This has the additional advantage, that it does not allow the excited state signal to have a direct influence on the result for the ground state form factors. Naturally, one has to pay for the increased number of fit parameters with a slightly larger statistical error for the ground state result -- a small price considering that one gets rid of one source of systematic uncertainty. In section~\ref{sec_correlation_function_fits} we will assess the validity of the ChPT predictions by comparing them to the results obtained from the fits. In particular we will be able to check whether the data is consistent with the parameter-free ChPT prediction for $d$ and whether the direct coupling of the smeared three-quark interpolating currents to the $N\pi$ state differs from the leading order ChPT prediction calculated for local currents.\par%
Note the elegance of the parametrization given in eqs.~\eqref{eq_res_A_noChPT} and~\eqref{eq_res_P_noChPT}. Even after relaxing the conditions~\eqref{eq_c} and~\eqref{eq_d}, it encodes the relative strength of the $N\pi$ excited state contribution in the different channels. The importance of this knowledge must not be underestimated. For instance, combining eq.~\eqref{eq_gs_trace_A} with eq.~\eqref{eq_res_A_noChPT} one can see that any determination of the axial form factor using solely the $\mathcal A_1$, $\mathcal A_2$, and $\mathcal A_3$ channels is not affected by these excited states at all.\par%
Finally, let us note that for the kinematics we use in the numerical analysis, setting the final state momentum to zero, $\vec{p^\prime} = \vec 0$, such that $\vec p = - \vec q$ (this setup is used in many lattice simulations), the parametrization becomes even simpler since one can replace $c^\prime p^i + d^\prime q^i=e^\prime q^i$ (with $e^\prime=d^\prime-c^\prime$) and $c \, p^{\prime i} + d \, q^i=d q^i$. In this kinematic situation, the $N\pi$ excited state energy corresponds to $E_N(\mathbf 0) + E_\pi(-\mathbf q)$ in the initial state, and $E_N(\mathbf p) + E_\pi(\mathbf q)$ in the final state.
\section{Data analysis\label{sec_data_analysis}}
\subsection{Lattice setup\label{sec_lattice_setup}}
In order to determine the axial and (induced) pseudoscalar form factors using the correlation functions described in section~\ref{sec_correlation_functions}, we have analyzed a large set of lattice ensembles generated within the CLS effort~\cite{Bruno:2014jqa}.\footnote{The ensembles rqcd021, and rqcd030 have been generated using the BQCD code~\cite{Nakamura:2010qh}.} The ensembles have been generated using a tree-level Symanzik improved gauge action and $N_f=2+1$ flavors of nonperturbatively order $a$ improved Wilson (clover) fermions. An efficient and stable hybrid Monte Carlo sampling is achieved by applying twisted-mass determinant reweighting~\cite{Luscher:2012av}, which avoids near-zero modes of the Wilson Dirac operator. The polynomial approximation of the strange quark determinant was corrected for by reweighting too, employing the method introduced in ref.~\cite{Mohler:2020txx}. The individual quarks in the nucleon interpolators at the source and the sink are Wupper\-tal-smeared~\cite{Gusken:1989qx}, employing spatially APE-smoothed~\cite{Falcioni:1984ei} gauge links. The corresponding smearing radii $r_{\rm sm}$ are defined via%
\begin{align} \label{eq_smearing_radii}
 r_{\rm sm}^2 &= \sum\limits_{\mathclap{n_x,n_y,n_z=-N_s/2}}^{N_s/2-1} \Psi^\dagger(\mathbf n a) \mathbf n^2 a^2 \Psi(\mathbf n a) \,,&
 \sum\limits_{\mathclap{n_x,n_y,n_z=-N_s/2}}^{N_s/2-1} \Psi^\dagger(\mathbf n a) \Psi(\mathbf n a) &= 1 \,,
\end{align}%
where $\Psi$ is the normalized smearing function.\par%
Some of the CLS ensembles (cf.\ table~\ref{tab_ensembles} for a full list of the ensembles used in this work) have been simulated employing very fine lattices down to $a=\unit{0.039}{\femto\meter}$. For these lattices we avoid large autocorrelation times by using open boundary conditions in the time direction~\cite{Luscher:2012av,Luscher:2011kk}. The latter allow the topological charge to flow into and out of the simulation volume through the temporal boundaries and thus topological freezing is avoided. While employing open boundary conditions is crucial for fine lattice spacings, we use lattices with both open and periodic boundary conditions for the coarser spacings. In total we have five different lattice spacings ranging from~$a=\unit{0.039}{\femto\meter}$ to~$a=\unit{0.086}{\femto\meter}$, see table~\ref{tab_spacings}.\par%
\begin{table}[tb]%
\centering
\caption{\label{tab_spacings}Lattice spacings~$a$, corresponding to the five different inverse couplings~$\beta$ used in this study. The lattice spacings have been obtained by determining the Wilson flow time at the \SU3 symmetric point in lattice units~$t_0^*/a^2$ and equating $t_0^*$ with the result $\mu_{\mathrm{ref}}^*=(8t_0^*)^{-1/2}\approx\unit{478}{\mega\electronvolt}$ of ref.~\cite{Bruno:2017gxd}.}%
\begin{widetable}{\textwidth}{lccccc}%
\toprule
$\beta$ & $3.40$ & $3.46$ & $3.55$ & $3.70$ & $3.85$\\
\midrule
$a\,[\femto\meter]$ & $0.086$ & $0.076$ & $0.064$ & $0.050$ & $0.039$\\
\bottomrule
\end{widetable}%
\end{table}%
\begin{figure}[t]%
\centering%
\includegraphics[width=.9\textwidth]{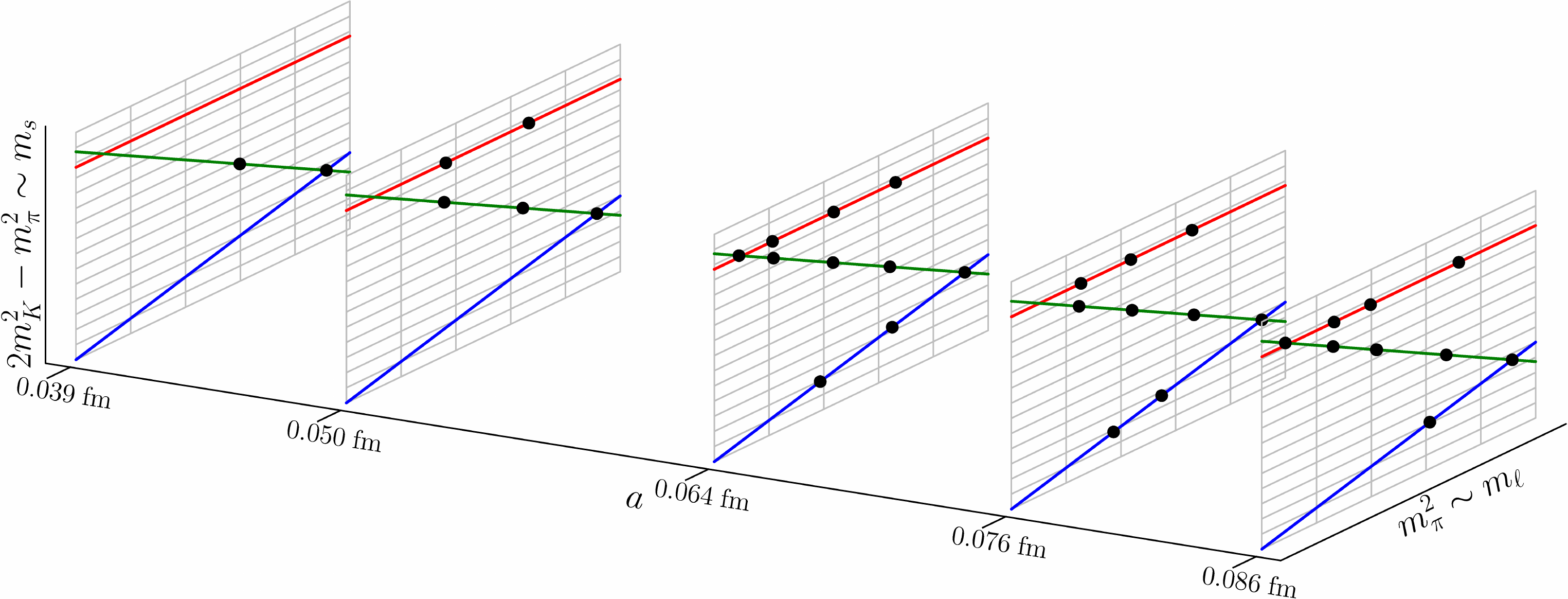}%
\caption{\label{fig_ensembles}Schematic visualization of the analyzed CLS ensembles in the space spanned by the lattice spacing and quark masses. On the flavor symmetric plane (blue), where $m_\ell=m_s$, flavor multiplets of hadrons have degenerate masses (e.g., $m_K^2=m_\pi^2$ and $m_N=m_\Sigma=m_\Xi=m_\Lambda$). The green lines are defined to have physical average quadratic meson mass ($2 m_K^2+m_\pi^2=\text{phys.}$). This corresponds to an approximately physical mean quark mass ($2m_\ell+m_s\approx\text{phys.}$). The red lines are defined by $2 m_K^2-m_\pi^2=\text{phys.}$ and indicate an almost physical strange quark mass ($m_s\approx\text{phys.}$). Physical masses are reached at the intersections of green and red lines.}%
\end{figure}%
As illustrated in figure~\ref{fig_ensembles}, the available ensembles have been generated along three different trajectories in the quark mass plane:\footnote{See also ref.~\cite{Bali:2016umi}. In practice the ensembles do not always lie exactly on top of the green and red trajectories shown in figure~\ref{fig_ensembles}.}
\begin{enumerate}[a)]
 \item sym, blue: trajectory with exact flavor symmetry, where the light and strange quark masses are degenerate ($m_\ell=m_s$)
 \item trM, green: ensembles created with $2m_\ell + m_s = \text{const.}$, such that $2 m_K^2+m_\pi^2\approx\text{phys.}$
 \item msc, red: ensembles created keeping the renormalized strange quark mass constant~\cite{Bali:2016umi}, so that $2 m_K^2-m_\pi^2\approx\text{phys.}$
\end{enumerate}%
Along trajectory a) observables do not depend on the quark mass splitting. Data from these ensembles thus enables a precise determination of the dependence on the average quark mass, and can also be used to obtain results in the three-flavor chiral limit. Trajectory b), where the average quark mass is kept approximately constant, yields complementary information on flavor symmetry breaking. The additional data along trajectory c) provides further insight into the dependence on the light quark mass. The physical point is close to the intersection of the latter two trajectories. Since we cover a large fraction of the relevant quark mass plane, any deviation of an ensemble from its target trajectory can be taken into account.\par%
The ensembles cover a range of volumes with $3.5\leq m_\pi L\leq6.4$ allowing us to investigate and control finite volume effects. The majority of the ensembles has $m_\pi L>4$. Having multiple quark mass trajectories with a wide range of lattice spacings and volumes enables us to simultaneously extrapolate to physical masses, to infinite volume, and to the continuum limit by means of a global fit to 37~ensembles. Our extrapolation strategy is explained in detail in section~\ref{sec_parametrization}.\par%
\afterpage{%
\begin{landscape}%
\setlength{\tabcolsep}{4.5pt}%
\begin{longtable}{rccrcllccccrcc}%
\captionsetup{width=1.485\textwidth}
\caption{\label{tab_ensembles}List of the ensembles used in this work, labeled by their identifier and sorted by the inverse coupling $\beta$ and the pion masses. We specify the geometries $N_t \times N_s^3$ as well as the boundary condition in time (periodic~(p) or open~(o)). The light and strange hopping parameters $\kappa_\ell$ and $\kappa_s$ used in the simulation and the resulting approximate meson masses are given in $\mega\electronvolt$. We also provide the root mean squared smearing radii $r_{\rm sm}$ for the light quark sources in $\femto\meter$ defined in eq.~\eqref{eq_smearing_radii}. $\#\text{conf.}$ gives the number of configurations analyzed. The column $t/a$ lists the source-sink distances in lattice units that have been analyzed on this lattice. The subscript \#meas.\ specifies how many measurements have been performed for the respective source-sink distance. In physical units these distances roughly correspond to $\unit{0.7}{\femto\meter}$, $\unit{0.9}{\femto\meter}$, $\unit{1.0}{\femto\meter}$, and $\unit{1.2}{\femto\meter}$. The last column specifies on which trajectories in the quark mass plane the ensemble lies, cf.\ figure~\ref{fig_ensembles}. An in-depth description of the ensemble generation can be found in ref.~\cite{Bruno:2014jqa}. Note that ensemble D201 was only used for the test with nonzero final momentum shown in figure~\ref{fig_finite_final_momentum_test}.}\\%
\toprule
\multicolumn{1}{c}{Ens.} & $\beta$ & $N_s$ & \multicolumn{1}{c}{$N_t$} & bc & \multicolumn{1}{c}{$\kappa_\ell$} & \multicolumn{1}{c}{$\kappa_s$} & $m_\pi$ & $m_K$ & $m_\pi L$ & $r_{\text{sm}}$ & \multicolumn{1}{c}{\!\!\!\#conf.} & \multicolumn{1}{c}{$t/a {}_{\text{\#meas.}}$} & traj.\\
\midrule
           U103 &  3.4 & 24 & 128 & o &        0.13675962 &        0.13675962 & 417 & 417 & 4.4 & 0.638 & 2473 & $8_{1}$, $10_{2}$, $12_{3}$, $14_{4}$ & trm, sym \\
           H101 &  3.4 & 32 &  96 & o &        0.13675962 &        0.13675962 & 420 & 420 & 5.9 & 0.643 & 2000 & $8_{2}$, $10_{2}$, $12_{2}$, $14_{2}$ & trm, sym \\
           H102 &  3.4 & 32 &  96 & o &          0.136865 &       0.136549339 & 352 & 439 & 4.9 & 0.669 & 1997 & $8_{1}$, $10_{2}$, $12_{3}$, $14_{4}$ & trm \\
           H105 &  3.4 & 32 &  96 & o &           0.13697 &        0.13634079 & 279 & 465 & 3.9 & 0.735 & 1996 & $8_{1}$, $10_{2}$, $12_{3}$, $14_{4}$ & trm \\
           N101 &  3.4 & 48 & 128 & o &           0.13697 &        0.13634079 & 279 & 463 & 5.8 & 0.722 &  320 & $8_{1}$, $10_{2}$, $12_{3}$, $14_{4}$ & trm \\
           C101 &  3.4 & 48 &  96 & o &           0.13703 &       0.136222041 & 220 & 472 & 4.6 & 0.772 & 2343 & $8_{1}$, $10_{2}$, $12_{3}$, $14_{4}$ & trm \\
           D101 &  3.4 & 64 & 128 & o &           0.13703 &       0.136222041 & 220 & 473 & 6.1 & 0.799 &  323 & $8_{1}$, $10_{2}$, $12_{3}$, $14_{4}$ & trm \\
           D150 &  3.4 & 64 & 128 & p &          0.137088 &        0.13610755 & 126 & 479 & 3.5 & 0.844 &  579 & $8_{1}$, $10_{2}$, $12_{3}$, $14_{4}$ & trm, msc \\
           H107 &  3.4 & 32 &  96 & o &  0.13694566590798 & 0.136203165143476 & 366 & 546 & 5.1 & 0.673 & 1564 & $8_{2}$, $10_{2}$, $12_{3}$, $14_{4}$ & msc \\
           H106 &  3.4 & 32 &  96 & o &    0.137015570024 &    0.136148704478 & 272 & 516 & 3.8 & 0.680 & 1553 & $8_{2}$, $10_{2}$, $12_{3}$, $14_{4}$ & msc \\
           C102 &  3.4 & 48 &  96 & o &           0.13703 &       0.136222041 & 222 & 501 & 4.6 & 0.779 & 1500 & $8_{2}$, $10_{2}$, $12_{3}$, $14_{4}$ & msc \\
\!\!\!\!rqcd021 &  3.4 & 32 &  32 & p &          0.136813 &          0.136813 & 338 & 338 & 4.7 & 0.676 & 1541 & $8_{2}$, $10_{2}$, $12_{4}$, $14_{4}$ & sym \\
\midrule
           B450 & 3.46 & 32 &  64 & p &           0.13689 &           0.13689 & 418 & 418 & 5.2 & 0.617 & 1594 & $9_{1}$, $11_{2}$, $13_{3}$, $16_{4}$ & trm, sym \\
           S400 & 3.46 & 32 & 128 & o &          0.136984 &       0.136702387 & 352 & 442 & 4.3 & 0.665 & 2872 & $9_{1}$, $11_{2}$, $13_{3}$, $16_{4}$ & trm \\
           N401 & 3.46 & 48 & 128 & o &         0.1370616 &      0.1365480771 & 285 & 461 & 5.3 & 0.721 & 1100 & $9_{1}$, $11_{2}$, $13_{3}$, $16_{4}$ & trm \\
           D450 & 3.46 & 64 & 128 & p &          0.137126 & 0.136420428639937 & 214 & 477 & 5.3 & 0.784 &  620 & $9_{4}$, $11_{4}$, $13_{4}$, $16_{4}$ & trm \\
\multicolumn{13}{c}{Continued on next page} \\ \pagebreak \toprule
\multicolumn{1}{c}{Ens.} & $\beta$ & $N_s$ & \multicolumn{1}{c}{$N_t$} & bc & \multicolumn{1}{c}{$\kappa_\ell$} & \multicolumn{1}{c}{$\kappa_s$} & $m_\pi$ & $m_K$ & $m_\pi L$ & $r_{\text{sm}}$ & \multicolumn{1}{c}{\!\!\!\#conf.} & \multicolumn{1}{c}{$t/a {}_{\text{\#meas.}}$} & traj.\\
\midrule
           B452 & 3.46 & 32 &  64 & p &         0.1370455 &       0.136378044 & 350 & 545 & 4.3 & 0.650 & 1944 & $9_{3}$, $11_{3}$, $13_{3}$, $16_{4}$ & msc \\
           N450 & 3.46 & 48 & 128 & p &         0.1370986 &       0.136352601 & 285 & 524 & 5.3 & 0.706 & 1132 & $9_{4}$, $11_{4}$, $13_{4}$, $16_{4}$ & msc \\
           D451 & 3.46 & 64 & 128 & p &           0.13714 &       0.136337761 & 217 & 503 & 5.4 & 0.784 &  532 & $9_{4}$, $11_{4}$, $13_{4}$, $16_{4}$ & msc \\
\!\!\!\!rqcd030 & 3.46 & 32 &  64 & p &         0.1369587 &         0.1369587 & 317 & 317 & 3.9 & 0.688 & 1224 & $9_{4}$, $11_{4}$, $13_{8}$, $16_{8}$ & sym \\
           X450 & 3.46 & 48 &  64 & p &          0.136994 &          0.136994 & 263 & 263 & 4.9 & 0.739 &  400 & $9_{2}$, $11_{2}$, $13_{4}$, $16_{4}$ & sym \\
\midrule
           N202 & 3.55 & 48 & 128 & o &             0.137 &             0.137 & 411 & 411 & 6.4 & 0.610 &  884 & $11_{1}$, $14_{2}$, $16_{2}$, $19_{4}$ & trm, sym \\
           N203 & 3.55 & 48 & 128 & o &           0.13708 &       0.136840284 & 345 & 442 & 5.4 & 0.660 & 1543 & $11_{1}$, $14_{2}$, $16_{3}$, $19_{4}$ & trm \\
           N200 & 3.55 & 48 & 128 & o &           0.13714 &        0.13672086 & 284 & 462 & 4.4 & 0.696 & 1712 & $11_{1}$, $14_{2}$, $16_{3}$, $19_{4}$ & trm \\
           D200 & 3.55 & 64 & 128 & o &            0.1372 &       0.136601748 & 201 & 481 & 4.2 & 0.786 & 1999 & $11_{1}$, $14_{2}$, $16_{3}$, $19_{4}$ & trm \\
           E250 & 3.55 & 96 & 192 & p &       0.137232867 &       0.136536633 & 130 & 489 & 4.1 & 0.829 &  490 & $11_{4}$, $14_{4}$, $16_{4}$, $19_{4}$ & trm, msc \\
           N204 & 3.55 & 48 & 128 & o &          0.137112 &       0.136575049 & 351 & 545 & 5.5 & 0.661 & 1500 & $11_{2}$, $14_{2}$, $16_{3}$, $19_{4}$ & msc \\
           N201 & 3.55 & 48 & 128 & o &        0.13715968 &       0.136561319 & 285 & 523 & 4.5 & 0.727 & 1522 & $11_{2}$, $14_{2}$, $16_{3}$, $19_{4}$ & msc \\
           D201 & 3.55 & 64 & 128 & o &         0.1372067 &       0.136546844 & 199 & 501 & 4.1 & 0.778 & 1078 & $11_{4}$, $14_{4}$, $16_{4}$, $19_{4}$ & msc \\
           X250 & 3.55 & 48 &  64 & p &           0.13705 &           0.13705 & 348 & 348 & 5.4 & 0.655 &  345 & $11_{2}$, $14_{2}$, $16_{4}$, $19_{4}$ & sym \\
           X251 & 3.55 & 48 &  64 & p &            0.1371 &            0.1371 & 267 & 267 & 4.2 & 0.719 &  436 & $11_{4}$, $14_{4}$, $16_{8}$, $19_{8}$ & sym \\
\midrule
           N300 &  3.7 & 48 & 128 & o &             0.137 &             0.137 & 422 & 422 & 5.1 & 0.591 &  760 & $14_{1}$, $17_{2}$, $21_{2}$, $24_{4}$ & trm, sym \\
           N302 &  3.7 & 48 & 128 & o &          0.137064 &   0.1368721791358 & 346 & 451 & 4.2 & 0.644 & 1383 & $14_{1}$, $17_{2}$, $21_{3}$, $24_{4}$ & trm \\
           J303 &  3.7 & 64 & 192 & o &          0.137123 &      0.1367546608 & 257 & 475 & 4.2 & 0.705 &  634 & $14_{1}$, $17_{2}$, $21_{6}$, $24_{8}$ & trm \\
           N304 &  3.7 & 48 & 128 & o & 0.137079325093654 & 0.136665430105663 & 351 & 554 & 4.3 & 0.620 & 1652 & $14_{2}$, $17_{2}$, $21_{3}$, $24_{4}$ & msc \\
           J304 &  3.7 & 64 & 192 & o &          0.137123 &      0.1367546608 & 260 & 523 & 4.2 & 0.708 & 1525 & $14_{3}$, $17_{3}$, $21_{3}$, $24_{4}$ & msc \\
\midrule
           J500 & 3.85 & 64 & 192 & o &          0.136852 &          0.136852 & 410 & 410 & 5.2 & 0.579 &  750 & $17_{1}$, $22_{2}$, $27_{3}$, $32_{4}$ & trm, sym \\
           J501 & 3.85 & 64 & 192 & o &         0.1369032 &       0.136749715 & 333 & 445 & 4.2 & 0.613 & 1507 & $17_{1}$, $22_{2}$, $27_{3}$, $32_{4}$ & trm \\
\bottomrule
\end{longtable}\vspace{-0.2cm}\setlength{\tabcolsep}{6pt}
\end{landscape}}
The local axial and pseudoscalar currents in our calculation have to be renormalized. We use the renormalization factors $Z_A$ from ref.~\cite{DallaBrida:2018tpn} (as recommended in this reference, we use the values $Z^l_{A,\rm sub}$ from their table~7), which have been determined using a new method based on the chirally rotated Schr\"odinger functional~\cite{Sint:2010eh}. In addition, we use the nonperturbative quark mass-dependent order
$a$ improvement coefficients described in ref.~\cite{Korcyl:2016ugy} (but with updated values from ref.~\cite{Korcyl:2019}). The isovector currents are multiplicatively renormalized using%
\begin{align}
\mathcal{A}^{\text{ren}} &= Z_A^{}(\beta)\Big[1+2am_\ell^{\text{bare}} b_A(\beta) + 2a(2m_\ell^{\text{bare}} + m_s^{\text{bare}})\tilde b_A(\beta)\Big] \mathcal{A}^{\text{imp}}\,, \\
m_\ell^{\text{ren}}\; \mathcal{P}^{\text{ren}} &= Z_A^{}(\beta)\Big[1+2am_\ell^{\text{bare}} b_A(\beta) + 2a(2m_\ell^{\text{bare}} + m_s^{\text{bare}})\tilde b_A(\beta)\Big] m_\ell^{\text{imp}}\; \mathcal{P}^{\text{imp}}\,,
\end{align}%
where $ m_\ell^{\text{imp}}$ is the PCAC light quark mass obtained from improved currents,
\begin{align}
m_\ell^{\text{imp}} &= \frac{ \la 0 | \partial^\mu\!\mathcal A_\mu^{\text{imp}} | \pi \ra }{2i \la 0 | \mathcal P^{\text{imp}} | \pi \ra}\,.
\end{align}
The bare quark mass%
\begin{align}
m_q^{\text{bare}} &= \frac{1}{2a} \Bigl( \frac{1}{\kappa_q} - \frac{1}{\kappa_{\rm crit}} \Bigr)
\end{align}
is calculated using the hopping parameter $\kappa_q$ (cf.\ table~\ref{tab_ensembles}) and its critical value $\kappa_{\rm crit}$~\cite{Bali:2023}. We exploit the fact that the product of quark mass and pseudoscalar current renormalizes in exactly the same way as the axialvector current. $\tilde b_A$ has been found to be zero within errors smaller than $0.1$~\cite{Korcyl:2019}. This corresponds to shifts of at most $4\permil$, depending on the ensemble, that decrease towards the continuum limit. We neglect this effect, which is small compared to the other sources of error, and proceed with continuum limit extrapolations that are quadratic in $a$. Within the Symanzik improvement program~\cite{Symanzik:1983dc,Symanzik:1983gh} also the currents themselves have to be $\mathcal O(a)$-improved. For the axialvector current this yields%
\begin{align}
\mathcal A_\mu^{\smash{\text{imp}}} = \mathcal A_\mu^{} + c_A a\partial_\mu \mathcal P \,,
\end{align}%
where we use the improvement coefficient $c_A$, nonperturbatively determined in ref.~\cite{Bulava:2015bxa}, and $\partial_\mu$ denotes the symmetrically discretized derivative. For the pseudoscalar current $\mathcal P^{\text{imp}}= \mathcal P$.\par%
\subsection{Fits to the correlation functions\label{sec_correlation_function_fits}}
To calculate the three-point functions~\eqref{eq_3ptdef} one has to evaluate all possible contractions. Disconnected diagrams do not contribute in our case, since we only consider isovector currents. The connected diagrams can be evaluated using sequential sources~\cite{Maiani:1987by}. Since each sink momentum requires new inversions, we restrict the numerical analysis to the case in which the final state three-momentum is set to zero ($\vec{p^\prime}=\vec{0}$). Note, however, that the parametrizations provided in section~\ref{sec_final_param} are applicable to all possible kinematic situations.\par%
On each ensemble we have analyzed $4$ source-sink separations that have been chosen such that they correspond roughly to the physical distances $\unit{0.7}{\femto\meter}$, $\unit{0.9}{\femto\meter}$, $\unit{1.0}{\femto\meter}$, and $\unit{1.2}{\femto\meter}$. The source-sink distance in lattice units and the corresponding number of measurement per configuration are specified in table~\ref{tab_ensembles}. On some ensembles we have reduced the computational cost by applying the coherent sink technique \cite{Bratt:2010jn}, where one inverts on multiple, temporally separated sequential sources simultaneously. For the statistical analysis we generate $500$~bootstrap samples per ensemble using a bin size of $20$~molecular dynamics units to eliminate autocorrelations.\par%
\begin{figure}[ptb]
\centering
\includegraphics[width=1.0\textwidth]{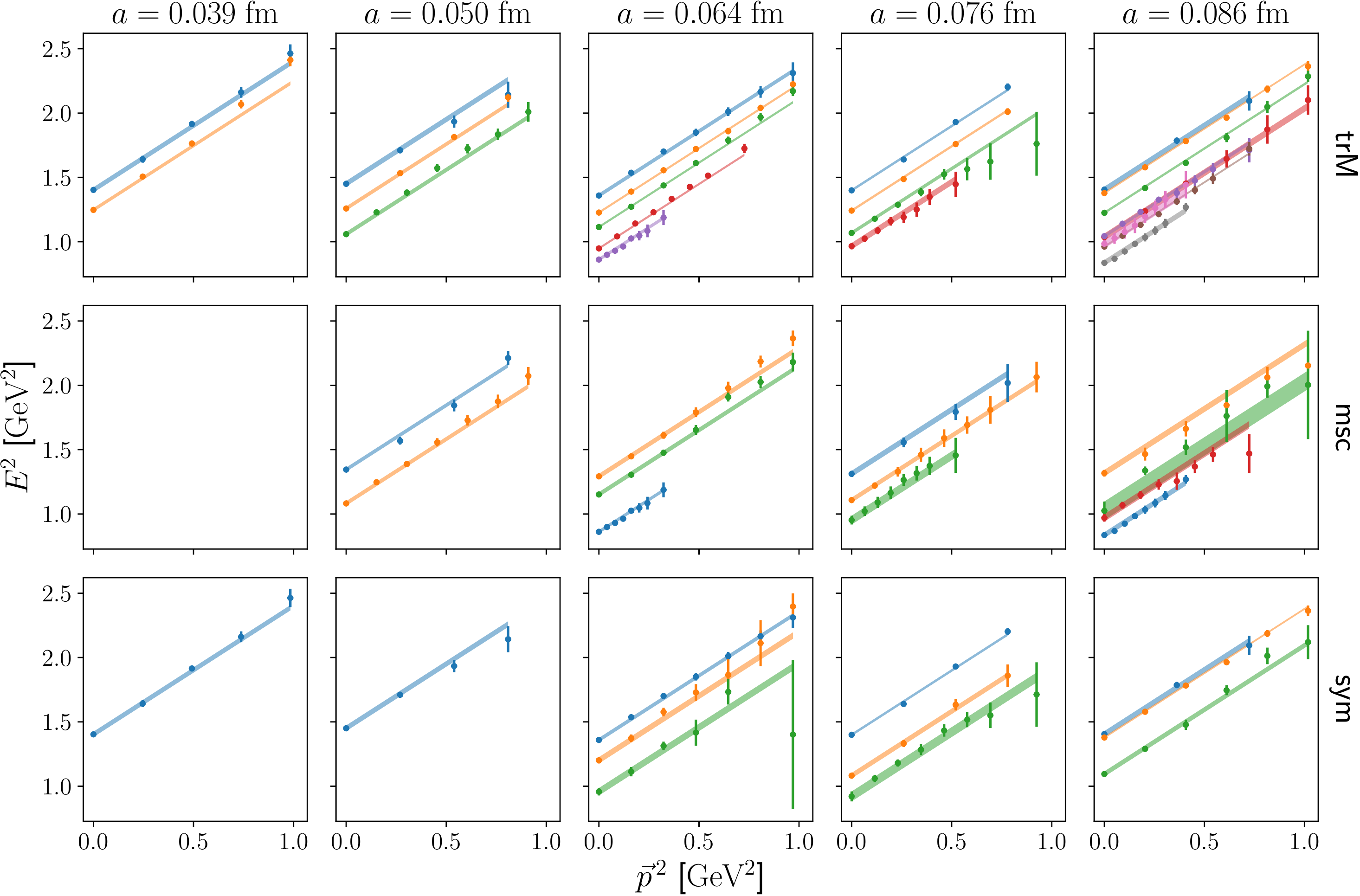}
\caption{\label{fig_dispersion_relation}Nucleon dispersion relation for the ensembles listed in table~\ref{tab_ensembles}. The data points show the squared ground state energies obtained from fits to two-point functions using the ansatz~\eqref{eq_2pt_spectral_decomposition} and treating the energies as free fit parameters. The lines correspond to $E^2=m^2+\vec p^2$ using the nucleon mass $m$ determined at zero momentum.}
\end{figure}
\begin{figure}[ptb]
\centering
\includegraphics[width=1.0\textwidth]{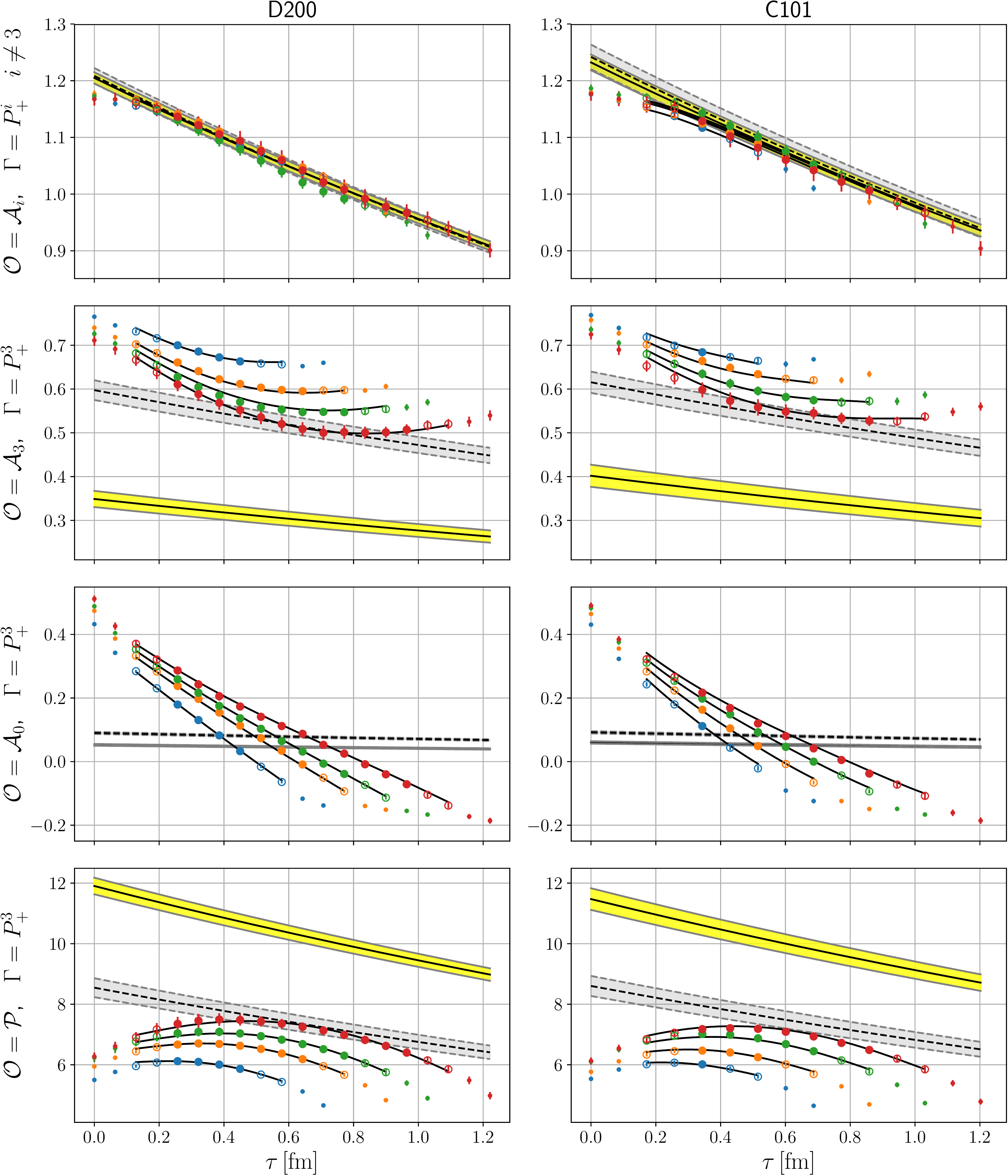}
\caption{\label{fig_correlation_function_fits}Fits to the ratio $R^{\vec 0, \vec p}_{\Gamma, \mathcal O}$ (defined in eq.~\eqref{eq_superior_ratio_def}) at a momentum transfer $\vec q = -\vec p = \frac{2\pi}{L} (0,0,-1)^T$ for ensemble D200 (left side) and C101 (right side) for various channels and spin projections, where we have exploited rotational symmetry to average over equivalent directions. The solid lines correspond to a simultaneous fit to all the channels taking into account the leading $N\pi$ contribution using eqs.~\eqref{eq_res_A_noChPT} and~\eqref{eq_res_P_noChPT}. The yellow band corresponds to the ground state. The gray band (dashed lines) shows the ground state extracted from a traditional fit using one generic excited state. The ground state contributions in the top (bottom) panels are sensitive to $G_A$ ($G_P$), exclusively, while those in the second and the third row yield linear combinations of $G_A$ and $G_\tP$ (see eqs.~\eqref{eq_gs_trace_row1}-\eqref{eq_gs_trace_row4}). The bands include the statistical error and an error due to a variation of the fit range.}
\end{figure}

%
The nucleon energies determined from fits to two-point functions using a spectral decomposition with one generic excited state~\eqref{eq_2pt_spectral_decomposition} agree with the continuum dispersion relation, see figure~\ref{fig_dispersion_relation}. With this justification, we employ the continuum dispersion relation for single nucleon energies in the subsequent analysis. The nucleon isovector form factors are obtained by a simultaneous fit to two-point functions and to the ratio%
\begin{align}\label{eq_superior_ratio_def}%
R_{\Gamma, \mathcal{O}}^{\vec{p}^{\mathrlap{\prime}}, \vec{p}} (t, \tau) &= \frac{C_{{\rm 3pt}, \Gamma}^{\vec{p}^{\mathrlap{\prime}}, \vec{p}, \mathcal{O}} (t, \tau) }{ C_{{\rm 2pt},P_+}^{\vec{p^\prime}} (t)}  \,,
\end{align}%
using the parametrizations given in section~\ref{sec_final_param}. In the literature also the ratio
\begin{align} \label{eq_usual_ratio_def}
\frac{ C_{{\rm 3pt}, \Gamma}^{\vec{p}^{\mathrlap{\prime}}, \vec{p}, \mathcal{O}} (t, \tau) }{ C_{{\rm 2pt},P_+}^{\vec{p^\prime}} (t) }
\sqrt{\frac{C_{{\rm 2pt},P_+}^{\vec{p^\prime}} (\tau) C_{{\rm 2pt},P_+}^{\vec{p^\prime}} (t) C_{{\rm 2pt},P_+}^{\vec{p} } (t-\tau)}{C_{{\rm 2pt},P_+}^{\vec{p}} (\tau) C_{{\rm 2pt},P_+}^{\vec{p} } (t) C_{{\rm 2pt},P_+}^{\vec{p^\prime}} (t-\tau)}}
\end{align}
is found, which is constructed such that the overlap factors drop out and the ground state contribution is time-independent. This is not the case for the ratio~\eqref{eq_superior_ratio_def}, where the ground state contribution is $\propto e^{-(E-E^\prime) \tau}$. Nevertheless, we find it to be advantageous for various reasons:
\begin{enumerate}
 \item It allows for a maximal cancellation of correlations, since the interpolating currents at the source and the sink occur at exactly the same spacetime positions with exactly the same phase factors in two- and three-point functions, cf.\ eqs.~\eqref{eq_2ptdef} and~\eqref{eq_3ptdef}.
 \item In contrast to eq.~\eqref{eq_usual_ratio_def} it does not introduce additional excited states from two-point functions at small separations $\tau$ or $t-\tau$.
 \item One avoids a technical problem of eq.~\eqref{eq_usual_ratio_def}: in the course of the error analysis one can encounter negative values for single bootstrap samples due to statistical fluctuations such that the argument of the square root is negative.
\end{enumerate}
Note that the argument in point 1 also explains why fitting the ratio~\eqref{eq_superior_ratio_def} is preferable to fitting the three-point function.\footnote{In principle, using the three-point function is of course equivalent. In practice, however, one would need even better statistics to enable fully correlated, simultaneous fits.} Results of the simultaneous fits using the ratio~\eqref{eq_superior_ratio_def} and the two-point functions are shown in figure~\ref{fig_correlation_function_fits}, where we have selected cases in which the effect due to the pion pole enhanced excited states is large, i.e., ensembles with small pion masses at small (but nonzero) momentum transfer. Note that for our kinematics the parametrization~\eqref{eq_res_A_noChPT} and~\eqref{eq_res_P_noChPT} only includes two additional fit parameters ($d$ and $e^\prime$) in addition to the usual excited state parametrization. These two parameters describe the $N\pi$ related excited state contributions for the axialvector and pseudoscalar channels simultaneously, for all spin-projections. That this is even possible strongly indicates that the results given in section~\ref{sec_final_param} are a very good approximation of the underlying physics.\par%
In order to take into account systematic uncertainties of our excited state analysis, we perform a fit range variation, where the minimal distance between the operators is varied between $2a$ and $4a$ in the ratios, and between $2a$ and $3a$ in the two-point functions. In figures~\ref{fig_correlation_function_fits} and~\ref{fig_comparison_with_subtraction} full circles (dots) correspond to data points that are always (never) part of the fitted window, while the open symbols indicate data points that are used only in some of the fits. The error bands of the extracted ground state contributions contain both the statistical error and the error related to the choice of the fit range.\par%
\begin{figure}[ptb]
\includegraphics[width=0.49\textwidth]{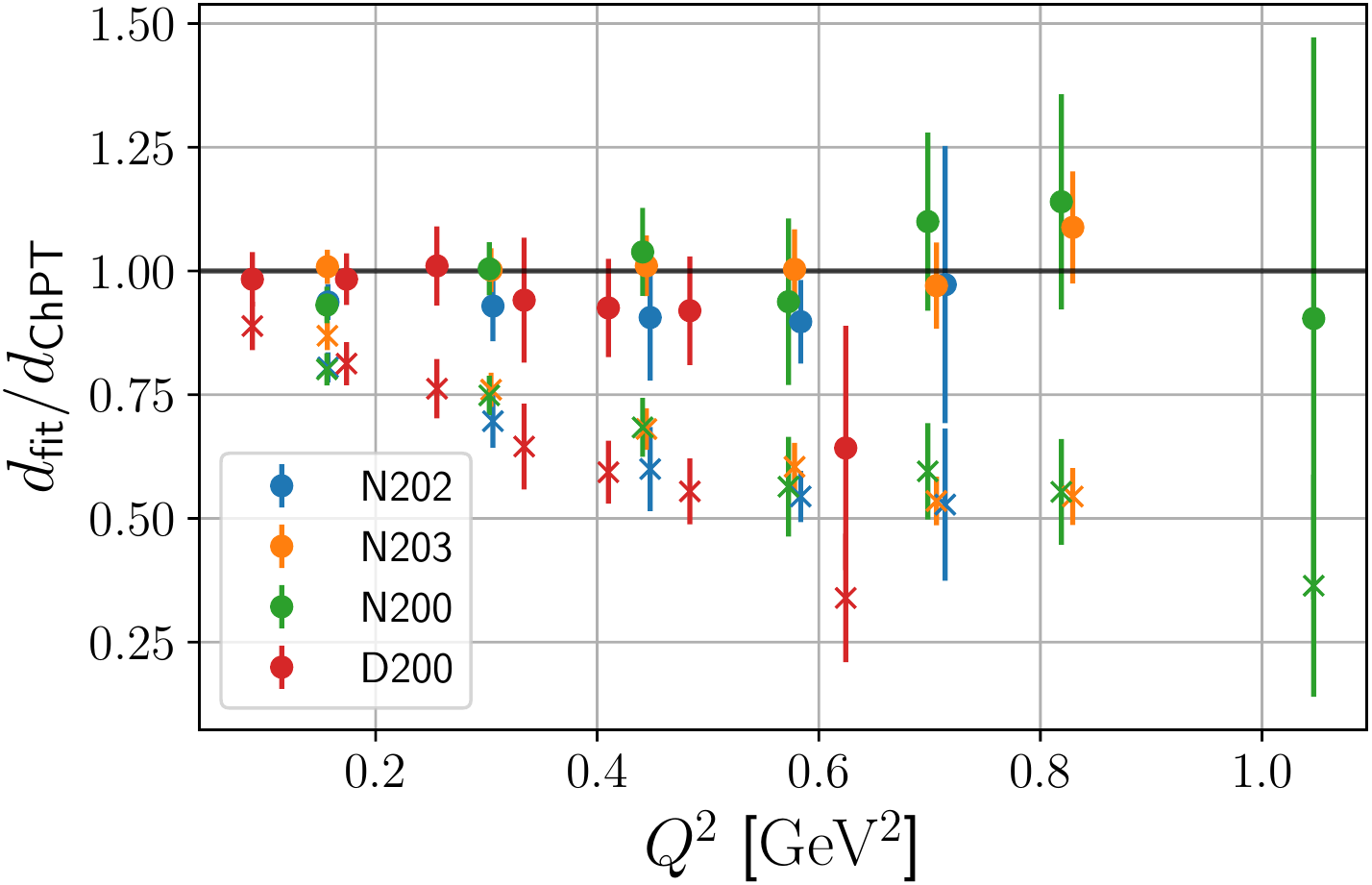}\hspace{0.01\textwidth}\includegraphics[width=0.49\textwidth]{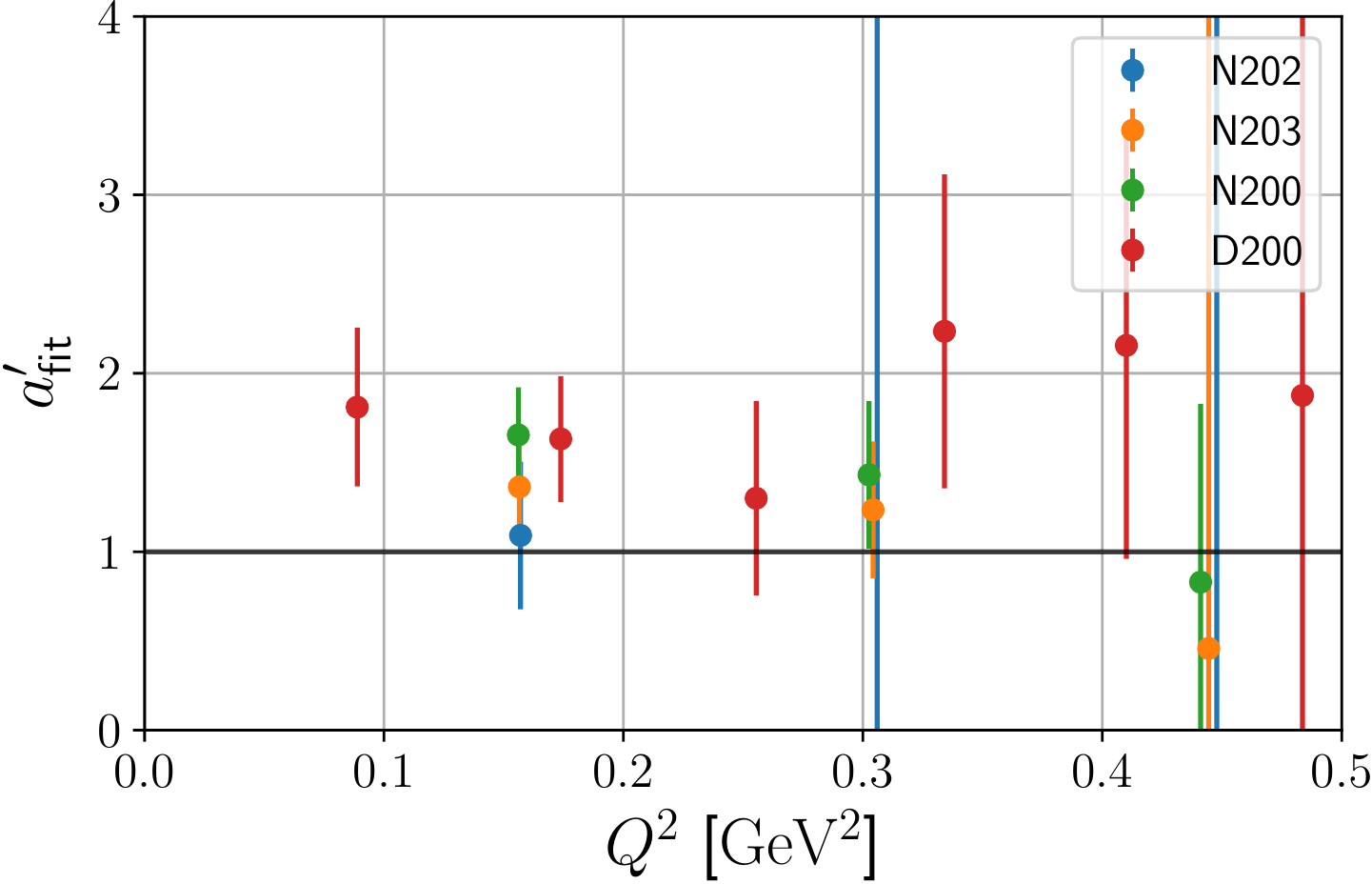}\\
\caption{\label{fig_eft_prediction}The plot on the left shows how well the parameter free tree-level ChPT prediction $d_{\text{ChPT}}$ (circles; see eq.~\eqref{eq_d}) describes the data obtained from the fit ($d_{\text{fit}}$). As anticipated in section~\ref{sec_eft}, the estimate using $g_A$ instead of $G_A$ (crosses) is not as good. This simply means that at nonzero momentum transfer the coupling of the pion to the nucleon is given by $G_A(Q^2)$ instead of $g_A$, as expected. In the plot on the right we show $a^\prime$ (cf.\ eq.~\eqref{eq_NPi_coupling}) obtained from our fit to the data. A value of $a^\prime=1$ would imply that the leading order ChPT estimate for the coupling of $N\pi$ to the three-quark operators is exact and that the operator smearing does not affect the coupling at all. As one can see, the data is not very sensitive to the value of $a^\prime$. We do not see any significant momentum dependence and no strong smearing effect.}%
\end{figure}%
In figure~\ref{fig_correlation_function_fits} the yellow bands correspond to the ground state contributions extracted from the EFT-inspired ansatz for the three-point function (eqs.~\eqref{eq_res_A_noChPT} and~\eqref{eq_res_P_noChPT}), while the gray band is the ground state signal obtained from a traditional multistate fit ansatz (also using eqs.~\eqref{eq_res_A_noChPT} and~\eqref{eq_res_P_noChPT}, but without the explicit $N\pi$ contribution, i.e., setting $c=c^\prime=d=d^\prime=0$). The decomposition of the ground state matrix elements in terms of form factors is determined by eq.~\eqref{eq_gs_trace}; see appendix~\ref{app_traces} for an explicit evaluation. As one can see, the ground state contribution can be disentangled from the huge signal of the $N\pi$ state (which fails to be resolved using the traditional ansatz with generic excited state contributions). Here, it is particularly advantageous that the coefficients of the $N\pi$ contributions are constrained for various channels and spin projections in our fit, which simplifies the determination of the corresponding fit parameters ($e^\prime = d^\prime - c^\prime$ and $d$, for our kinematics). To this end, the seemingly linear behavior in $\mathcal A_0$ (i.e., row $3$ in figure~\ref{fig_correlation_function_fits}, where the spin projection is aligned with the momentum) is actually helpful and it is noteworthy that this data can be described very well by our fit ansatz. The ratio shown in the top panels (which is sensitive to $G_A$ but independent of $G_\tP$) is not affected by the pole enhanced $N\pi$ excited state contribution.\footnote{The small shift within errors occurs because we perform a simultaneous fit such that the determined energy of the generic excited state is influenced by the fit in the other channels.} Indeed, we do not see any evidence in our numerical data for $N\pi$ or other low-lying multiparticle state contributions in this channel. This supports the choice in previous lattice calculations to determine the axial form factor using this channel in combination with traditional excited state fits.\par%
The ansatz including the $N\pi$ excited states explicitly allows for a much better description of the data. In the case of D200 for instance, fits using block-correlated covariance matrices yield $\chi^2/\text{d.o.f.}\approx1.31$ (including $N\pi$) versus $\chi^2/\text{d.o.f.} \approx 7.17$ (excluding $N\pi$). Note, however, that we have decided to use uncorrelated fits to extract the results. This avoids instabilities in the covariance matrix and prevents an underestimation of the statistical errors.\par%
We find that almost the complete excited state contamination can be attributed to this $N\pi$ state, and that there are only very mild additional contributions at the sink (where $\vec p^\prime = \vec 0$). Nevertheless, we refrain from removing the additional generic excited states from the parametrization, in order to exclude an underestimation of the error in the extracted ground state contribution. Actually, one can also obtain a very good description of the data with even smaller statistical errors if one would use the ChPT-biased parametrizations discussed in section~\ref{sec_eft}, which may indicate that possible higher order corrections are small. Nevertheless, the latter would entail a systematic uncertainty that we intend to avoid.\par%
However, we can confront the results of our fits with the corresponding ChPT prediction, see figure~\ref{fig_eft_prediction}. In particular for the parameters $d$ and $d^\prime$ in eqs.~\eqref{eq_res_A_noChPT} and~\eqref{eq_res_P_noChPT} ChPT yields a parameter free prediction, see eq.~\eqref{eq_d}. Since $d$ corresponds to one of our fit parameters, a direct comparison is possible (left plot in figure~\ref{fig_eft_prediction}). As anticipated in section~\ref{sec_eft}, the prediction using $G_A(Q^2)$ (circles) as the pion-nucleon coupling, instead of $g_A=G_A(0)$ (crosses), agrees well with our data, even at large $Q^2$, where one would usually not expect ChPT to work. For our kinematics, the $N\pi$ excitation in the final state can also couple directly to the three-quark operator (this corresponds to the diagrams on the left and right in the second row of figure~\ref{fig_diagrams}). Therefore, we can try to determine $a^\prime$ (defined in eq.~\eqref{eq_NPi_coupling}) directly from the data. A value $a^\prime=1$ means that the leading order ChPT estimate for the coupling of $N\pi$ to the three-quark operators calculated for local currents is exact in spite of the smearing. As one can see from the large statistical errors in the right plot of figure~\ref{fig_eft_prediction}, our data is not very sensitive to $a^\prime$. This is expected, since $c$ and $c^\prime$ (which contains $a^\prime$) are suppressed compared to $d$ and $d^\prime$ by one factor of $\mathcal O(\frac{m_\pi}{m})$. We neither see a significant momentum dependence nor a strong smearing effect. If anything, the direct coupling of the three-quark operators to $N\pi$ seems to be slightly enhanced by the smearing.\par%
\begin{figure}[ptb]
\includegraphics[height=0.280784\textheight]{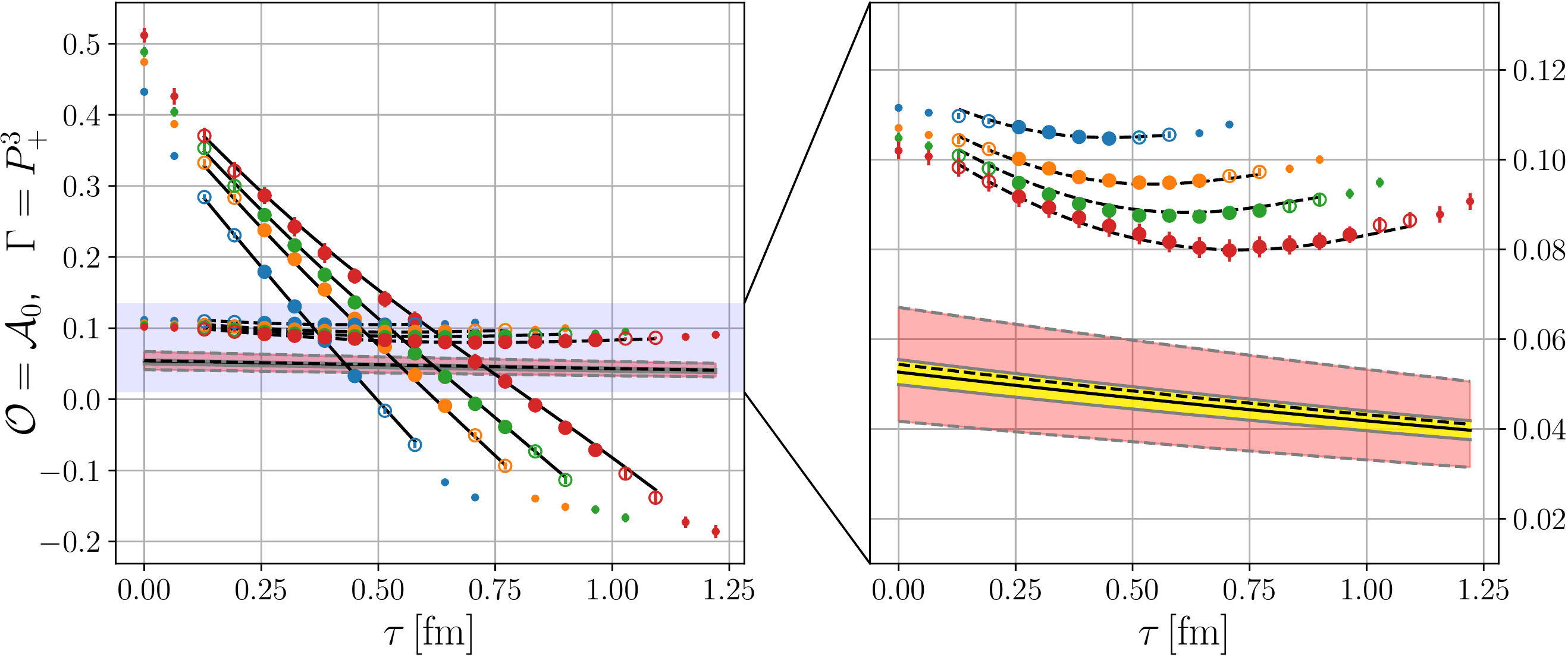}\\[\baselineskip]
\begin{minipage}{0.51\textwidth}
\includegraphics[height=0.280784\textheight,trim=0 0 390.3 0, clip]{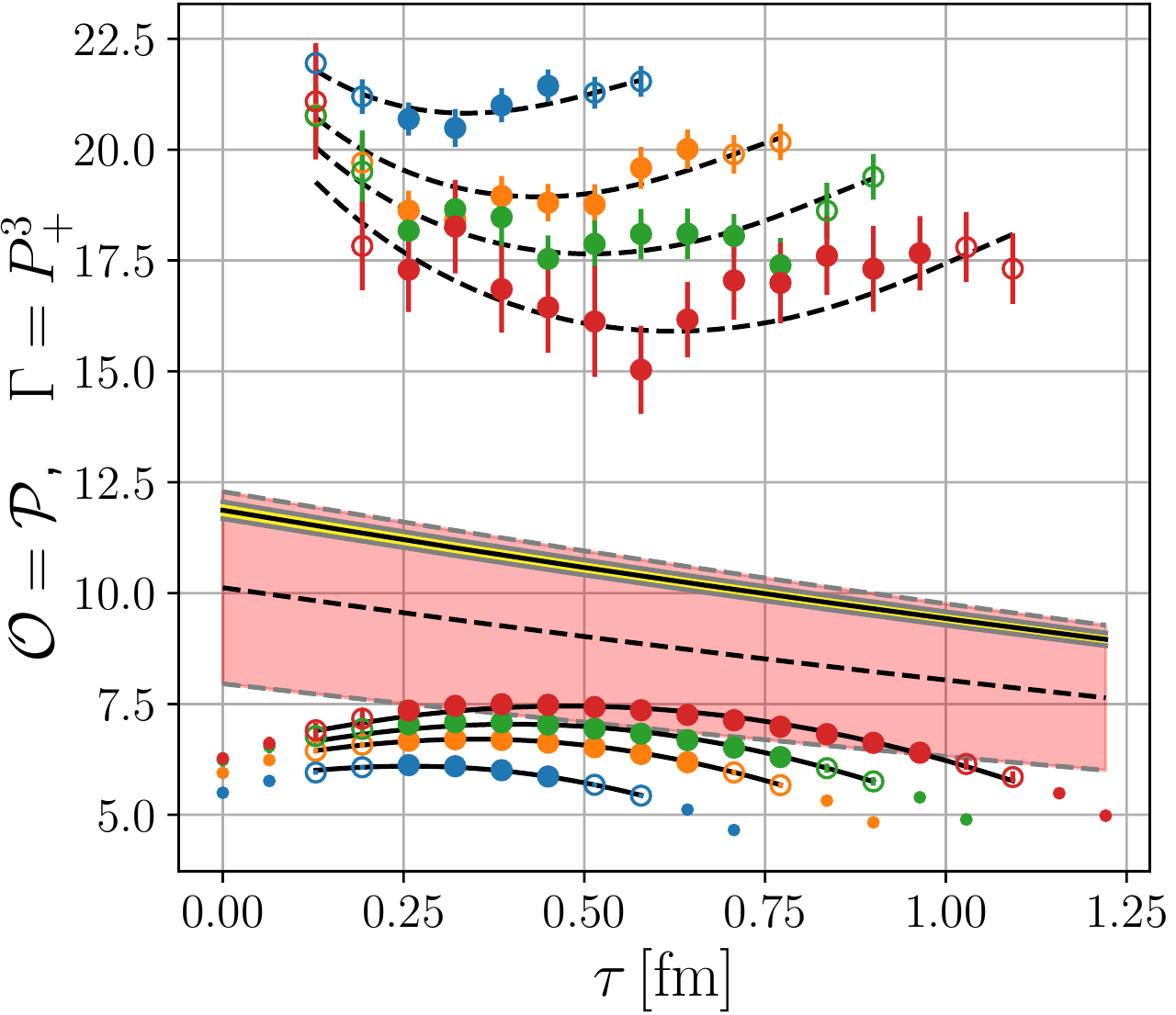}
\end{minipage}\begin{minipage}[]{0.49\textwidth}%
\vbox to 0.280784\textheight{\caption{\label{fig_comparison_with_subtraction} Comparison to the subtraction method proposed in ref.~\cite{Bali:2018qus} for the ratio $R^{\vec 0, \vec p}_{\Gamma, \mathcal O}$ (defined in eq.~\eqref{eq_superior_ratio_def}) at the momentum transfer $\vec q = -\vec p = \frac{2\pi}{L} (0,0,-1)^T$ for ensemble D200. The solid and dashed lines show fits to the unsubtracted and subtracted data, respectively, where the yellow and red bands show the corresponding ground state signals. In both cases we have taken into account the leading $N\pi$ contribution. For the subtracted current the fit ansatz has to be adapted, cf.\ appendix~\ref{app_ansatz_sub}.}}%
\end{minipage}%
\end{figure}%
In figure~\ref{fig_comparison_with_subtraction} we reinvestigate the subtraction method that some of us have proposed in ref.~\cite{Bali:2018qus}. As one can clearly see in the upper panels of figure~\ref{fig_comparison_with_subtraction}, it almost entirely removes the seemingly linear behavior in the $\mathcal A_0$ channel caused by the $N\pi$ states. We find that the results for the ground state obtained from fits to the unsubtracted (solid lines; ground state yellow) and the subtracted (dashed lines; ground state red) data are mutually compatible, once we take into account the leading $N\pi$ contribution.\footnote{Note that the subtraction method in combination with traditional excited state fits (as used in ref.~\cite{Bali:2018qus}) does not yield the correct ground state. In particular in the pseudoscalar channel the correction overshoots and yields too large values. This has strong effects on $G_\tP$ and $G_P$, while $G_A$ is unaffected. For a detailed study of this topic see also ref.~\cite{Bar:2019igf}} For the subtracted correlation functions, the fit ansatz given in section~\ref{sec_final_param} has to be adapted appropriately, cf.\ appendix~\ref{app_ansatz_sub}. However, the ground state extracted from the subtracted data has a much larger statistical uncertainty. A closer look shows that the subtraction method here has fallen victim to its own success: since the largest and clearest excited state contaminations (in $\mathcal A_0$) have been subtracted successfully, the corresponding parameters cannot be determined as reliably, which in turn leads to a large error in the ground state. One can conclude that a combination of the analysis method proposed here (taking into account the relevant $N\pi$ excitation explicitly in the fit to the correlation function) and the subtraction method proposed in ref.~\cite{Bali:2018qus} is not advantageous.\par
\begin{figure}[ptb]
\centering
\includegraphics[width=\textwidth]{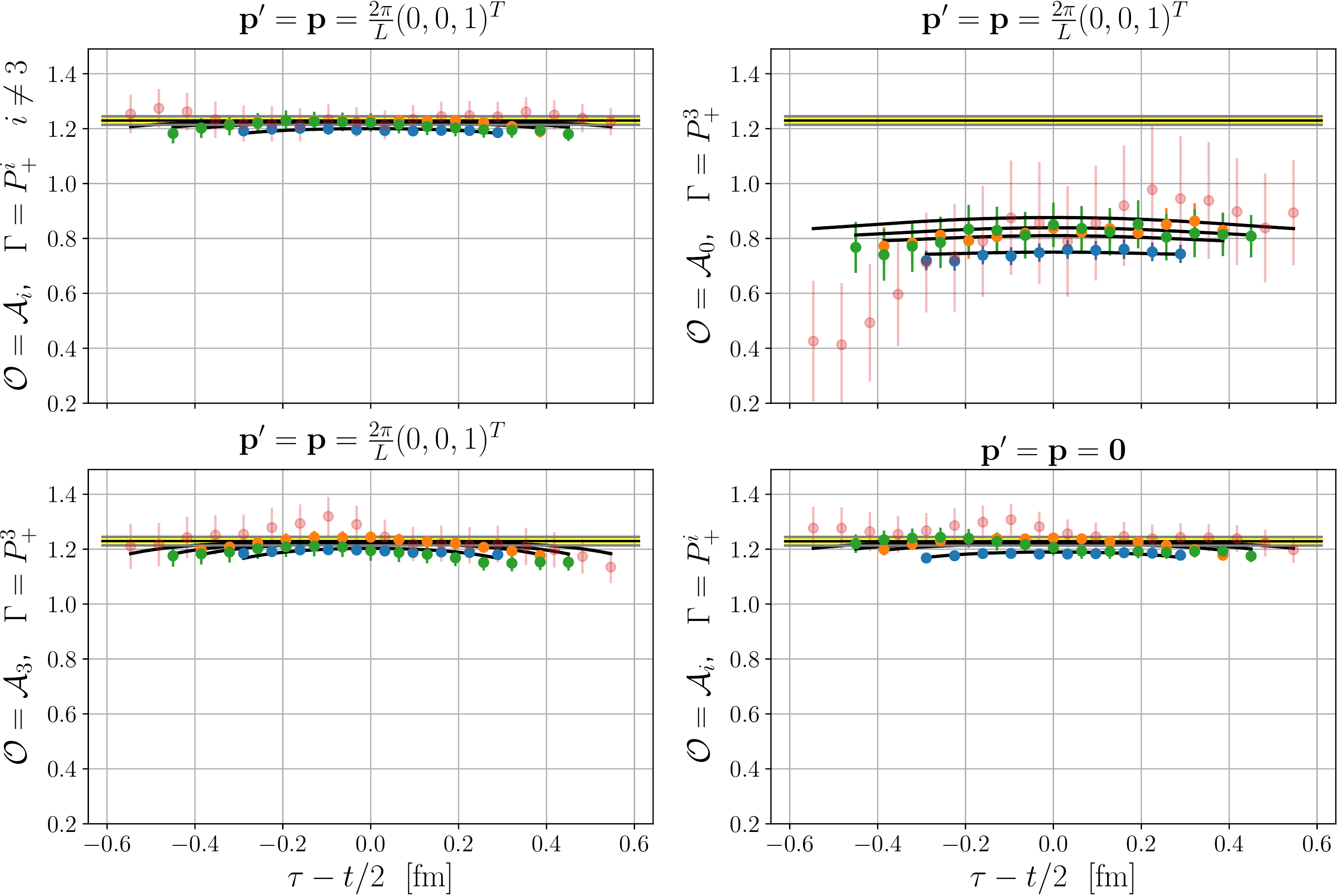}
\caption{\label{fig_finite_final_momentum_test}Fits to the ratio $R^{\vec p, \vec p}_{\Gamma, \mathcal O}$ (as defined in eq.~\eqref{eq_superior_ratio_def}, but rescaled such that the ground state contribution in all the channels corresponds to $g_A$) at the momentum transfer $\vec q = \vec 0$, with $\vec p^\prime = \vec p = \frac{2\pi}{L} (0,0,1)^T$ and with $\vec p^\prime = \vec p =  \vec 0$ for the contributing axial channels. This analysis has been performed on ensemble D201. The solid lines correspond to a simultaneous fit to all the channels taking into account the leading $N\pi$ contribution using eqs.~\eqref{eq_res_A_noChPT} and~\eqref{eq_res_P_noChPT}, where the yellow band corresponds to the ground state. The bands include the statistical error and an error due to a variation of the fit range.}
\end{figure}%
As a consistency check, we have also considered the case $\mathbf q = 0$ with $\mathbf p^\prime = \mathbf p \neq \mathbf 0$ on one of our ensembles (D201). In this situation eq.~\eqref{eq_res_A_noChPT} predicts that the correlation functions of $\mathcal A_1$, $\mathcal A_2$, and $\mathcal A_3$ are not affected by the $N\pi$ excited state, while $\mathcal A_0$ gets a contribution $\propto \exp\bigl(-(E_N + m_\pi/2) t \bigr) \cosh \bigl( m_\pi (\tau -t/2 ) \bigr)$ in the three-point function. In figure~\ref{fig_finite_final_momentum_test} we show that this is indeed the case and that a simultaneous fit using eq.~\eqref{eq_res_A_noChPT} yields a consistent description of the data for all the channels. This suggest that the observation in ref.~\cite{Liang:2016fgy} (see also ref.~\cite{Liang:2018pis}), that a determination of $g_A$ from the $\mathcal A_0$ channel in a moving frame (at $Q^2=0$) gives results different from those obtained using $\mathcal A_1$, $\mathcal A_2$, and $\mathcal A_3$, can be attributed to the same $N\pi$ excited state contaminations that have been problematic at nonzero~$Q^2$ in other studies.\par%
\subsection{Excited state energies\label{sec_excited_energies}}%
\begin{figure}[tb]\centering%
\includegraphics[width=.6\textwidth]{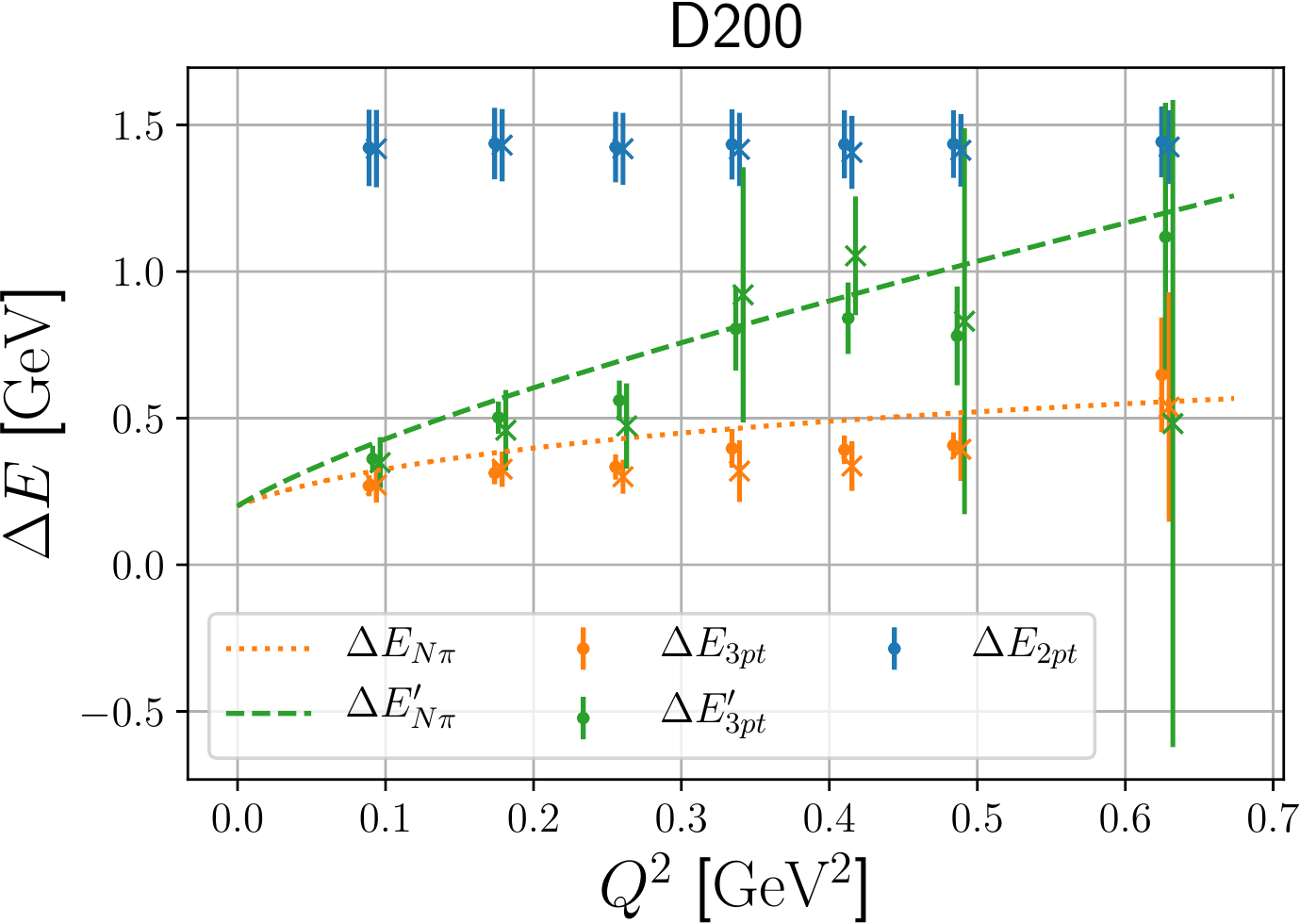}%
\caption{\label{fig_excitedenergies}Energy gaps between the ground state and the excited states on the ensemble D200. The crosses have been obtained from a fit using the ansatz from eqs.~\eqref{eq_res_A_noChPT} and~\eqref{eq_res_P_noChPT} but taking $\Delta E_{\rm 3pt}=\Delta E_{N\pi}$ and $\Delta E^\prime_{\rm 3pt}=\Delta E^\prime_{N\pi}$ as free fit parameters, while $\Delta E_{\rm 2pt} = \Delta E$ corresponds to the energy of the generic excited state determined from two- and three-point functions. The dots have been obtained from a fit without an explicit $N\pi$ state (i.e., $c=c^\prime=d=d^\prime=0$ in eqs.~\eqref{eq_res_A_noChPT} and~\eqref{eq_res_P_noChPT}) but relaxing the condition that the excited state energies in two- and three-point function have to match (i.e., $\Delta E_{\rm 3pt}=\Delta E$ and $\Delta E^\prime_{\rm 3pt}=\Delta E^\prime$ from the three-point function and $\Delta E_{\rm 2pt} = \Delta E$ from the two-point function). The orange, dotted line and the green, dashed line show the energy gaps for a noninteracting nucleon-pion system in the initial and the final state, respectively, as obtained from the diagrams in the left and the right column of figure~\ref{fig_diagrams}. For our kinematics the energys are $E_{N\pi} = E_\pi(\mathbf q) + E_N(\mathbf 0)$ and $E_{N\pi}^\prime = E_\pi(\mathbf q) + E_N(-\mathbf q)$.}%
\end{figure}%
In ref.~\cite{Jang:2019vkm} it has been proposed to use the signal of the timelike axialvector channel to determine the energy of the low-lying $N\pi$ excitation. The main difference with respect to the traditional excited state fit method is that one does not impose that the leading excited states in the two- and three-point functions have the same energy. In figure~\ref{fig_excitedenergies} (which roughly reproduces Fig.~3 of ref.~\cite{Jang:2019vkm}\footnote{Figure number from the arXiv v2 version.}) we show the energy gaps to the various excited states obtained from two different fits to the correlation functions on ensemble D200 (with $m_\pi \approx \unit{201}{\mega\electronvolt}$). The dots (fit~$1$) have been obtained using the method proposed in ref.~\cite{Jang:2019vkm} (with the slight difference that we perform a simultaneous fit to all the channels instead of the two-step method presented in ref.~\cite{Jang:2019vkm}), while the crosses (fit~$2$) have been obtained using our fit ansatz from eqs.~\eqref{eq_res_A_noChPT} and~\eqref{eq_res_P_noChPT} but leaving $\Delta E_{N\pi}$ and $\Delta E^\prime_{N\pi}$ as free fit parameters. In contrast to fit~$1$, fit~$2$ contains the additional excited states known from the two-point function, which leads to larger statistical uncertainties, in particular when the energy levels of the $N\pi$ state and the excited state from the two point function (blue data points) get close to each other. Both kinds of fits lead to energies for the nucleon-pion states that approximately correspond to those of a noninteracting system (cf.\ the diagrams in the left and the right column of figure~\ref{fig_diagrams}), which for our kinematics means that $E_{N\pi} = E_\pi(\mathbf q) + E_N(\mathbf 0)$ in the initial state (orange, dotted line) and $E_{N\pi}^\prime = E_\pi(\mathbf q) + E_N(-\mathbf q)$ in the final state (green, dashed line). The fact that both methods result in compatible values for the $N\pi$ excited state energies is encouraging and suggests that the physical interpretation obtained using EFT (cf.\ section~\ref{sec_eft}) is correct.\par%
In particular for the low-lying $N\pi$ state (which for our kinematics occurs in the initial state) at intermediate $Q^2$ one can see that the energies obtained from the fits slightly undershoot those of the noninteracting system. This effect is found to be a bit more significant in ref.~\cite{Jang:2019vkm}. One may speculate that this small deviation is due to an interaction between the nucleon and the pion. For the time being we have chosen to ignore these small deviations in our fits.\par%
\section{Form factors\label{sec_formfactors}}
\subsection{Approximate restoration of PCAC and PPD\label{sec_approx_pcac_ppd}}
\begin{figure}[tb]\centering%
\includegraphics[width=0.49\textwidth]{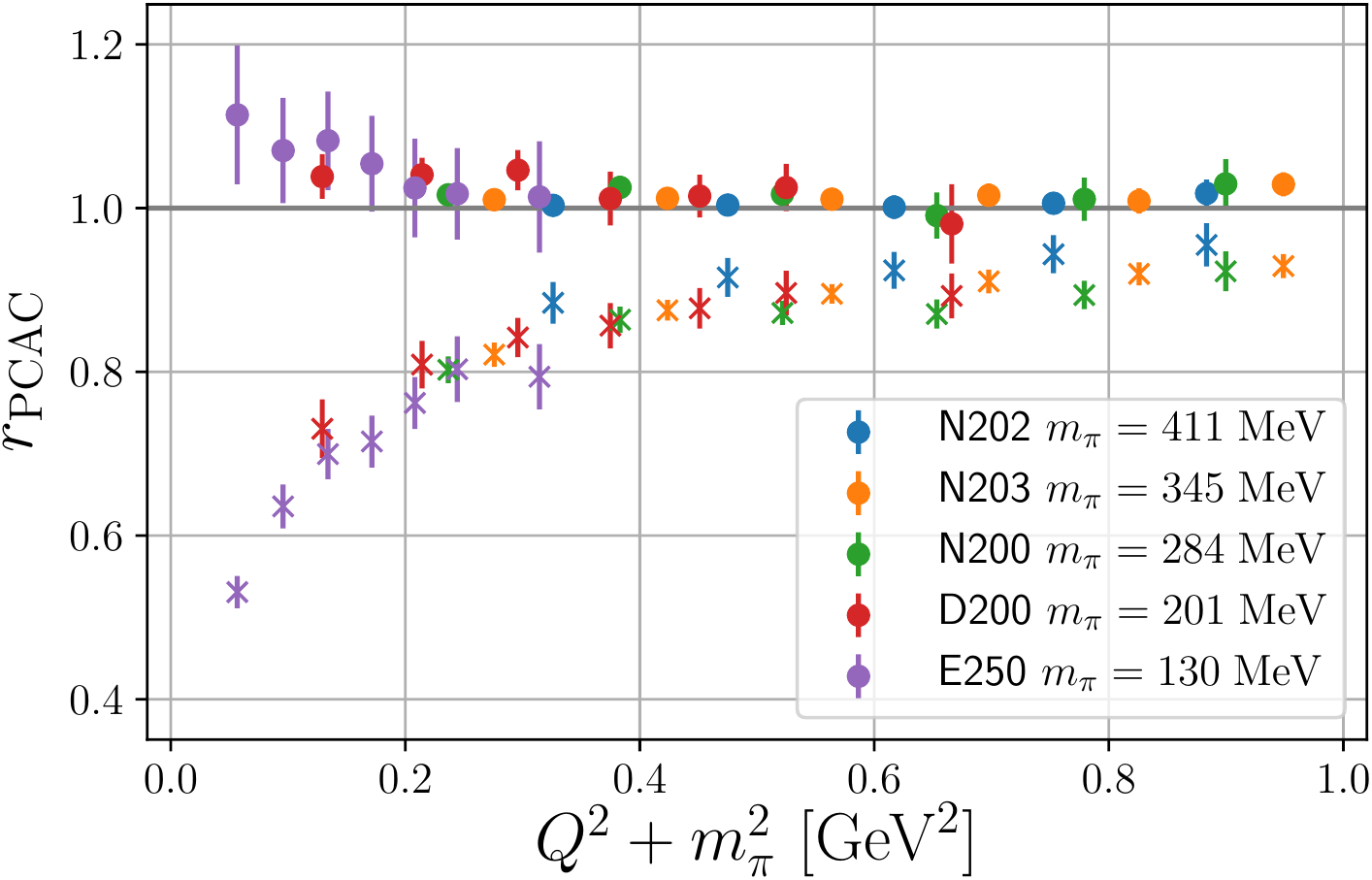}\hfill\includegraphics[width=0.49\textwidth]{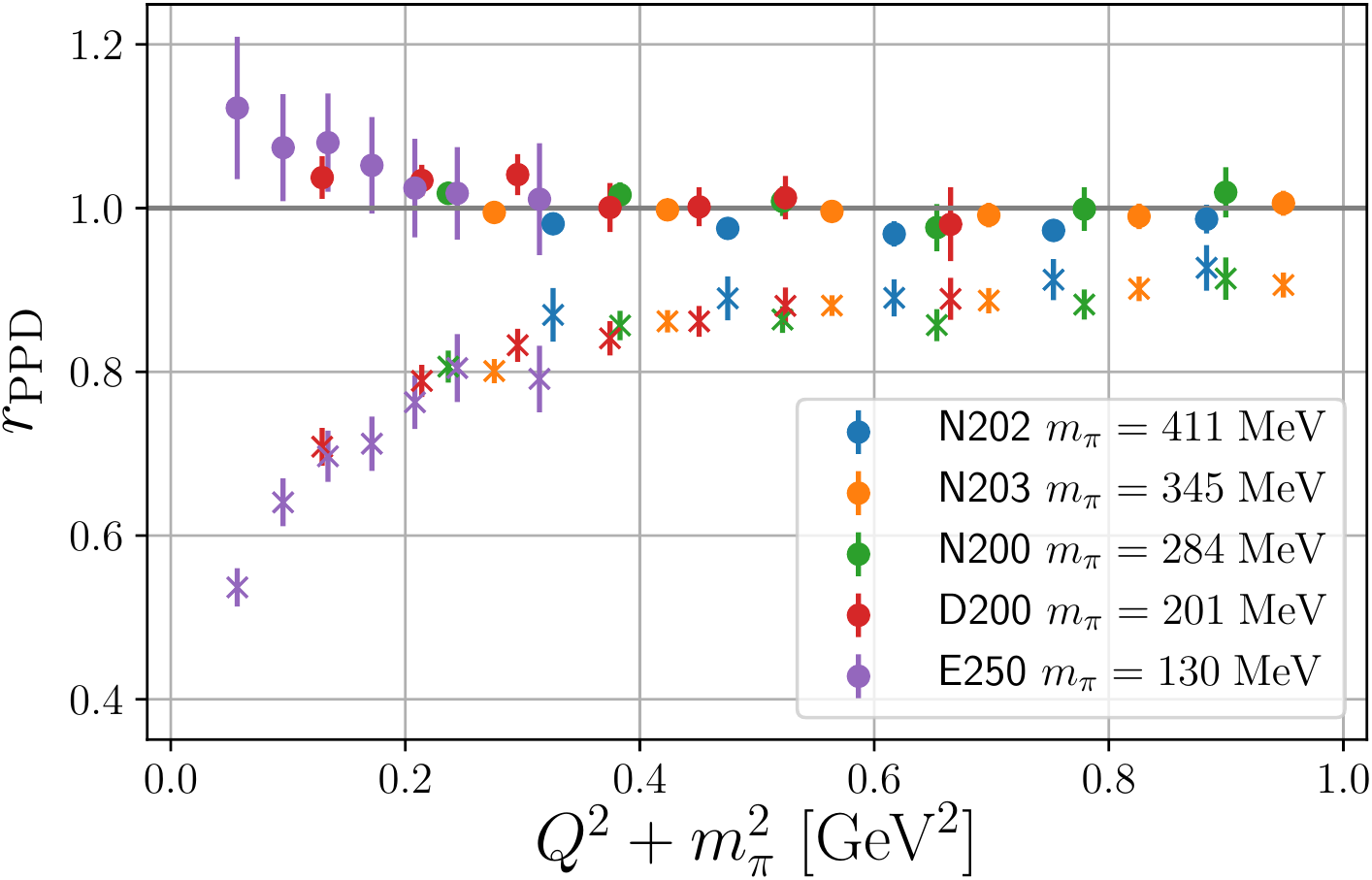}%
\caption{\label{fig_r_mpi}Violation of PCAC (left panel) and PPD (right panel) displayed for various ensembles along the trajectory with constant average quark mass (green lines in figure~\ref{fig_ensembles}) at $\beta=3.55$. The plots show the ratios defined in eqs.~\eqref{eq_ratioPCAC} and~\eqref{eq_ratioPPD}. The filled circles are obtained taking into account the pion pole enhanced excited states directly in the fit functions, while the crosses were obtained using a traditional fit ansatz (with the same excited state energies in the two- and three-point functions).}
\end{figure}

As mentioned in the introduction, form factors extracted from data using a traditional fit ansatz (with the same excited state energies in the two- and the three-point functions) show strong violations of PCAC and PPD. In particular in the case of PCAC this result was puzzling since the latter is fulfilled at the correlation function level (up to small, expected discretization effects). In order to quantify the violation of the PCAC relation at the form factor level (cf.\ eq.~\eqref{eq_NucleonFFPCAC}), we define the ratio (cf.\ also ref.~\cite{Rajan:2017lxk})%
\begin{align}
r_{\rm{PCAC}} &= \frac{\frac{m_\ell}{m} G_P (Q^2)+ \frac{Q^2}{4m^2} G_\tP(Q^2)}{G_A(Q^2)} \,, \label{eq_ratioPCAC}
\end{align}%
where $r_{\rm{PCAC}}=1$ if PCAC holds exactly. As the panel on the left-hand side of figure~\ref{fig_r_mpi} demonstrates, using the parametrization of excited state contributions described in section~\ref{sec_final_param}, the PCAC relation is now fulfilled reasonably well on all ensembles, in particular on the ensembles with small pion masses, which previously exhibited the largest deviations. We emphasize that our fit ansatz does not impose PCAC on the ground state. While we see a significant improvement for all ensembles, small deviations of~$\mathord{\sim}5\%$ remain in some cases.\par%
The induced pseudoscalar form factor is often estimated by%
\begin{align}
G_\tP &\overset{?}{\approx} \frac{4 m^2 G_A}{m_\pi^2+Q^2} & &\Rightarrow & r_{\rm{PPD} } &= \frac{(m_\pi^2+Q^2) G_\tP(Q^2)}{4 m^2 G_A(Q^2)} \overset{?}{=} 1 \,, \label{eq_ratioPPD}
\end{align}
which is usually referred to as the pion pole dominance (PPD) assumption. Note that this relation does not have to hold exactly, even in the continuum. However, one would expect it to be satisfied at least approximately for small pion masses. The panel on the right-hand side of figure~\ref{fig_r_mpi} shows that this is indeed the case if one explicitly takes into account the pion pole enhanced excited states in the spectral decomposition of the correlation function.\par%
As reported in ref.~\cite{Jang:2019vkm} the problem can also be resolved (though within larger statistical uncertainties), if one uses a traditional multi-state fit ansatz, but relaxes the condition that the excited state energies of the two- and three-point functions have to match. One can exploit the huge excited state signal in the timelike axialvector channel to determine the energy gaps quite precisely (cf.\ also section~\ref{sec_excited_energies}). This can be seen as further confirmation that the previously observed large deviations from PCAC and PPD were indeed caused by unresolved, pion pole enhanced excited states. Note, however, that our ansatz (shown in eqs.~\eqref{eq_res_A_noChPT} and~\eqref{eq_res_P_noChPT}) conveys insight into the structure of the excited state contamination. For instance, it is clear that, for $\mathcal A^\mu$ with $\mu=1,2,3$, the result for $G_A$ will not be affected by the leading $N\pi$ excited state contribution. Heuristically speaking, this is because $G_A$ is not subject to pion pole dominance.\par%
\subsection{Parametrization and extrapolation\label{sec_parametrization}}
In this section we will explore two common form factor parametrizations: the traditional dipole ansatz and the $z$-expansion, which has become fashionable lately. In both cases we also consider parametrizations that are consistent with PCAC in the continuum (section~\ref{sec_param_pcac}) and we will use a generic ansatz for the combined continuum, chiral and volume extrapolation explained in section~\ref{sec_param_extrapolation}.%
\subsubsection{Dipole ansatz\label{sec_param_dipole}}%
Motivated by eqs.~\eqref{eq_GA}, \eqref{eq_GPtilde}, and~\eqref{eq_GP}, we rewrite the form factors as
\begin{align}
G_A & \equiv A(Q) \,, & 
G_\tP& \equiv \frac{4m^2}{Q^2+m_\pi^2} \tP(Q) \,,
&G_P & \equiv \frac{m}{m_\ell}\frac{m_\pi^2}{Q^2+m_\pi^2} P(Q)  \,,\label{eq_residualFF}
\end{align}%
where the pion pole is isolated (cf.\ also ref.~\cite{Green:2017keo}) such that one can use similar parametrizations for the residual form factors $X(Q)$, $X \in  \{A,\, \tP,\, P\}$. The prefactors not only ensure that all the functions $X(Q)$ have the same mass dimension, but also enable us to obtain the correct chiral behavior of the form factors at small $Q^2$ despite using the same generic ansatz for all the form factors, see section~\ref{sec_param_extrapolation} below.\par%
One can consider various parametrizations for the residual form factors. For instance, one can use a dipole ansatz
\begin{align}
X(Q) &= \frac{g_X}{\bigl(1+Q^2/M_X^2\bigr)^2} \,, \label{eq_Dipole}
\end{align}
which reproduces the traditional dipole form for the axial form factor with the axial coupling $g_A$ and the axial dipole mass $M_A$. This parametrization not only yields the correct low-energy behavior (if one uses a generic parametrization for the pion mass, volume and lattice spacing dependence of $g_X$ and $M_X$, cf.\ section~\ref{sec_param_extrapolation} below), but also yields the correct asymptotic limit $G_A \propto 1/Q^4$, $G_\tP \propto 1/Q^6$, and $G_P \propto 1/Q^6$~\cite{Alabiso:1974ye}, at large momentum transfer.
\subsubsection{\texorpdfstring{$z$}{z}-expansion\label{sec_param_zexp}}
One may also parametrize the residual form factors using the $z$-expan\-sion~\cite{Hill:2010yb,Bhattacharya:2011ah}, which automatically imposes analyticity constraints. This corresponds to an expansion of the form factors in the variable%
\begin{align}
z &= \frac{\sqrt{t_{\rm{cut}}+Q^2}-\sqrt{t_{\rm{cut}}-t_0\vphantom{Q^2}}}{\sqrt{t_{\rm{cut}}+Q^2}+\sqrt{t_{\rm{cut}}-t_0\vphantom{Q^2}}} \,, \label{eq_zDefinition}
\end{align}
where $t_{\rm{cut}}=9 m_\pi^2$ is the particle production threshold and $t_0$ is a tunable parameter.\footnote{We have set $t_0$ to $-t_{\rm{cut}}^{\text{phys}}=-9 m_{\pi,\text{phys}}^2$ in our analysis. By choosing a negative value one can avoid the erratic behavior at $t_{\rm{cut}}=t_0$, while approaching the chiral limit.} We then parametrize%
\begin{align}
X(Q) &= \sum_{n=0}^N a^X_n z(Q)^n \,, \label{eq_zExpansion}
\end{align}%
where the $X(Q)$ are defined as in section~\ref{sec_param_dipole}. Without additional constraints this parametrization has $N+1$ free parameters and is usually called a $z^{(N+1)}$ ansatz. Again, the generic parametrization discussed in section~\ref{sec_param_extrapolation} will yield the correct chiral behavior. However, eq.~\eqref{eq_zExpansion} does not incorporate any constraints at large momentum transfer. In order to reproduce the correct asymptotic behavior one has to enforce restrictions of the type%
\begin{align}
\lim_{Q \rightarrow \infty} & Q^k X(Q) \overset{!}{=} 0 \,, \text{ for } 0\leq k\leq n \,, \label{eq_asymptotic_constraint}
\end{align}%
which can be implemented (as long as $n<N$) by demanding%
\begin{align}
0&=\sum_{l=0}^N l^k a^X_l \,, \text{ for } 0\leq k\leq n \,.
\end{align}%
These can be incorporated, e.g., by fixing%
\begin{align}
a^X_k &= \frac{(-1)^{k+n+1}}{k!(n-k)!} \sum_{l=n+1}^N \frac{l! }{(l-(n+1))!(l-k)} a^X_l \,, \text{ for } 0\leq k\leq n \,. \label{eq_ZExpConstraints}
\end{align}%
Alternatively, one can solve the problem recursively by setting
\begin{align}
a^X_k &= \frac{(-1)^{2k+1}}{k!} \sum_{l=k+1}^N \frac{l! }{(l-(k+1))!(l-k)} a^X_l \,, \text{ for } 0\leq k\leq n \,. \label{eq_ZExpConstraints_rekursive}
\end{align}%
To enforce the correct scaling in the asymptotic limit, $G_A \propto 1/Q^4$, $ G_\tP \propto 1/Q^6$, and $G_P \propto 1/Q^6$~\cite{Alabiso:1974ye}, we have to apply the formulas above for $n=3$, thereby fixing $a^X_k$ for $k=0,1,2,3$, such that $4$ coefficients are fixed and only $N-3$ coefficients are free parameters.\footnote{We neglect possible $\mathcal O(Q^2a^2)$ lattice artifacts since we only have lattice data with $Q^2\ll a^{-2}$. Such effects could be implemented by relaxing the constraint~\eqref{eq_asymptotic_constraint} at nonzero lattice spacing.} This parametrization with the correct asymptotic behavior is usually referred to as the $z^{4+(N-3)}$ ansatz.%
\subsubsection{Consistency with PCAC in the continuum\label{sec_param_pcac}}
Let us assume the following ansatz for the extrapolation to the physical point ($m_\pi\to m_\pi^{\text{phys}}$, $a\to0$, $L\to\infty$),%
\begin{align}
x &= x^\na(m_\pi,m_K,L) x^a(a,m_\pi,m_K) \,, \label{eq_extrapolation}
\end{align}%
where we have factorized the dependence on the lattice spacing into $x^a$ with
\begin{align} \label{eq_xa_continuum}
x^a(0,m_\pi,m_K) = 1
\end{align}%
for all parameters in the form factor decompositions, i.e., $x \in \{g_A, M_A, g_\tP, M_\tP, g_P, M_P\}$ for the dipole ansatz, and $x \in \{a^A_n, \smash{a^\tP_n}, a^P_n\}$, $n=4,5\dots,N$ for the $z$-expansion. This allows us to perform a combined fit to all ensembles for each form factor. The expressions used for $x^\na$ and $x^a$ will be given below in section~\ref{sec_param_extrapolation}.\par%
Since we know that the partial conservation of the axial current has to be satisfied exactly in the continuum limit, we can use eq.~\eqref{eq_NucleonFFPCAC} to obtain $G_P$ from $G_A$ and $G_\tP$:
\begin{align}%
\frac{m_\ell}{m} G_P(Q^2) &= G_A(Q^2) - \frac{Q^2}{4m^2} G_\tP(Q^2) + \mathcal{O}(a^2) \,. \label{eq_NucleonFFPCAC_same_but_different}
\end{align}%
However, one then has to impose the additional constraints%
\begin{align} \label{eq_constraint_PCAC_asymptotic}
\lim_{Q \rightarrow \infty} & Q^n \biggl(G_A-\frac{Q^2}{4 m^2} G_\tP\biggr)\biggr|_{a=0} \overset{!}{=} 0 & 
 &\hat= &
\lim_{Q \rightarrow \infty} & Q^n \biggl(A(Q)-\tP(Q)\biggr)\biggr|_{a=0} \overset{!}{=} 0 \,, \text{ for } n=4,5 \,,
\end{align}
in order to preserve the correct asymptotic behavior of $G_P$, cf.\ also eq.~\eqref{eq_residualFF}. For the dipole parametrizations one gets%
\begin{align}
 g_A M_A^4 \biggr|_{a=0} \overset{!}{=} g_\tP M_\tP^4 \biggr|_{a=0} \,.
\end{align}
The equivalent constraints for the $z$-expansion can be obtained using eq.~\eqref{eq_ZExpConstraints} and read%
\begin{align}
\bigl(a^A_k-a^\tP_k \bigr) \biggr|_{a=0} &= \frac{(-1)^k}{k!(5-k)!} \sum_{l=6}^N \frac{l! }{(l-6)!(l-k)} (a^A_l-a^\tP_l) \biggr|_{a=0} \,, \text{ for } k=4,5 \,.
\end{align}\par%
Let us now parametrize the pseudoscalar form factor using%
\begin{align} \label{eq_residualFF_P_alternative}
 P(Q) &= \biggl( 1 + \frac{Q^2}{m_\pi^2} \biggr) P_1(Q) - \frac{Q^2}{m_\pi^2} P_2(Q) \,.
\end{align}%
The ansatz~\eqref{eq_residualFF_P_alternative} becomes consistent with PCAC in the continuum limit once we demand that
\begin{align} \label{eq_PCAC_constraint_P}
 P_1(Q) \biggr|_{a=0}  &= A(Q) \biggr|_{a=0}  \,, &  P_2(Q) \biggr|_{a=0}  &= \tP(Q) \biggr|_{a=0} \,.
\end{align}%
Unfortunately, PCAC is broken on the lattice by discretization effects, such that $P_1(Q)$ and $P_2(Q)$ differ from $A(Q)$ and $\tP(Q)$ at nonzero lattice spacing. Hence, we use the same ansatz for both (e.g., the dipole form~\eqref{eq_Dipole} or the $z$-expansion~\eqref{eq_zExpansion}), but we start with independent parameters. Here, the asymptotic constraints yield%
\begin{align} \label{eq_asymptotic_constraint_P}
\lim_{Q \rightarrow \infty} & Q^n \biggl(P_1(Q)-P_2(Q)\biggr) \overset{!}{=} 0 \,, \text{ for } n<6 \,,
\end{align}%
independent of $a$. Note, that eq.~\eqref{eq_PCAC_constraint_P} and~\eqref{eq_asymptotic_constraint_P} can only be fulfilled simultaneously if the axial and induced pseudoscalar form factors meet the requirement~\eqref{eq_constraint_PCAC_asymptotic}. For the two parametrizations (cf.\ sections~\ref{sec_param_dipole} and~\ref{sec_param_zexp}) that we consider, the constraints for $n<4$ hold automatically. Similar to the above, the remaining two constraints can be satisfied by%
\begin{align}
 g_{P_1} M_{P_1}^4 \overset{!}{=} g_{P_2} M_{P_2}^4
\end{align}
when using the dipole ansatz, and by%
\begin{align}
\bigl(a^{P_1}_k-a^{P_2}_k \bigr) &= \frac{(-1)^k}{k!(5-k)!} \sum_{l=6}^N \frac{l! }{(l-6)!(l-k)} (a^{P_1}_l-a^{P_2}_l) \,, \text{ for } k=4,5 \,,
\end{align}%
when using the $z$-expansion.\par%
To summarize, if we wish our form factor parametrizations to obey PCAC in the continuum limit, we start by parametrizing $P(Q)$ as in eq.~\eqref{eq_residualFF_P_alternative}, thereby introducing more parameters at first. However, as discussed above, these parameters are highly constrained such that the ansatz enforcing PCAC will have less free fit parameters in the end. Using the dipole ansatz, we have $g_A, M_A, g_\tP, M_\tP, g_{P_1}, M_{P_1}, g_{P_2}, M_{P_2}$, which can be factorized in a lattice spacing dependent and a lattice spacing independent part as shown in eq.~\eqref{eq_extrapolation}. The constraints discussed above can be incorporated by setting
\begin{align}
  g_{P_2}  &= g_{P_1} \biggl(\frac{M_{P_1}}{M_{P_2}}\biggr)^4 \,, &
  g_{P_1}^{\na} &= g_A^{\na} \,,  &
  \bigg[ g_{P_2}^{\na} &= g_\tP^{\na} \,, \biggr]   \\
  g_\tP^{\na}  &= g_A^{\na} \biggl(\frac{M_A^{\na}}{M_\tP^{\na}}\biggr)^4 \,,  &
  M_{P_1}^{\na} &= M_{A}^{\na} \,,  &
  M_{P_2}^{\na} &= M_\tP^{\na} \,,
\end{align}
where the constraint in brackets is not independent of the others. If one uses the $z$-expansion, one starts with $a^A_n, a^\tP_n, a^{P_1}_n, a^{P_2}_n$, $n=4,5\dots,N$. Again, we assume these coefficients to be factorized as in eq.~\eqref{eq_extrapolation}. Here, the constraints discussed above can be implemented by setting
\begin{align}
a^{P_2}_k &= a^{P_1}_k + \frac{(-1)^k}{k!(5-k)!} \sum_{l=6}^N \frac{l! }{(l-6)!(l-k)} (a^{P_2}_l-a^{P_1}_l) \,, \text{ for } k=4,5 \,, \\
a^{\tP,\na}_k &= a^{A,\na}_k + \frac{(-1)^k}{k!(5-k)!} \sum_{l=6}^N \frac{l! }{(l-6)!(l-k)} (a^{\tP,\na}_l-a^{A,\na}_l) \,, \text{ for } k=4,5 \,, \\
a^{P_1,\na}_k &= a^{A,\na}_k \,, \text{ for } k=4,5,\dots,N \,, \\
a^{P_2,\na}_k &= a^{\tP,\na}_k \,, \text{ for } k=\biggl[4,5,\!\biggr]\, 6,\dots,N \,.
\end{align}%
As above, the constraints in brackets are not independent of the others.
\subsubsection{Continuum, quark mass, and volume extrapolation\label{sec_param_extrapolation}}
In our combined analysis of all the ensembles we will consider four kinds of fits: the dipole ansatz ($\text{2P}$), the $z$-expansion with the correct asymptotic behavior ($z^{4+(N-3)}$), and the two corresponding parametrizations where PCAC holds automatically in the continuum ($!\text{2P}$ and $!z^{4+(N-3)}$, respectively). They are listed in table~\ref{tab_FF_params}. We have factorized the occurring parameters $x=x^{\na} x^a$ (see eq.~\eqref{eq_extrapolation}) into a continuum limit part $x^\na$, and a part which describes discretization effects $x^a$, where $x^a\rightarrow 1 (a \rightarrow 0)$, see eq.~\eqref{eq_xa_continuum}. In the parametrizations that respect PCAC, the number of parameters is reduced due to the constraints derived in section~\ref{sec_param_pcac} (see also table~\ref{tab_FF_params}). We perform a combined continuum, quark mass, and volume extrapolation using the generic ansatz%
\begin{align}
\begin{split}
x^\na(m_\pi,m_K,L) &= c_1^x + c_2^x \bar m^2 + c_3^x \delta m^2\\
&\quad+ c_4^x \frac{m_\pi^2}{\sqrt{m_\pi L}} e^{-m_\pi L} + c_5^x \frac{m_K^2}{\sqrt{m_K L}} e^{-m_K L} + c_6^x \frac{m_\eta^2}{\sqrt{m_\eta L}} e^{-m_\eta L} \,, 
\end{split} \label{eq_extrapolation_cont}\\
 x^a(a,m_\pi,m_K) &= 1 + a^2 \bigl( d_1^x + d_2^x \bar m^2 + d_3^x \delta m^2 \bigr) \,, \label{eq_extrapolation_disc}
\end{align}
where we set $m_\eta^2=(4 m_K^2-m_\pi^2)/3$ using the Gell-Mann--Oakes--Renner relation~\cite{GellMann:1968rz}. The functional form of the finite volume terms is motivated by the leading contribution found in ChPT calculations of the axial coupling, cf.\ refs.~\cite{Beane:2004rf,Khan:2006de}. To parametrize the quark mass plane we have defined the linear combinations%
\begin{align}%
\begin{split}
\delta m^2 &= m_K^2-m_\pi^2 \approx B(m_s-m_\ell)\,,\\
\bar{m}^2  &= (2m_K^2+m_\pi^2)/3 \approx 2B(m_s+2m_\ell)/3\,, \taghere
\end{split}
\end{align}%
such that $\delta m=0$ corresponds to exact flavor symmetery, i.e., the blue line in figure~\ref{fig_ensembles}, while the green line with physical average masses is defined by $\bar{m}=\text{phys.}\approx\unit{411}{\mega\electronvolt}$. Along the line of an approximately physical strange quark mass, i.e., the red line in figure~\ref{fig_ensembles}, the average mass varies; all ensembles used in this study have $\bar{m}<\unit{500}{\mega\electronvolt}$. Note that our additional ensembles with exact flavor symmetry (along the blue line in figure~\ref{fig_ensembles}) facilitate the determination of the parameters $c_1^x$, $c_2^x$, $d_1^x$, and $d_2^x$.\par%
\begin{table}[t]%
\centering%
\caption{\label{tab_FF_params}Overview of the form factor parametrizations. We will use the dipole ansatz (2P) and the $z$-expansion with the correct asymptotic behavior ($z^{4+(N-3)}$) as described in sections~\ref{sec_param_dipole} and~\ref{sec_param_zexp}, respectively. For both cases we also consider parametrizations where PCAC is fulfilled in the continuum limit (marked by a preceding~!\ in the identifier), cf.\ section~\ref{sec_param_pcac}. In the rightmost column, we give the total number of fit parameters used for the combined continuum, quark mass, and volume extrapolation per form factor, assuming that formulas~\eqref{eq_extrapolation_cont} and~\eqref{eq_extrapolation_disc} are used for the extrapolation of $x^\na$ and $x^a$, respectively.}%
\begin{widetable}{\textwidth}{lclll}%
\toprule
id & PCAC & $x^\na$ & $x^a$ & \#params per FF \\
\midrule
$\text{2P}$	&\pcacOF& $g_A^\na$, $g_\tP^\na$, $g_P^\na$, 				& $g_A^a$, $g_\tP^a$, $g_P^a$, 				& $18$\\ 
		&	& $M_A^\na$, $M_\tP^\na$, $M_P^\na$				& $M_A^a$, $M_\tP^a$, $M_P^a$				&\\ \midrule
$!\text{2P}$   	&\pcacON& $g_A^\na$, 			 						& $g_A^a$, $g_\tP^a$, $g_{P_1}^a$,			& $13$\\ 
		&	& $M_A^\na$, $M_\tP^\na$ 						& $M_A^a$, $M_\tP^a$, $M_{P_1}^a$, $M_{P_2}^a$		&\\ \midrule 
$z^{4+(N-3)}$  	&\pcacOF& $a^{A,\na}_4$, $a^{A,\na}_5$, \dots, $a^{A,\na}_N$,   	& $a^{A,a}_4$, $a^{A,a}_5$, \dots, $a^{A,a}_N$,		&$9N-27$\\ 
		&	& $a^{\tP,\na}_4$, $a^{\tP,\na}_5$, \dots, $a^{\tP,\na}_N$,	& $a^{\tP,a}_4$, $a^{\tP,a}_5$, \dots, $a^{\tP,a}_N$,	&\\
	        &	& $a^{P,\na}_4$, $a^{P,\na}_5$, \dots, $a^{P,\na}_N$		& $a^{P,a}_4$, $a^{P,a}_5$, \dots, $a^{P,a}_N$		&\\ \midrule
$!z^{4+(N-3)}$  &\pcacON& $a^{A,\na}_4$, $a^{A,\na}_5$, \dots, $a^{A,\na}_N$,   	& $a^{A,a}_4$, $a^{A,a}_5$, \dots, $a^{A,a}_N$, 	&$8N-30$\\ 
		&	& $a^{\tP,\na}_6$, $a^{\tP,\na}_7$, \dots, $a^{\tP,\na}_N$	& $a^{\tP,a}_4$, $a^{\tP,a}_5$, \dots, $a^{\tP,a}_N$,	&\\
	        &	&											& $a^{P_1,a}_4$, $a^{P_1,a}_5$, \dots, $a^{P_1,a}_N$,	&\\
	        &	&											& $a^{P_2,a}_6$, $a^{P_2,a}_7$, \dots, $a^{P_2,a}_N$	&\\
\bottomrule
\end{widetable}%
\end{table}%
\subsection{Results\label{sec_results}}
\begin{figure}[pt]
\centering
\includegraphics[width=\textwidth]{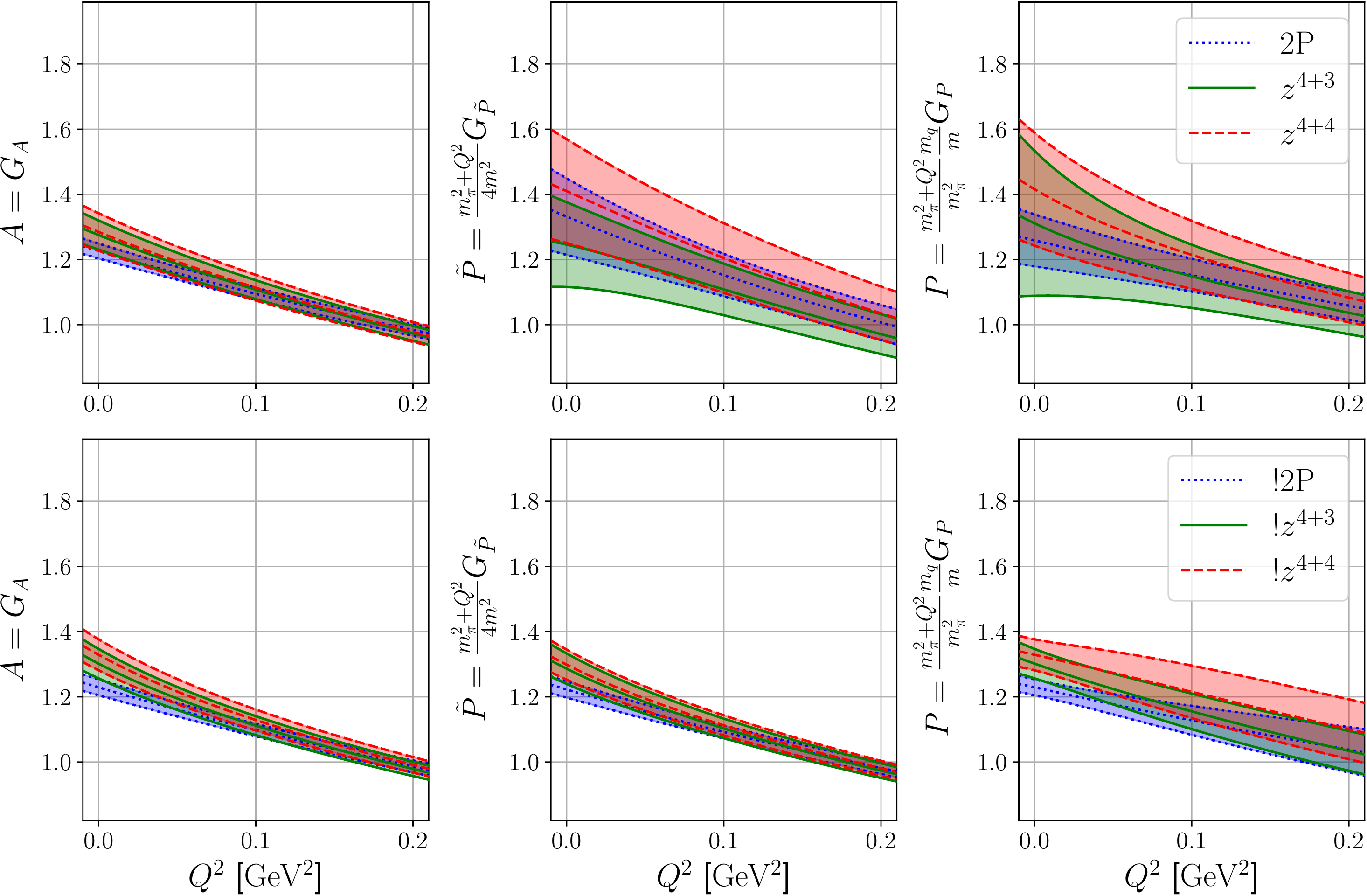}%
\caption{\label{fig_continuum_all_fits}Comparison of continuum results at the physical point for the residual form factors obtained using the different fits, cf.\ table~\ref{tab_FF_params}. The fits enforcing PCAC in the continuum (lower panels) yield significantly smaller statistical errors. The mean values of the plotted curves can be reproduced using the parameters provided in table~\ref{tab_mean_param_results}.}
\end{figure}
\begin{table}
\centering%
\caption{\label{tab_mean_param_results}Results for the parameters at the physical point in the continuum for the dipole ansatz~\eqref{eq_Dipole} and the $z$-expansion~\eqref{eq_zExpansion}, together with the uncorrelated $\chi^2$ per degree of freedom of the corresponding fit. For convenience, we also provide the values for the parameters, which are entirely fixed by constraints.}%
\begin{widetable}{\textwidth}{lcccccccccc}%
\toprule
id		& $X$		&$\chi^2/\text{d.o.f.}$	& $g_X$		& \!\!\!\!$M_X\,[\giga\electronvolt]$\!\!\!\! \\\midrule
2P        	 & $A$        & 0.80    &  1.226		 &  1.311		 \\
          	 & $\tP$      & 0.65    &  1.332		 &  1.154		 \\
          	 & $P$        & 0.66    &  1.259		 &  1.487		 \\\midrule
!2P       	 & $A=P_1$    & 0.71    &  1.229		 &  1.312		 \\
          	 & $\tP=P_2$  &         &  1.222		 &  1.313		 \\\midrule
%
id		& $X$		&$\chi^2/\text{d.o.f.}$	& $a_0^X$	& $a_1^X$	& $a_2^X$	& $a_3^X$	& $a_4^X$	& $a_5^X$	& $a_6^X$	& $a_7^X$  \\\midrule
$z^{4+3}$ 	 & $A$        & 0.94   &  1.009		& \M 1.756		& \M 1.059		&  1.621		&  3.919		& \M 5.739		&  2.005		 \\
          	 & $\tP$      & 0.66   &  1.008		& \M 1.831		& \M 1.713		&  4.994		& \M 1.522		& \M 1.984		&  1.047		 \\
          	 & $P$        & 0.66   &  1.066		& \M 1.461		& \M 1.053		& \M 2.504		& 12.446		& \M 12.260		&  3.766		 \\\midrule
!$z^{4+3}$	 & $A=P_1$    & 0.83   &  1.013		& \M 1.713		& \M 0.591		& \M 0.771		&  7.790		& \M 8.418		&  2.689		 \\
          	 & $\tP=P_2$  &        &  1.007		& \M 1.678		& \M 0.680		& \M 0.653		&  7.701		& \M 8.382		&  2.684		 \\\midrule
$z^{4+4}$ 	 & $A$        & 0.97   &  1.014		& \M 1.777		& \M 1.026		&  1.596		&  3.928		& \M 5.740		&  2.005		& \M 0.00003		 \\
          	 & $\tP$      & 0.61   &  1.080		& \M 2.211		& \M 0.920		&  4.201		& \M 1.164		& \M 2.016		&  1.031		& 0.00001		 \\
          	 & $P$        & 0.66   &  1.117		& \M 1.692		& \M 0.641		& \M 2.858		& 12.583		& \M 12.271		&  3.762		& \M 0.00012		 \\\midrule
!$z^{4+4}$	 & $A=P_1$    & 0.79   &  1.027		& \M 1.773		& \M 0.488		& \M 0.854		&  7.818		& \M 8.418		&  2.688		& 0.00002		 \\
          	 & $\tP=P_2$  &        &  1.015		& \M 1.703		& \M 0.662		& \M 0.625		&  7.649		& \M 8.352		&  2.678		& \M 0.00031		 \\
\bottomrule
\end{widetable}%
\end{table}
Figure~\ref{fig_continuum_all_fits} provides a compilation of (continuum, quark mass, and finite volume extrapolated) form factors that have been obtained from the parametrizations discussed in the previous sections. The parameters producing the central values can be taken from table~\ref{tab_mean_param_results}. Surprisingly, even the fits using a dipole ansatz (2P) give a reasonable description of the data (actually, it has in most cases the smallest $\chi^2/\text{d.o.f.}$ of all fits, cf.\ table~\ref{tab_mean_param_results}), despite the fact that the functional form is very constrained. However, the latter may lead to an underestimation of the error, and it may also induce a smaller slope at zero momentum transfer. In order to reduce this bias one may relax the constraints due to the choice of parametrization. The currently most popular and probably best suited ansatz for this task is the $z$-expansion described in section~\ref{sec_param_zexp}. To this end, we have performed $z^{4+3}$, and $z^{4+4}$ fits (and the corresponding fits that are constrained to be consistent with PCAC in the continuum limit). While the $z^{4+3}$ fit is almost as restrictive as the dipole ansatz ($27$ vs.~$18$ parameters per form factor), expansions with a larger number of parameters ($z^{4+4}$, $z^{4+5}$, etc.) introduce less and less parametrization bias. In practice, however, the choice will always be a balancing act between reducing the parametrization bias and being able to control the systematics of all occurring parameters. Therefore, the statistical quality of the data and its coverage of lattice spacings, quark masses, and volumes are a deciding factor.\par%
We emphasize that PCAC was not enforced when extracting the form factors from fits to the correlators. Nevertheless, due to the advances in the understanding of excited state contaminations in the correlation functions, we are now able to resolve the ground state contributions such that the resulting form factors agree with PCAC (and also PPD) reasonably well. This enables us to perform combined fits to all form factors using parametrizations that automatically obey PCAC in the continuum limit. As one can easily see in table~\ref{tab_FF_params}, the resulting parametrizations are much more restrictive than their counterparts. For example, the dipole fit (!2P) has in total three free parameters (at the physical point in the continuum limit) for all form factors. However, in contrast to the parametrization bias discussed above, the PCAC constraints do not evoke any kind of systematic uncertainty, since they only reflect an exactly known symmetry. Unsurprisingly, we find that the continuum extrapolation is more stable when using these PCAC-consistent parametrizations. Overall, we find that both the !2P and the $!z^{4+3}$ fit yield very good descriptions of the data ($\chi^2/\text{d.o.f.}=0.71$ and $\chi^2/\text{d.o.f.}=0.83$, respectively), while still allowing for a controlled extrapolation to the physical point. Our final results are therefore based on these fits. The $!z^{4+4}$ fit also provides a very good description of our data ($\chi^2/\text{d.o.f.}=0.79$). However, it is less trustworthy since it relies on an excessive number of parameters, which leads to larger systematic uncertainties in the combined continuum, quark mass, and volume extrapolation.\par%
\DeclareRobustCommand{\captiontail}{obtained using the $!z^{4+3}$ ansatz fitted to all available ensembles. This is a combined fit to all form factors with $\chi^2/\text{d.o.f.}=0.83$. The panels correspond to different lattice spacings and quark mass trajectories (see section~\ref{sec_lattice_setup}), where the yellow band corresponds to the form factor obtained from the fit, evaluated at physical masses, at infinite volume, but at the lattice spacing corresponding to the particular row.}%
\begin{figure}[pt]
\centering
\includegraphics[width=\textwidth]{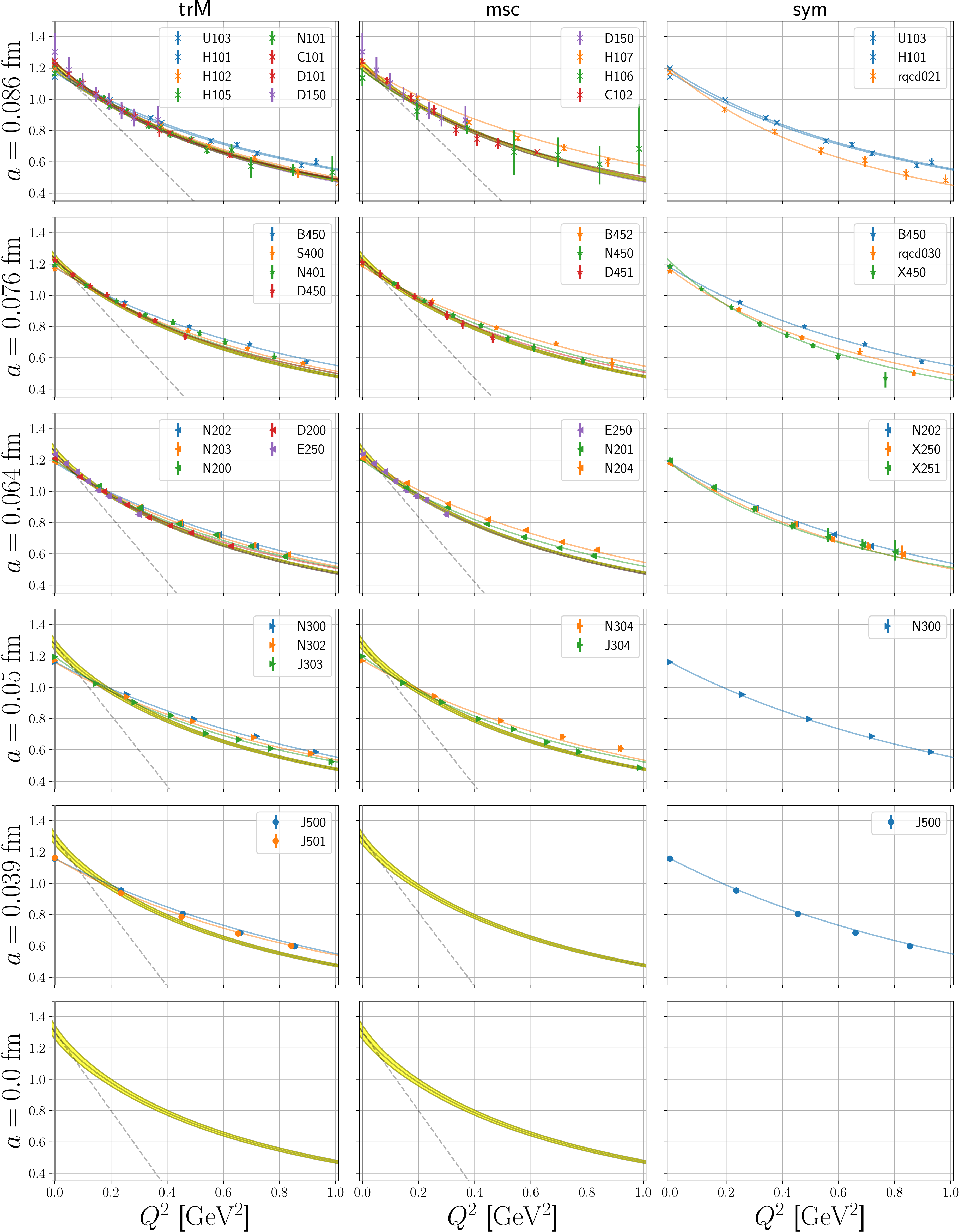}%
\caption{\label{fig_final_GA}The axial form factor $G_A(Q^2)$ \captiontail}
\end{figure}
\begin{figure}[pt]
\centering
\includegraphics[width=\textwidth]{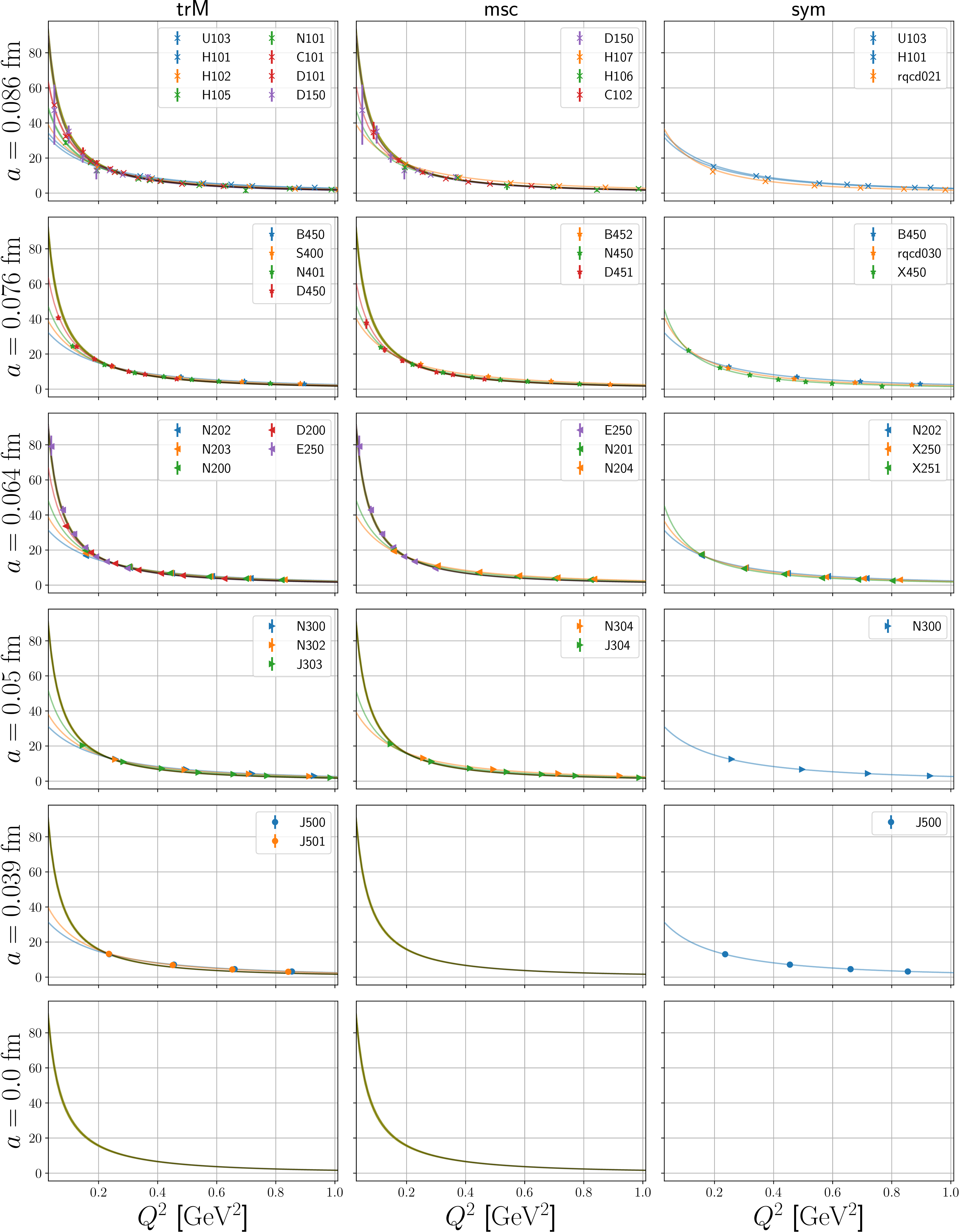}%
\caption{\label{fig_final_GPtilde}The induced pseudoscalar form factor $G_\tP(Q^2)$ \captiontail}
\end{figure}
\begin{figure}[pt]
\centering
\includegraphics[width=\textwidth]{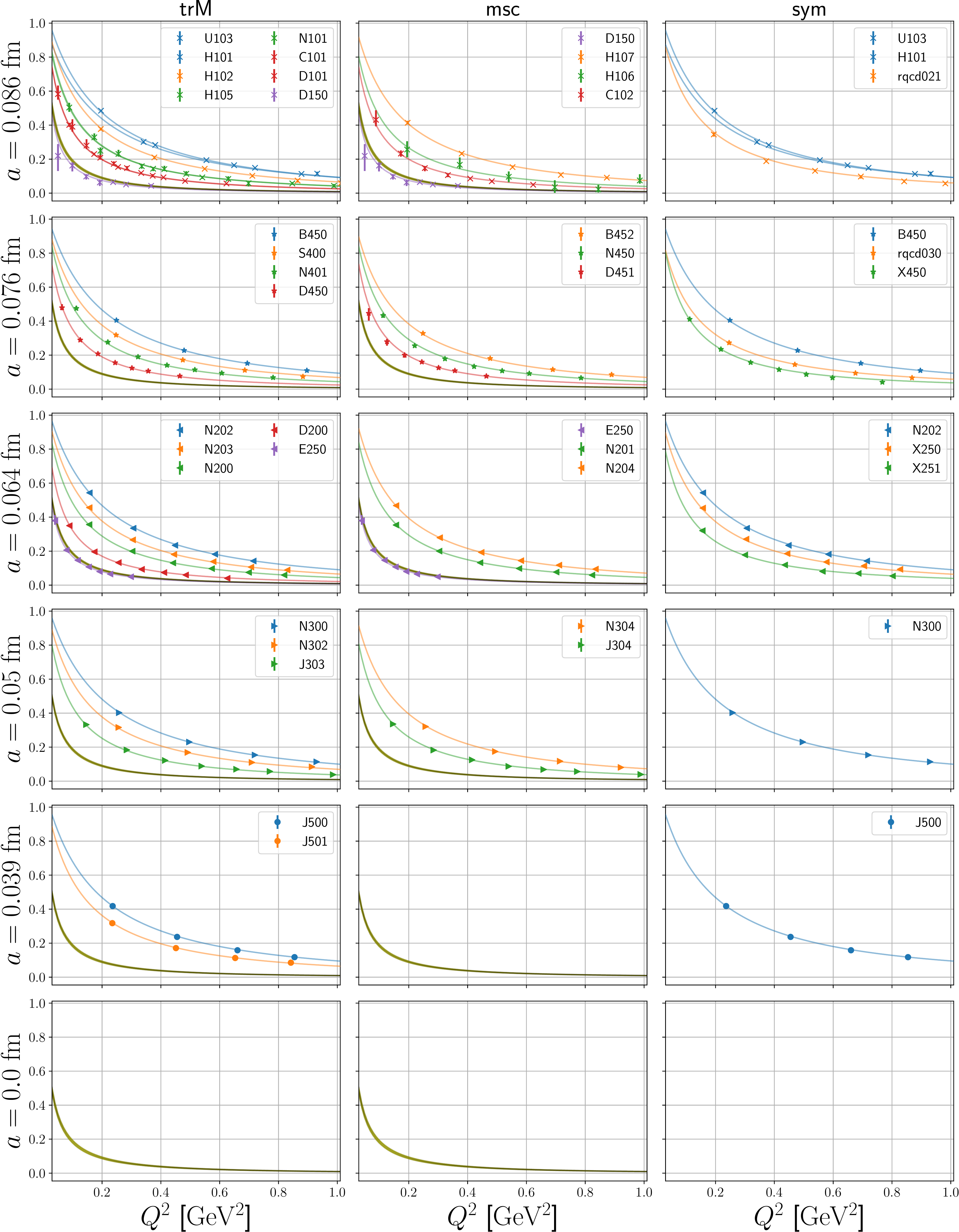}%
\caption{\label{fig_final_GP}The pseudoscalar form factor $\tfrac{m_\ell}{m} G_P(Q^2)$ \captiontail}
\end{figure}
In figures~\ref{fig_final_GA}, \ref{fig_final_GPtilde} and~\ref{fig_final_GP}, we show our data and how well it is described by the $!z^{4+3}$ fit. (For the !2P fit such plots look equally convincing.) The $6$ rows in each figure correspond to the five available lattice spacings and to the continuum limit, while the columns correspond to the different quark mass trajectories, see the explanation in section~\ref{sec_lattice_setup}. Along the trM and msc trajectories, some of the ensembles have close to physical masses (C101, C102, D200, D450, D451, with $m_\pi \approx \unit{200}{\mega\electronvolt}$ and, in particular, D150 and E250, with $m_\pi\approx\unit{130}{\mega\electronvolt}$). Note that the sym trajectory with exact flavor symmetry does not approach the physical point in the quark mass plane. The colored curves show the mean fit result evaluated at the masses, volume, and lattice spacing of the respective ensemble, while the yellow band corresponds to the extrapolated result at physical masses, in infinite volume, and at the lattice spacing for the particular row. The data show that the form factors exhibit an increasing slope (in $Q^2$) for decreasing pion masses (as one would expect) and lattice spacings. In figure~\ref{fig_final_GA} one can see that also the data for $g_A=G_A(0)$ is well described by the fit. However, in particular for large pion masses, the data at $Q^2=0$ lies significantly below the extrapolated value, which highlights the importance of the extrapolation to physical masses. In this context one should note that the $z$-expansion (shown here) exhibits a different mass dependence as the dipole ansatz, since the pion mass directly enters the definition of $z$ in eq.~\eqref{eq_zDefinition}. What is harder to see from the figures is the increase of the slope towards the smaller lattice spacings. In order to provide some way for the reader to appreciate how big this effect is, we indicate the slope of $G_A$ at $Q^2=0$ in figure~\ref{fig_final_GA} by a dashed line. In figures~\ref{fig_final_GPtilde} and~\ref{fig_final_GP} it is particularly encouraging that the data for our physical mass ensemble at small lattice spacing (E250) nicely reproduces the expected pion pole structure in the (induced) pseudoscalar form factor (cf.\ eq.~\eqref{eq_residualFF}).\par%
\begin{figure}[tb]\centering%
\centering%
\includegraphics[width=\textwidth]{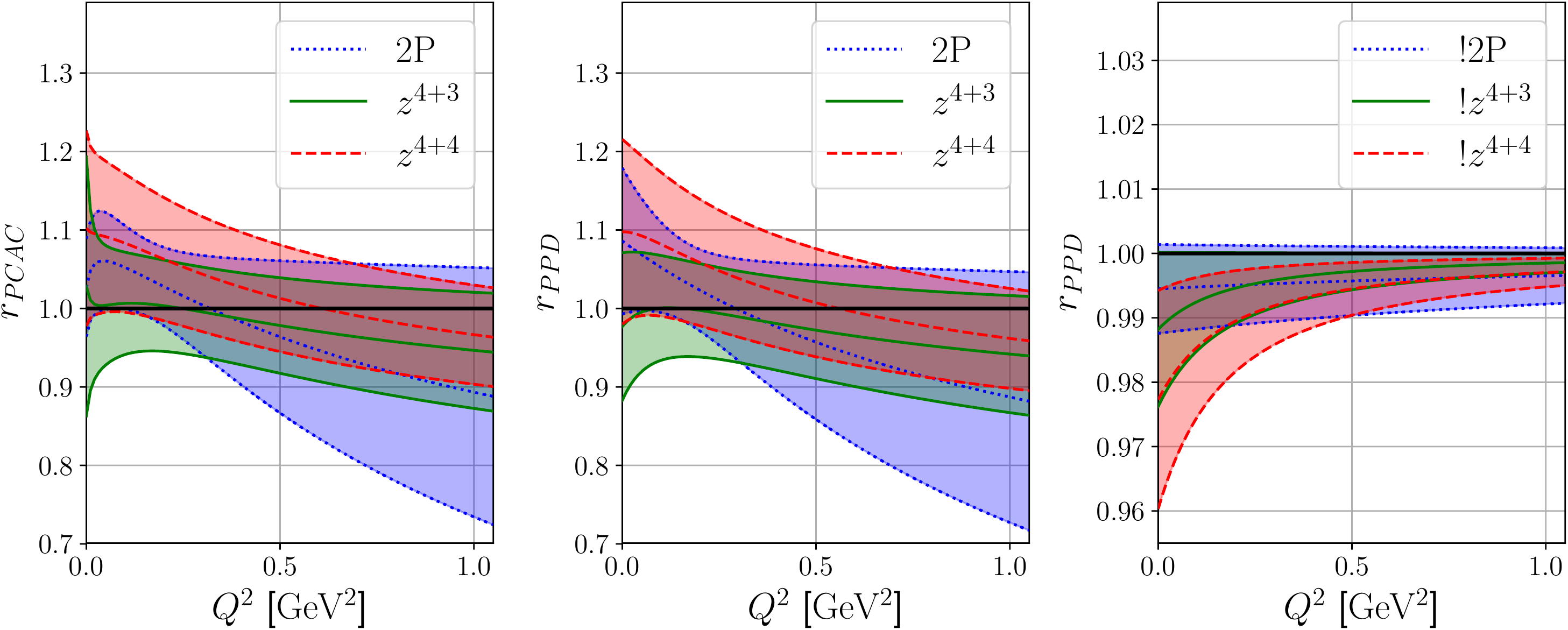}%
\caption{\label{fig_rpcac_rppd_physical_point}The $r_{\rm{PCAC}}$ (left panel) and $r_{\rm{PPD}}$ (center panel) ratios (defined in eqs.~\eqref{eq_ratioPCAC} and~\eqref{eq_ratioPPD}) at the physical point in the continuum limit. These are obtained using a $\text{2P}$ (dotted, blue), $z^{4+3}$ (solid, green) or a $z^{4+4}$ (dashed, red) fit ansatz. In the case of $r_{\rm{PPD}}$, we also show results of the corresponding fits that are constrained to be consistent with PCAC in the continuum limit (right panel), cf.\ table~\ref{tab_FF_params}.}
\end{figure}%
Above, in figure~\ref{fig_r_mpi}, we have demonstrated that the nucleon form factor data extracted from the correlation functions using the results presented in section~\ref{sec_final_param} agree reasonably well with PCAC and PPD. In figure~\ref{fig_rpcac_rppd_physical_point} we show the result for the ratios $r_{\rm{PCAC}}$ (left panel) and $r_{\rm{PPD}}$ (center panel) after the extrapolation, using the previously discussed form factor parametrizations that do not enforce PCAC. We find that both PCAC and PPD are fulfilled within large statistical errors. As one can see by comparing the center and the right panel (note the difference in the scale between the two plots), the dipole and $z$-expansion fits with enforced PCAC relation allow for a much better resolution of possible deviations from the pion pole dominance assumption for the induced pseudoscalar form factor. We find the PPD assumption to be valid at the 1\%--2\% level at all momentum transfers, independent of the parametrization.\par%
\begin{table}
\centering%
\caption{\label{tab_results}Results for the form factors $G_X(0)$ at zero momentum transfer and for the mean squared radii $r_X^2 = -6 G^\prime_X(0)/G_X(0)$ obtained from fits using various form factor parametrizations. We also provide results for the pion-nucleon coupling $g_{\pi NN}$ and for the induced pseudoscalar coupling at the muon capture point $g^\star_P$, which can be directly compared to the experimental value $g^\star_P=8.01(55)$ from muon capture~\cite{Andreev:2012fj,Andreev:2015evt}.}%
\begin{widetable}{\textwidth}{lllllllll}%
\toprule
id		&$G_A(0)$	&$r_A^2\,[\femto\meter\squared]$	&$G_\tP(0)$	&$r_\tP^2\,[\femto\meter\squared]$	&$\tfrac{m_\ell}{m}G_P(0)$	&$r_P^2\,[\femto\meter\squared]$	& $g^\star_P$	& $g_{\pi NN}$	\\\midrule
2P         &    1.226(23) &    0.272(21) &      246(22) &    11.98(12) &    1.259(80) &     11.85(7) &      9.02(76) &   15.55(3.00) \\
!2P        &    1.229(24) &    0.272(21) &       226(5) &     11.91(2) &    1.229(24) &     11.84(8) &      8.30(17) &     12.93(55) \\
$z^{4+3}$  &    1.275(45) &    0.351(58) &      231(24) &    11.85(22) &   1.311(222) &    12.04(36) &      8.48(84) &   13.23(3.06) \\
!$z^{4+3}$ &    1.302(45) &    0.449(42) &       238(9) &     12.06(4) &    1.302(45) &    11.94(14) &      8.68(30) &   14.78(1.16) \\
$z^{4+4}$  &    1.285(58) &    0.357(47) &      261(30) &    11.99(12) &   1.416(173) &    12.10(14) &    9.54(1.04) &   17.41(4.31) \\
!$z^{4+4}$ &    1.329(48) &    0.465(24) &       240(9) &     12.06(3) &    1.329(48) &    11.83(19) &      8.76(30) &   15.07(1.14) \\
\bottomrule
\end{widetable}%
\end{table}
The results for the form factors at zero momentum transfer and for the mean squared radii are given in table~\ref{tab_results}, where we also provide the induced pseudoscalar coupling at the muon capture point%
\begin{align} \label{eq_gPt_muon_capture}
g^\star_P &= \frac{m_\mu}{2 m} G_\tP(0.88 m_\mu^2) \,,
\end{align}%
with the muon mass $m_\mu=\unit{105.6}{\mega\electronvolt}$, and for the pion-nucleon coupling constant%
\begin{align}%
 g_{\pi NN} &= \lim_{Q^2\rightarrow - m_\pi^2} \frac{m_\pi^2+Q^2}{4 m F_\pi} G_\tP(Q^2) = \frac{m}{F_\pi} \tP(-m_\pi^2) \,,
\end{align}%
where we use the PDG value of $F_\pi=\unit{92.07}{\mega\electronvolt}$~\cite{Tanabashi:2018oca}. As a general trend we find that the fits which ensure that PCAC is satisfied in the continuum limit yield smaller statistical uncertainties. We find reasonable values for $g^\star_P$ that are in agreement with the approximate realization of PPD in nature.\footnote{In earlier work~\cite{Bali:2018qus} we found much smaller values that would have suggested a strong violation of the PPD assumption.} From table~\ref{tab_results} one can actually read off that the different parametrizations yield compatible results, with the exception of the axial radius, where the dipole fits give significantly smaller values, and the pion-nucleon coupling, where the $!z^{4+4}$ fit seems to be an outlier.\par%
In our opinion, the !2P and the $!z^{4+3}$ yield the most reliable results (for the fits with more free parameters the chiral and continuum extrapolation is less stable). However, given our set of available data, we cannot decide whether the !2P or the $!z^{4+3}$ fit is better. We have therefore decided to perform an analysis of systematic uncertainties for both of these fits. In table~\ref{tab_results_error} we provide, in addition to the statistical error $()_s$, estimates for the systematic uncertainties due to the quark mass extrapolation $()_m$ and the continuum extrapolation $()_a$. To this end, we have performed additional fits with cuts in the fit ranges ($\bar m < \unit{450}{\mega\electronvolt}$ and $a<\unit{0.08}{\femto\meter}$, respectively). We then take the difference between the results from these fits and our main result as an estimate of the corresponding systematic uncertainties. As discussed in section~\ref{sec_excited_energies}, our main analysis is performed using the fit ansatz with the energies of the nucleon-pion states fixed to the noninteracting value. To estimate the systematic uncertainty due to this choice, we have performed additional fits, where the energies for the nucleon-pion states are free fit parameters.\footnote{In these fits, we did not allow for the contributions of additional generic excited states to the three-point functions. Keeping these, without fixing the $N\pi$ energies turned out not to be feasible for the statistics presently available on most of our ensembles.} The $N\pi$ energies obtained from these fits are consistent with those presented in figure~\ref{fig_excitedenergies}. The difference between our main result and the result obtained from this alternative fit is given as an estimate for the systematic uncertainty of our excited state analysis $()_{ex}$.\par%
\begin{table}
\centering%
\caption{\label{tab_results_error}Results obtained from the !2P and the $!z^{4+3}$ fit including the statistical error $()_s$ and estimates of the systematic uncertainties due to quark mass extrapolation $()_m$, due to the continuum extrapolation $()_a$, and due to additional excited state effects $()_{ex}$. The systematics are specific to the particular fits and do not reflect differences between fitans\"atze. Since both fits satisfy PCAC in the continuum, $G_A(0) = \tfrac{m_\ell}{m}G_P(0)$ holds automatically.}%
\begin{widetable}{\textwidth}{lll}%
\toprule
		&!2P				&!$z^{4+3}$ \\\midrule
$G_A(0) = \tfrac{m_\ell}{m}G_P(0)$ & $1.229 (24)_s (6)_{ex} (3)_m (17)_a$           & $1.302 (45)_s (42)_{ex} (38)_m (46)_a$         \\
$G_\tP(0)$                         & $226 (5)_s (4)_{ex} (2)_m (2)_a$               & $238 (9)_s (5)_{ex} (7)_m (5)_a$               \\
$r_A^2\,[\femto\meter\squared]$    & $0.272 (21)_s (6)_{ex} (7)_m (24)_a$           & $0.449 (42)_s (42)_{ex} (42)_m (49)_a$         \\
$M_A\,[\giga\electronvolt]$        & $1.312 (50)_s (15)_{ex} (16)_m (54)_a$         & $1.020 (50)_s (52)_{ex} (44)_m (52)_a$         \\
$r_\tP^2\,[\femto\meter\squared]$  & $11.91 (2)_s (0)_{ex} (1)_m (2)_a$             & $12.06 (4)_s (3)_{ex} (4)_m (3)_a$             \\
$r_P^2\,[\femto\meter\squared]$    & $11.84 (8)_s (24)_{ex} (6)_m (2)_a$            & $11.94 (14)_s (8)_{ex} (3)_m (12)_a$           \\
$g^\star_P$                        & $8.30 (17)_s (14)_{ex} (6)_m (8)_a$            & $8.68 (30)_s (18)_{ex} (23)_m (16)_a$          \\
$g_{\pi N N}$                      & $12.93 (55)_s (44)_{ex} (20)_m (32)_a$         & $14.78 (1.16)_s (72)_{ex} (98)_m (67)_a$       \\
$\Delta_{\rm GT}$                  & $0.86 (2.39)_s (3.71)_{ex} (1.21)_m (88)_a \%$ & $6.53 (4.26)_s (1.30)_{ex} (2.90)_m (53)_a \%$ \\
\bottomrule
\end{widetable}%
\end{table}
\subsection{Discussion\label{sec_discussion}}

\begin{figure}
\begin{minipage}{.3\textwidth}%
\includegraphics[width=.915\textwidth]{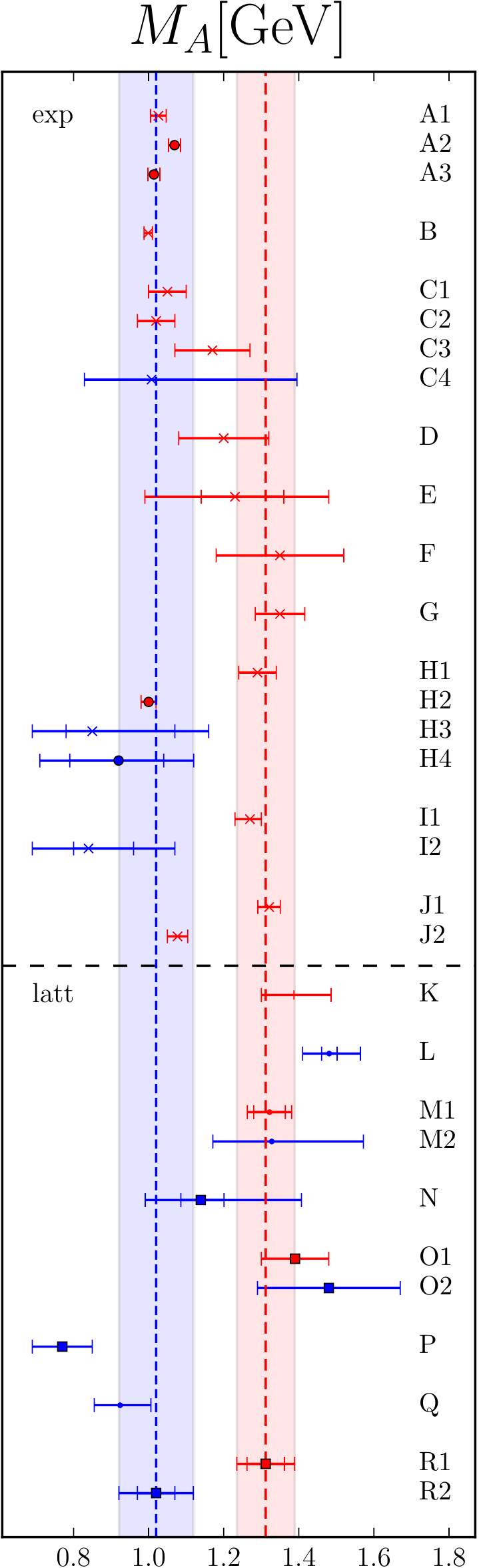}\hfill%
\end{minipage}\begin{minipage}[b]{.7\textwidth}\tiny%
\renewcommand{\arraystretch}{1.1}%
\begin{widetable}{\textwidth}{lll}
\toprule
id & ref. & description \\
\midrule
A	 & \cite{Bernard:2001rs}		 & reanalysis of experimental data (year $\leq 1999$)	\\
A1	 & 				 & \quad $\nu$ scattering; various targets; world avg.\ year $\leq 1990$	\\
A2	 & 				 & \quad $\pi$ electroproduction; world avg.\ year $\leq 1999$	\\
A3	 & 				 & \quad $\pi$ electroproduction; world avg.\ year $\leq 1999$; HBChPT corrected	\\
B	 & \cite{Kuzmin:2007kr}		 & $\nu$ scattering; reanalysis of ANL, BNL, FNAL, CERN, and IHEP data;	\\
	 & 				 & \ various targets; RFG model; dipole ansatz	\\
C	 & \cite{Meyer:2016oeg}		 & reanalysis of $\nu$ scattering data (from BNL~\cite{Baker:1981su}, ANL~\cite{Miller:1982qi}, FNAL~\cite{Kitagaki:1983px})	\\
C1	 & 				 & \quad BNL data; dipole ansatz	\\
C2	 & 				 & \quad ANL data; dipole ansatz	\\
C3	 & 				 & \quad FNAL data; dipole ansatz	\\
C4	 & 				 & \quad combined analysis of BNL, ANL, and FNAL data; $z$-exp	\\
D	 & \cite{Gran:2006jn}		 & $\nu$ scattering; K2K (SciFi); oxygen target; dipole ansatz	\\
E	 & \cite{Adamson:2014pgc}		 & $\nu$ scattering; MINOS; iron target; dipole ansatz	\\
F	 & \cite{AguilarArevalo:2010zc}		 & $\nu$ scattering; MiniBooNE; carbon target; assuming RFG model; dipole ansatz	\\
G	 & \cite{Juszczak:2010ve}		 & reanalysis of \cite{AguilarArevalo:2010zc}; RFG model and spectral function model; dipole ansatz	\\
H	 & \cite{Bhattacharya:2011ah}		 & reanalysis of MiniBooNE~\cite{AguilarArevalo:2010zc} and $\pi$ electroproduction data	\\
H1	 & 				 & \quad MiniBooNE~\cite{AguilarArevalo:2010zc} data; dipole ansatz	\\
H2	 & 				 & \quad $\pi$ electroproduction data (from refs.~\cite{Amaldi:1972vf,Brauel:1973cw,DelGuerra:1975uiy,DelGuerra:1976uj,Esaulov:1978ed}); dipole ansatz	\\
H3	 & 				 & \quad MiniBooNE~\cite{AguilarArevalo:2010zc} data; $z$-exp	\\
H4	 & 				 & \quad $\pi$ electroproduction data (from refs.~\cite{Amaldi:1972vf,Brauel:1973cw,DelGuerra:1975uiy,DelGuerra:1976uj,Esaulov:1978ed}); $z$-exp	\\
I	 & \cite{Bhattacharya:2015mpa}		 & analysis of MiniBooNE~\cite{Aguilar-Arevalo:2013dva} $\bar \nu$ scattering data	\\
I1	 & 				 & \quad dipole ansatz	\\
I2	 & 				 & \quad $z$-exp	\\
J	 & \cite{Nieves:2011yp}		 & reanalysis of MiniBooNE data~\cite{AguilarArevalo:2010zc}	\\
J1	 & 				 & \quad LFG model; dipole ansatz	\\
J2	 & 				 & \quad LFG model + multi-nucleon reactions + RPA, etc., see~\cite{Nieves:2011pp}	\\
\midrule
K	 & \cite{Yamazaki:2009zq}		 & $N_f=2+1$ DWF; RBC/UKQCD; $a=\unit{0.114}{\femto\meter}$	\\
L	 & \cite{Green:2017keo}		 & $N_f=2+1$ Wilson (clover) fermions; $a=\unit{0.114}{\femto\meter}$	\\
M	 & \cite{Alexandrou:2017hac}		 & $N_f=2$ Wilson (clover) fermions; ETMC; $a=\unit{0.0938}{\femto\meter}$	\\
M1	 & 				 & \quad dipole ansatz	\\
M2	 & 				 & \quad $z$-exp	\\
N	 & \cite{Capitani:2017qpc}		 & $N_f=2$ Wilson (clover) fermions; CE	\\
O	 & \cite{Rajan:2017lxk}		 & $N_f=2+1+1$ Wilson (clover-on-HISQ) fermions; PNDME; CE	\\
O1	 & 				 & \quad dipole ansatz	\\
O2	 & 				 & \quad $z$-exp	\\
P	 & \cite{Bali:2018qus}		 & $N_f=2$ Wilson (clover) fermions; RQCD; subtraction method; CE; $z$-exp	\\
Q	 & \cite{Jang:2019vkm}		 & $N_f=2+1+1$ Wilson (clover-on-HISQ) fermions; PNDME; $a=\unit{0.0871}{\femto\meter}$;	\\
	 & 				 & \ takes into account $N\pi$ state;  $z$-exp	\\
R	 & 		 & {\bf This work;} $N_f=2+1$ Wilson (clover) fermions; RQCD;	\\
	 & 				 & \ full resolution of $N\pi$ state; CE	\\
R1	 & 				 & \quad dipole ansatz	\\
R2	 & 				 & \quad $z$-exp	\\
\bottomrule
\end{widetable}
\end{minipage}
\caption{\label{fig_comparison_MA}Compilation of results for the axial dipole mass $M_A$ from experiment (A-J) and lattice simulations (K-R). Extractions based on a dipole ansatz are colored red, while those using any variant of the $z$-expansion are colored blue. The error bands show the results of our !2P (red) and our $!z^{4+3}$ (blue) fits, with all errors added in quadrature.\\
{\bf Symbols:} crosses:~$\nu$ scattering; circles:~$\pi$ electroproduction; tic:~not continuum extrapolated; dot:~single ensemble; square:~continuum extrapolated.\\
{\bf Abbreviations:} RFG:~relativistic Fermi gas~\cite{Smith:1972xh}; LFG:~local Fermi gas; RPA:~random phase approximation~\cite{Bohm:1951zz,Pines:1952zz,Bohm:1953zza}; DWF:~domain wall fermions; HISQ:~highly improved staggered quarks; CE:~continuum extrapolated.}
\end{figure}
\begin{figure}
\begin{minipage}{.3\textwidth}%
\includegraphics[width=.915\textwidth]{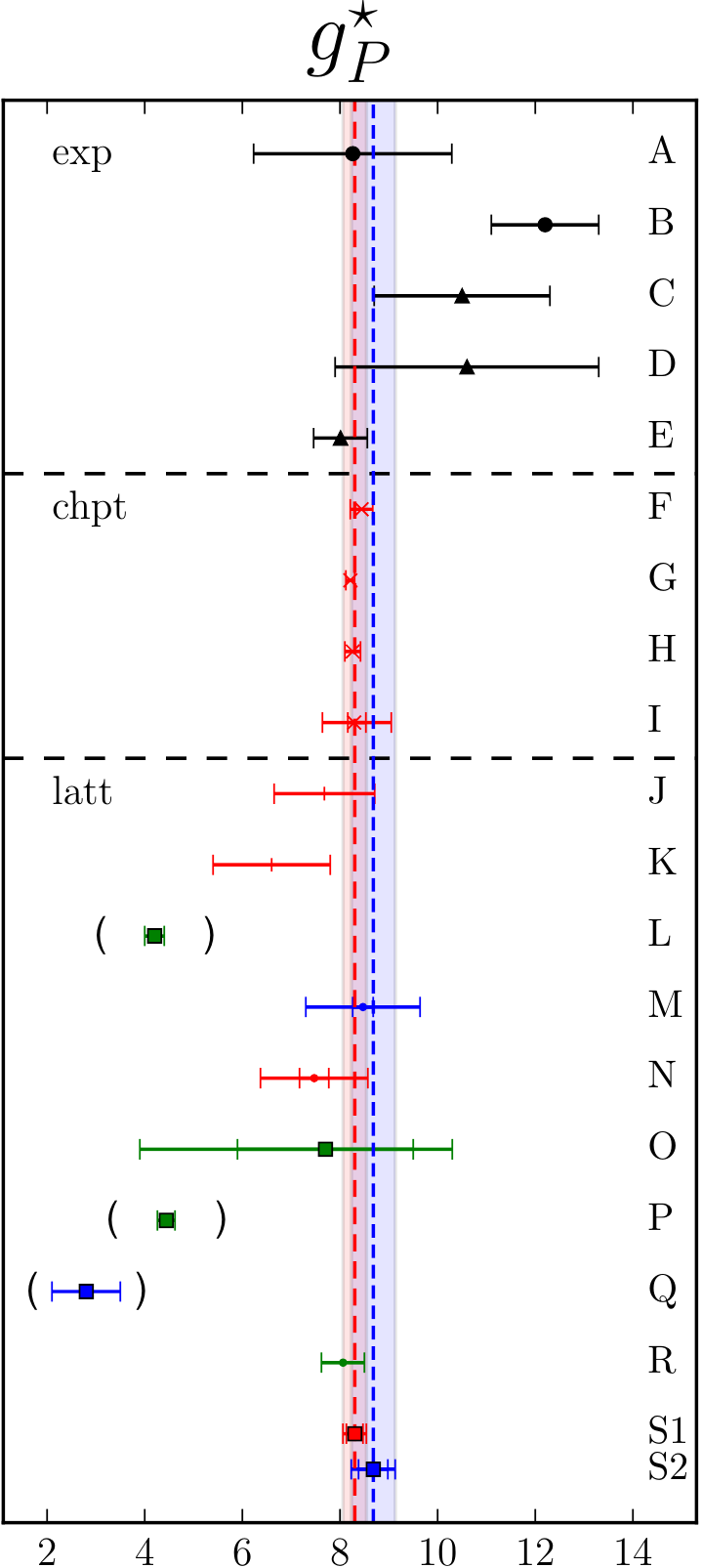}\hfill%
\end{minipage}\begin{minipage}[b]{.7\textwidth}\tiny%
\renewcommand{\arraystretch}{1.1}%
\begin{widetable}{\textwidth}{lll}
\toprule
id & ref. & description \\
\midrule
A	 & \cite{Hart:1977zz}		 & RMC on calcium; $g_P^\star=6.5(1.6)g_A$;	\\
	 & 				 & \ point in plot obtaind by multiplying with $g_A=1.27$	\\
B	 & \cite{Jonkmans:1996my,Wright:1998gi}		 & RMC on hydrogen; TRIUMF; updated value from~\cite{Gorringe:2002xx}	\\
C	 & \cite{Gorringe:2002xx}		 & OMC world avg.\ (year $\leq 1981$)	\\
D	 & \cite{Bardin:1980mi}		 & OMC in hydrogen; Saclay; updated value from~\cite{Gorringe:2002xx}	\\
E	 & \cite{Andreev:2012fj,Andreev:2015evt}		 & OMC in hydrogen gas; MuCap	\\
\midrule
F	 & \cite{Bernard:1994wn}		 & HBChPT; $M_A$ from $\nu$ scattering; assuming $g_{\pi NN}=13.31$	\\
G	 & \cite{Fearing:1997dp}		 & HBChPT; $M_A$ from $\pi$ electroproduction~\cite{DelGuerra:1976uj,Esaulov:1978ed,Choi:1993vt}; assuming $g_{\pi NN}=13.0$	\\
H	 & \cite{Bernard:2001rs}		 & HBChPT; $M_A$ from $\nu$ scattering; assuming $g_{\pi NN}=13.10$	\\
I	 & \cite{Schindler:2006it}		 & covariant BChPT (EOMS); $M_A$ from $\nu$ scattering; assuming $g_{\pi NN}=13.21$~\cite{Schroder:2001rc}	\\
\midrule
J	 & \cite{Lin:2008uz}		 & $N_f=2$ DWF; $a=\unit{0.116}{\femto\meter}$; dipole ansatz	\\
K	 & \cite{Yamazaki:2009zq}		 & $N_f=2+1$ DWF; RBC/UKQCD; $a=\unit{0.114}{\femto\meter}$; dipole ansatz	\\
L	 & \cite{Bali:2014nma}		 & $N_f=2$ Wilson (clover) fermions; RQCD; CE; EFT ansatz	\\
	 & 				 & \ corrected by missing factor of 2	\\
M	 & \cite{Green:2017keo}		 & $N_f=2+1$ Wilson (clover) fermions; $a=\unit{0.114}{\femto\meter}$; $z$-exp	\\
N	 & \cite{Alexandrou:2017hac}		 & $N_f=2$ Wilson (clover) fermions; ETMC; $a=\unit{0.0938}{\femto\meter}$; dipole ansatz	\\
O	 & \cite{Capitani:2017qpc}		 & $N_f=2$ Wilson (clover) fermions; CE; EFT ansatz	\\
P	 & \cite{Rajan:2017lxk}		 & $N_f=2+1+1$ Wilson (clover-on-HISQ) fermions; PNDME; CE; EFT ansatz	\\
Q	 & \cite{Bali:2018qus}		 & $N_f=2$ Wilson (clover) fermions; RQCD; subtraction method; CE; $z$-exp	\\
R	 & \cite{Jang:2019vkm}		 & $N_f=2+1+1$ Wilson (clover-on-HISQ) fermions; PNDME; $a=\unit{0.0871}{\femto\meter}$;	\\
	 & 				 & \ takes into account $N\pi$ state;  $z$-exp	\\
S	 & 		 & {\bf This work;} $N_f=2+1$ Wilson (clover) fermions; RQCD;	\\
	 & 				 & \ full resolution of $N\pi$ state; CE	\\
S1	 & 				 & \quad dipole ansatz	\\
S2	 & 				 & \quad $z$-exp	\\
\bottomrule
\end{widetable}
\end{minipage}
\caption{\label{fig_comparison_gPs}Compilation of data for the pseudoscalar coupling at the muon capture point $g_P^\star$ from experiment (A-E), BChPT (F-I), and lattice simulations (J-S). Extractions based on a dipole ansatz are colored red, while those using any variant of the $z$-expansion are colored blue. Some lattice calculations use an EFT ansatz colored green (pion pole term combined with Taylor expansion). The error bands correspond to the result of our !2P (red) and our $!z^{4+3}$ (blue) fits, with all errors added in quadrature. The lattice results in parentheses are outdated, since they are strongly affected by the pion pole enhanced excited states treated in this article, cf.\ also the discussion in ref.~\cite{Jang:2019vkm}.\\
{\bf Symbols:} circle:~radiative muon capture; triangle:~ordinary muon capture; tic:~not continuum extrapolated; dot:~single ensemble; square:~continuum extrapolated.\\
{\bf Abbreviations:} RMC:~radiative muon capture; OMC:~ordinary muon capture; HBChPT:~heavy baryon ChPT; EOMS:~extended on-mass-shell scheme; DWF:~domain wall fermions; HISQ:~highly improved staggered quarks; CE:~continuum extrapolated.}
\end{figure}
Both the !2P and the $!z^{4+3}$ fit describe the data well~(with similar values for the $\chi^2/d.o.f.$) and, as one can see in table~\ref{tab_results_error}, yield compatible results for almost all observables. For definiteness we choose to quote the values from the $!z^{4+3}$ fit as our final result in these cases, merely because it might have less parametrization bias and because the slightly larger statistical uncertainty is more conservative. In the case of the axial radius, which is directly linked to the axial dipole mass $M_A=\sqrt{12}/r_A$, however, we find that the dipole fit and the $z$-expansion yield significantly different results. Our main conclusion here has to be that $r_A$ (and the small $Q^2$ behavior of the form factors in general, cf.\ figure~\ref{fig_continuum_all_fits}) is highly parametrization dependent -- a nuisance which also plagues determinations from experiment, cf.\ below. It is consistent that we also find a parametrization dependence of the axial coupling constant, where the value $g_A=1.302(86)$ ($z$-exp) is higher than $g_A=1.229(30)$ (dipole). In this case one can compare to the value from an analysis that only takes into account data at zero momentum transfer, which is in agreement with the result obtained from the dipole fit. Further details will be given in a future publication. Note, that this parametrization dependence of the form factors gradually disappears at increasing momentum transfer~$Q^2$.\par%
In figure~\ref{fig_comparison_MA} we show a compilation of experimental data and lattice data for the axial dipole mass. While the $20^{\rm th}$ century world average (cf.\ ref.~\cite{Bernard:2001rs}) supports a value of $M_A$ around $\unit{1}{\giga\electronvolt}$, newer experiments by K2K~\cite{Gran:2006jn}, MINOS~\cite{Adamson:2014pgc}, and, in particular, \mbox{MiniBooNE~\cite{AguilarArevalo:2010zc,Aguilar-Arevalo:2013dva}} yield larger values. This has fueled some discussions lately. One possible explanation is that the discrepancy is caused by nuclear effects. In ref.~\cite{Nieves:2011yp} it has been demonstrated that, using a local Fermi gas (LFG) model combined with multi-hadron interactions and the random phase approximation (RPA), one can recover smaller values for $M_A$ from MiniBooNE data. As argued in ref.~\cite{Butkevich:2013vva}, larger values for $M_A$ in MiniBooNE may also be a consequence of transverse enhancement (TE) due to meson exchange currents (MEC), cf.\ ref.~\cite{Bodek:2011ps}.\par%
Another line of inquiry is persued, e.g., in refs.~\cite{Bhattacharya:2011ah,Bhattacharya:2015mpa} and~\cite{Meyer:2016oeg}. It is based on the suspicion that the dipole ansatz may be too restrictive. Using the $z$-expansion one finds smaller values and much larger errors for $M_A$. In ref.~\cite{Bhattacharya:2011ah} it is shown that the MiniBooNE data is consistent with old $\pi$ electroproduction data under these circumstances. Our analysis supports this picture. The results for the axial radii obtained from the dipole fit (!2P) and the $z$-expansion ($!z^{4+3}$) correspond to the axial pole masses of $M_A=\unit{1.31(8)}{\giga\electronvolt}$ (dipole) and $M_A=\unit{1.02(10)}{\giga\electronvolt}$ ($z$-exp). The situation we find is thus very similar to the one reported in ref.~\cite{Bhattacharya:2011ah}, where extractions using a dipole ansatz yield $M_A=\unit{1.29(5)}{\giga\electronvolt}$ (dipole, \cite{Bhattacharya:2011ah}), while the $z$-expansion yields a smaller value $M_A=\unit{(0.85^{+0.22}_{-0.07} \pm 0.09)}{\giga\electronvolt}$ ($z$-exp, \cite{Bhattacharya:2011ah}), see also ref.~\cite{Bhattacharya:2015mpa}. It is notable that the $z$-expansion coefficients we obtain from our fits (see table~\ref{tab_mean_param_results}) approximately satisfy the constraints that are imposed in ref.~\cite{Bhattacharya:2011ah}.\par%
For the dipole ansatz our result is in good agreement with previous lattice determinations. In particular the agreement with the continuum extrapolated value from ref.~\cite{Rajan:2017lxk} is encouraging. For the $z$-expansion the situation is not so clear, since the lattice results scatter over a wide range. In part this may be caused by the use of different variants of the $z$-expansion (number of parameters, use of priors, choice of $t_0$ in eq.~\eqref{eq_zDefinition}, implementation of constraints, etc.).\par%
\begin{figure}
\begin{minipage}{.3\textwidth}%
\includegraphics[width=.915\textwidth]{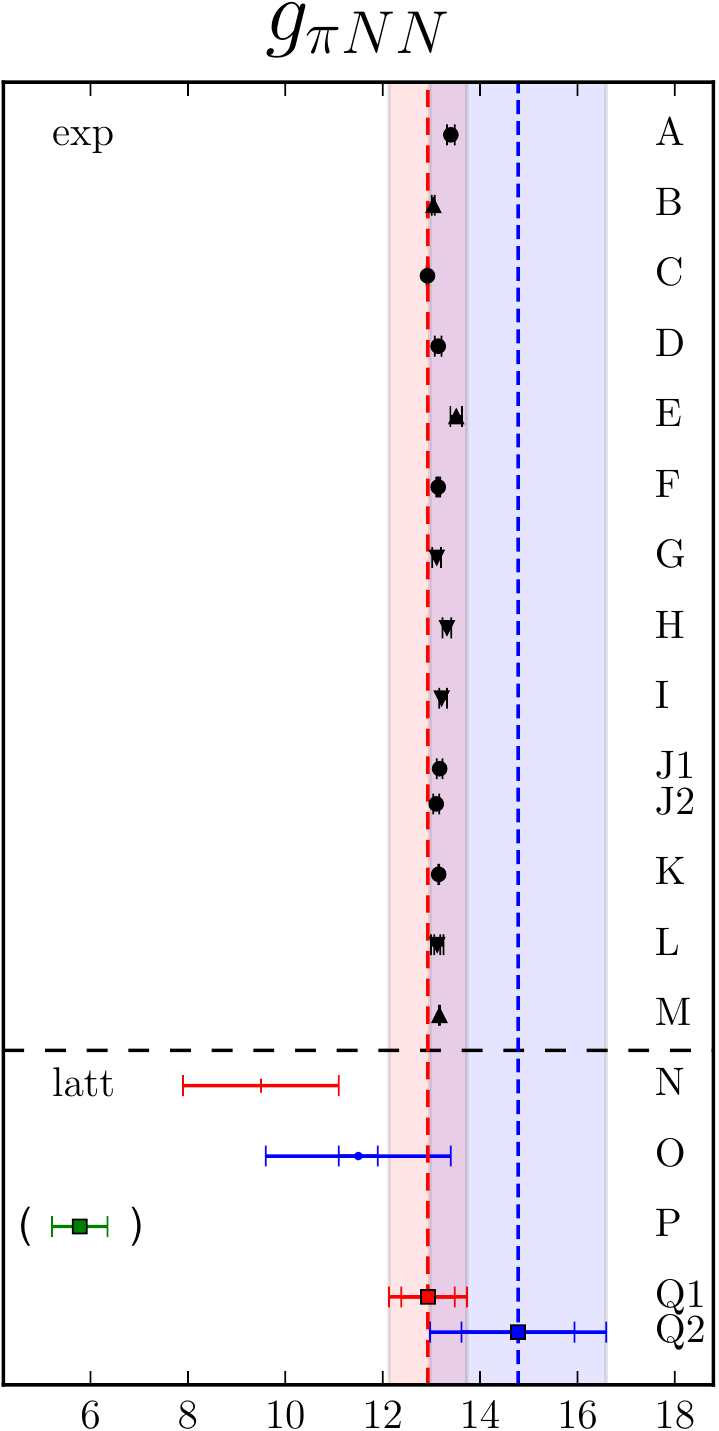}\hfill%
\end{minipage}\begin{minipage}[b]{.7\textwidth}\tiny%
\renewcommand{\arraystretch}{1.15}%
\begin{widetable}{\textwidth}{lll}
\toprule
id & ref. & description \\
\midrule
A	 & \cite{Koch:1980ay}		 & $\pi N$ scattering; PWA	\\
B	 & \cite{Klomp:1991vz,Stoks:1992ja,deSwart:1997ep}		 & $np$, $pp$ scattering; PWA	\\
C	 & \cite{Arndt:1991vf,Arndt:1990bp}		 & $\pi N$ scattering; PWA	\\
D	 & \cite{Arndt:1994bu}		 & $\pi N$ scattering; PWA; GMO	\\
E	 & \cite{Rahm:1998jt}		 & $np$ backward cross section	\\
F	 & \cite{Pavan:1999cr}		 & $\pi N$ scattering; PWA; DR	\\
G	 & \cite{Schroder:1999uq}		 & $\pi^- p$ and $\pi^- d$ pionic atoms; GMO	\\
H	 & \cite{Ericson:2000md}		 & $\pi^- p$ and $\pi^- d$ pionic atoms; GMO	\\
I	 & \cite{Schroder:2001rc}		 & $\pi^- p$ and $\pi^- d$ pionic atoms; GMO	\\
J	 & \cite{Bugg:2004cm}		 & $\pi N$ scattering; DR;	\\
J1	 & 				 & \quad CERN data	\\
J2	 & 				 & \quad TRIUMF data	\\
K	 & \cite{Arndt:2006bf}		 & $\pi N$ scattering; PWA; DR	\\
L	 & \cite{Baru:2010xn,Baru:2011bw,Hoferichter:2015hva}		 & $\pi^- p$ and $\pi^- d$ pionic atoms; GMO; including third-order ChPT corrections	\\
M	 & \cite{Perez:2016aol}		 & $np$, $pp$ scattering; PWA	\\
\midrule
N	 & \cite{Yamazaki:2009zq}		 & $N_f=2+1$ DWF; RBC/UKQCD; $a=\unit{0.114}{\femto\meter}$; dipole ansatz	\\
O	 & \cite{Green:2017keo}		 & $N_f=2+1$ Wilson (clover) fermions; $a=\unit{0.114}{\femto\meter}$; $z$-exp	\\
P	 & \cite{Rajan:2017lxk}		 & $N_f=2+1+1$ Wilson (clover-on-HISQ) fermions; PNDME; CE; EFT ansatz	\\
Q	 & 		 & {\bf This work;} $N_f=2+1$ Wilson (clover) fermions; RQCD;	\\
	 & 				 & \ full resolution of $N\pi$ state; CE	\\
Q1	 & 				 & \quad dipole ansatz	\\
Q2	 & 				 & \quad $z$-exp	\\
\bottomrule
\end{widetable}
\end{minipage}
\caption{\label{fig_comparison_gPiNN}Compilation of data for the pion-nucleon coupling constant $g_{\pi NN}$ from experiment~\mbox{(A-M)} and from lattice simulations (N-Q). We do not discriminate between charged and neutral pion-nuclon couplings here, which can be slightly different. In the lattice section we have only listed direct determinations, ignoring all results that are merely based on the Goldberger--Treiman relation~\cite{Goldberger:1958vp}. Extractions based on a dipole ansatz are colored red, while those using any variant of the $z$-expansion are colored blue. Some lattice calculations use an EFT ansatz colored green (a pion pole term combined with a Taylor expansion). The error bands show the results of our !2P (red) and our $!z^{4+3}$ (blue) fits, with all errors added in quadrature. The lattice result in parentheses is outdated, cf.\ the discussion in ref.~\cite{Jang:2019vkm}. For a recent review, see ref.~\cite{Matsinos:2019kqi}.\\
{\bf Symbols:} circle:~$N \pi$ scattering; triangle~(up):~$NN$ scattering; triangle~(down):~pionic atoms; tic:~not continuum extrapolated; dot:~single ensemble; square:~continuum extrapolated.\\
{\bf Abbreviations:} PWA:~partial wave analysis; GMO:~Goldberger--Miyazawa--Oehme sum rule~\cite{Goldberger:1955zza}; DR:~dispersion relation; DWF:~domain wall fermions; HISQ:~highly improved staggered quarks; CE:~continuum extrapolated.}
\end{figure}%
In figure~\ref{fig_comparison_gPs} we have compiled results for the induced pseudoscalar coupling at the muon capture point $g_P^\star$ from experiment, ChPT, and lattice QCD. The ChPT predictions\footnote{Heavy baryon ChPT actually reproduces the Adler--Dothan--Wolfenstein formula~\cite{Adler:1966gc,Wolfenstein:1970zz}, cf.\ ref.~\cite{Bernard:2001rs}.} are based on measurements of the axial radius and experimental data for $g_{\pi NN}$. They persistantly call for a value slightly above $8$. While older measurements of ordinary muon capture (OMC) were in agreement with this prediction (within large errors), the TRIUMF measurement~\cite{Jonkmans:1996my,Wright:1998gi} lies significantly higher. It has to be seen as a success of BChPT that the new OMC measurement by MuCap~\cite{Andreev:2012fj,Andreev:2015evt} is spot on with a small error. Independent of the choice of parametrization our results are in agreement with both the ChPT prediction and the MuCap result. In particular recent lattice results that include a chiral and a continuum extrapolation using ensembles with close to physical pion masses have yielded much smaller values. In retrospect, it is clear that these findings were caused by the pion pole enhanced $N\pi$ excited state contribution, which was not fully under control. See also ref.~\cite{Jang:2019vkm}, where the same conclusion has been drawn.\par%
Results for the pion-nucleon coupling constant $g_{\pi NN}$ are collected in figure~\ref{fig_comparison_gPiNN}. The experimental results from $\pi N$ scattering, $NN$ scattering, and pionic atoms have reached a high precision, and in particular recent determinations are in quite good agreement with each other. The discussion is now centering on the understanding of charge and isospin breaking effects (see, e.g., refs.~\cite{Hoferichter:2009ez,Hoferichter:2009gn}) --- a question that is out of reach of current lattice QCD analyses of nuclean structure, which usually ignore QED effects and use degenerate light quark masses. Also the experimental precision is not yet within reach.\footnote{There are a number of indirect estimates based on the Goldberger--Treiman relation, see, e.g., refs.~\cite{Lin:2008uz,Yamazaki:2009zq,Bali:2014nma,Alexandrou:2017hac}. While such estimates can have quite small statistical errors and may serve as consistency checks, they should not be considered as independent measurements of $g_{\pi NN}$.} However, a comparison of the lattice values with the experimental results and, in particular, with the analysis of refs.~\cite{Baru:2010xn,Baru:2011bw,Hoferichter:2015hva}, which includes higher order ChPT corrections and an estimate of systematic uncertainties, can serve as a consistency check. It is thus quite encouraging that our results for $g_{\pi NN}$ from both, the !2P and the $!z^{4+3}$ fit, are in agreement with these determinations. As one can see in table~\ref{tab_results_error}, a meaningful prediction of the Goldberger--Treiman discrepancy $\Delta_{\rm GT}=1-\frac{m g_A}{F_\pi g_{\pi NN}}$ is not possible with our current accuracy.\par%
\section{Summary\label{sec_summary}}%
In this article we have presented a method that can control pion pole enhanced excited state contributions that occur in the axial and pseudoscalar channels. The technique is based on EFT considerations similar to refs.~\cite{Bar:2015zwa, Tiburzi:2015tta, Tiburzi:2015sra, Bar:2017kxh, Bar:2017gqh, Bar:2018akl, Bar:2018xyi, Bar:2019gfx}, but simultaneously reduces the ChPT input. The EFT analysis presented in section~\ref{sec_eft} is mainly used to understand the general structure of the pole enhanced $N\pi$ contribution, which then can be taken into account explicitly in the spectral decomposition of the three-point functions, see section~\ref{sec_final_param}. The fits give amplitudes consistent with EFT expectations, however, we do not constrain these in the analysis. Our numerical analysis presented in section~\ref{sec_data_analysis} demonstrates that, using our new technique, the ground state can be extracted reliably, even at small pion masses where the pole enhanced excited state constitutes (at currently available source-sink distances) the largest contribution in some channels.\par%
We find that the nucleon form factors extracted at nonvanishing lattice spacings satisfy constraints from PCAC up to small deviations of roughly~$5\%$, which can be attributed to discretization effects. We find the PPD assumption to be fulfilled to the same degree. Note, however, that the pion pole dominance assumption for the pseudoscalar form factors is only a (seemingly very good) estimate and is not expected to be satisfied exactly, even in the continnum. PCAC, however, \emph{has} to hold exactly in the continuum. We leverage the latter information in our form factor analysis: in addition to the usual dipole ansatz and the $z$-expansion, we have derived (for both cases) parametrizations that are consistent with PCAC in the continuum, cf.\ section~\ref{sec_param_pcac}. The latter stabilize the continuum extrapolation considerably, without adding any parametrization bias.\par%
\begin{figure}[tb]\centering%
\centering%
\includegraphics[width=0.49\textwidth]{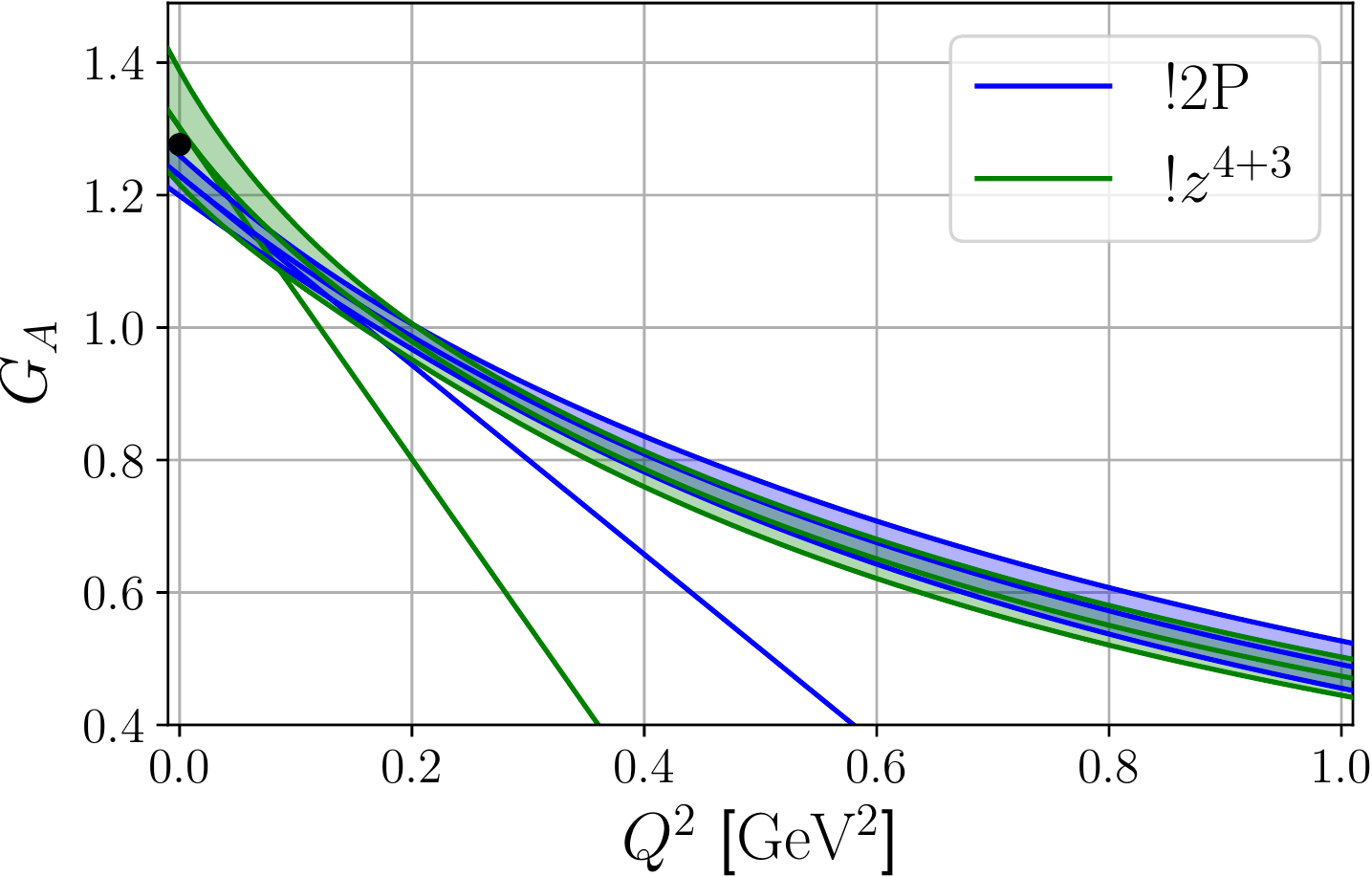}\hfill\includegraphics[width=0.49\textwidth]{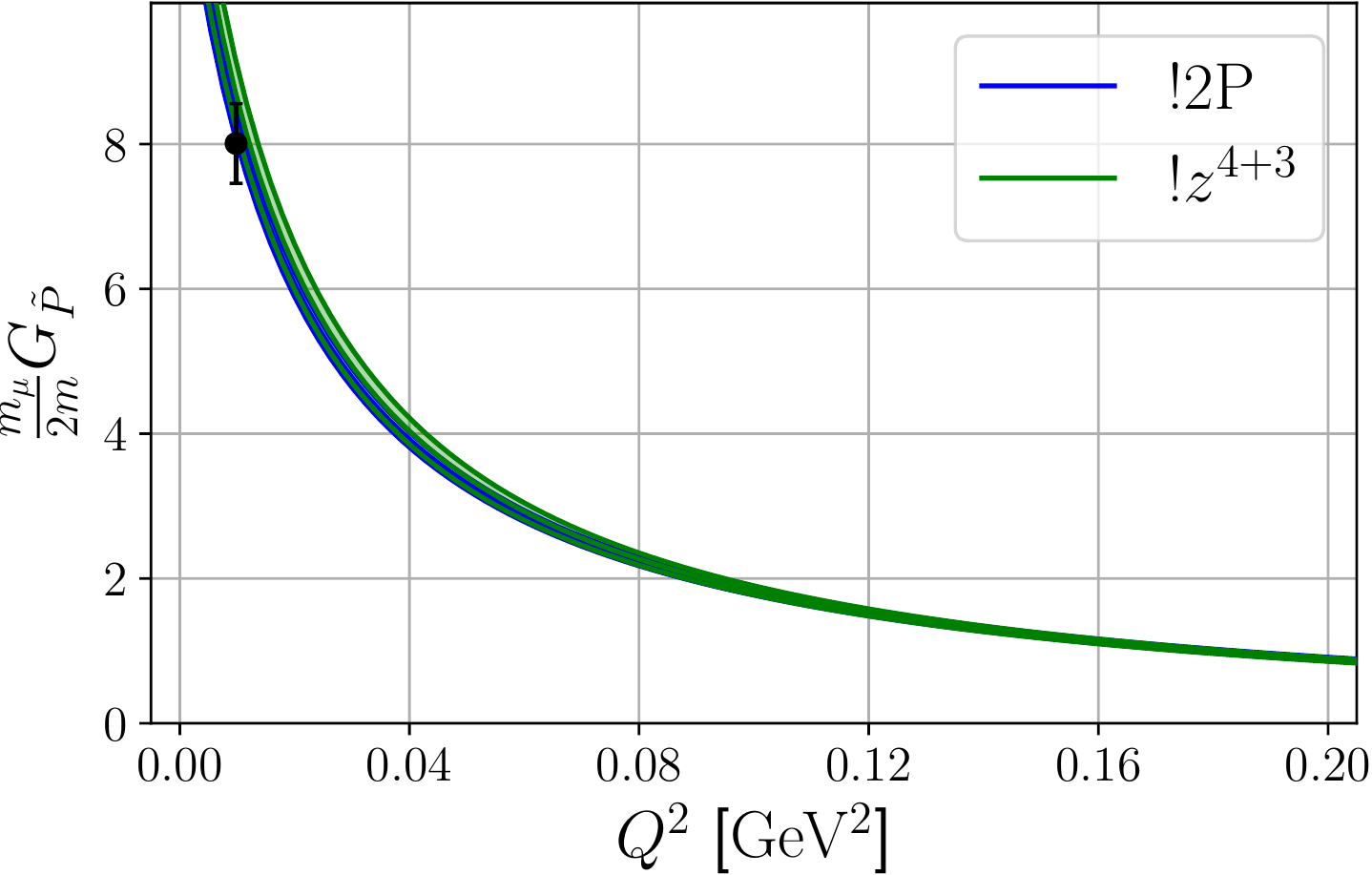}%
\caption{\label{fig_ff_summary}Results for the form factors obtained from the !2P (blue) and the $!z^{4+3}$ (green) fits. The bands show the statistical and systematic errors added in quadrature. The left panel shows the axial form factor $G_A(Q^2)$. At $Q^2=0$ the black circle indicates the experimental result for $g_A$~\cite{Tanabashi:2018oca} (see also refs.~\cite{Brown:2017mhw,Markisch:2018ndu,Gonzalez-Alonso:2018omy,Hayen:2019nic}), while the lines indicate the slope of the corresponding fit. On the right panel we plot the results for $\frac{m_\mu}{2 m} G_\tP(Q^2)$, which can be compared to the experimental value for the induced pseudoscalar coupling $g^\star_P$ (cf.\ eq.~\eqref{eq_gPt_muon_capture}) from OMC~\cite{Andreev:2012fj,Andreev:2015evt} (black circle).}
\end{figure}%
Using a large set of CLS ensembles, we are able to take all the relevant limits (continuum limit, infinite volume limit, and extrapolation to physical quark masses) in a controlled fashion. To this end, we use generic extrapolation formulas (see section~\ref{sec_param_extrapolation}) for the parameters occurring in the form factor parametrization. The results at the physical point (in the continuum and for infinite volume) obtained from various form factor parametrizations are given in tables~\ref{tab_mean_param_results} and~\ref{tab_results}. Within present errors, our form factor data are well represented both by the dipole parametrization and by $z$-expansion fits. The final numbers, including estimates of systematic uncertainties due to the quark mass and the continuum extrapolation, can be taken from table~\ref{tab_results_error}. In figure~\ref{fig_ff_summary} we show the results for the form factors. One can see that the deviations between the dipole fit and the $z$-expansion mainly affect the small $Q^2$ region, and gradually disappear at increasing momentum transfer~$Q^2$. Files containing the data used to create this figure are included as supplementary material.\par%
In particular the slope of the axial form factor at zero momentum transfer, which is proportional to the axial radius (i.e., inversely proportional to the so-called axial mass), exhibits a substantial parametrization dependence, as can be seen in figure~\ref{fig_ff_summary}. To reduce this ambiguity and to eventually rule out one of the parametrizations one would have to improve the resolution of the form factor in the region of small momentum transfer. This can be achieved by increasing the number of data points at very small values of $Q^2$ (one could also compute the derivative of the form factor at $Q^2=0$~\cite{Alexandrou:2015spa,Alexandrou:2020aja}) or by substantially reducing the errors of the data in this region.\footnote{Certainly such data would be most helpful at small lattice spacings and at physical quark masses in order to control the necessary extrapolations.} Interestingly, the tendency of obtaining a larger radius from the $z$-expansion also applies to the analysis of experimental data, which do not cover the very low-$Q^2$ region well either. In fact both our $z$-expansion and our dipole fit results for the axial radius are in agreement with the respective findings from recent quasi-elastic (anti-)neutrino nucleon scattering data (MiniBooNE, \cite{Bhattacharya:2011ah,Bhattacharya:2015mpa}), where the same parametrization bias has been reported. We emphasize that within the $Q^2$ regime that is of interest regarding terrestrial long baseline neutrino experiments the two parametrization of our data overlap within a fraction of a standard deviation so that both parametrizations can be used equally well for neutrino phenomenology. In contrast to most determinations from experiment (in particular the more precise ones), our method does not rely on any assumptions regarding nuclear effects. Therefore, the results can also be used to benchmark nuclear models.\par%
In figure~\ref{fig_rpcac_rppd_physical_point}, we plot the ratios $r_{\rm{PCAC}}$ and $r_{\rm{PPD}}$ at the physical point, where deviations from unity correspond to a violation of PCAC and deviations from the PPD assumption, respectively. In particular the fits with exact PCAC in the continuum (i.e., $r_{\rm{PCAC}}=1$ automatically) allow us to draw conclusions with respect to the pion pole dominance ansatz for the pseudoscalar form factors. We find that our results are consistent with the PPD ansatz independent of the choice of parametrization of the form factor. The values we extract for the induced pseudoscalar coupling at the muon capture point are in good agreement with the experimental value obtained from muon capture~{\cite{Andreev:2012fj,Andreev:2015evt}.\par%
\acknowledgments
We are grateful to O.~B\"ar, R.~Gupta, M.~Hoferichter, K.-F.~Liu, D.~Mohler, and S.~Schaefer for fruitful discussions. The authors also would like to express their gratitude towards B.~Gl{\"a}{\ss}le, P.~Georg, D.~Richtmann, and J.~Simeth for support. This work was funded by the Deut\-sche For\-schungs\-ge\-mein\-schaft (collaborative research centre SFB/TRR\nobreakdash-55), the European Union's Horizon 2020 Research and Innovation programme under the Marie Sk\l{}odowska-Curie grant agreement no.~813942 (ITN EuroPLEx), and the STRONG-2020 project under grant agreement no.~824093. S.~Weish\"aupl received support from the German BMBF grant~05P18WRFP1.\par%
We used a modified version of the {\sc Chroma}~\cite{Edwards:2004sx} software package along with the {\sc LibHadronAnalysis} library~\cite{Bali:2017mft} and improved inverters~\cite{Nobile:2010zz,Luscher:2012av,Frommer:2013fsa,Heybrock:2015kpy}. The configurations were generated as part of the CLS effort~\cite{Bruno:2014jqa,Bali:2016umi} using {\sc openQCD} (\url{https://luscher.web.cern.ch/luscher/openQCD/})~\cite{Luscher:2012av}. We thank all our CLS colleagues for the joint generation of the gauge ensembles. Additional $m_\ell=m_s$ ensembles were generated with {\sc openQCD} by members of the Mainz group on the Wilson and Clover HPC Clusters of IKP Mainz as well as by RQCD on the QPACE computer using the BQCD code~\cite{Nakamura:2010qh}.\par%
The computation of observables was carried out on the QPACE~2 and QPACE~3 systems of the SFB/TRR\nobreakdash-55, on the Regensburg QPACE~B machine, the Regensburg HPC-cluster ATHENE~2, and at various supercomputer centers. In particular, the authors gratefully acknowledge computing time granted by the John von Neumann Institute for Computing (NIC), provided on the Booster partition of the supercomputer JURECA~\cite{jureca} at J\"ulich Supercomputing Centre (JSC, \url{http://www.fz-juelich.de/ias/jsc/}).\par%
Regarding the generation of recent gauge ensembles, the authors gratefully acknowledge the Gauss Centre for Supercomputing (\kern-.6ptGCS\kern-.3pt) for providing computing time for GCS Large-Scale Projects on the GCS share of the two supercomputers JUQUEEN~\cite{juqueen} and JUWELS~\cite{juwels} at JSC as well as on SuperMUC at Leibniz Supercomputing Centre (LRZ, \url{https://www.lrz.de}). GCS is the alliance of the three national supercomputing centres HLRS (Universit\"at Stuttgart), JSC (For\-schungs\-zen\-trum J\"ulich), and LRZ (Bayerische Akademie der Wissenschaften), funded by the German Federal Ministry of Education and Research (BMBF) and the German State Ministries for Research of Baden-W{\"u}rttemberg (MWK), Bayern (StMWFK) and Nordrhein-Westfalen (MIWF).\par%
\appendix
\section{Traces\label{app_traces}}
For the ground state contributions defined in eq.~\eqref{eq_gs_trace} one finds
\begin{align}%
  B_{P_+^i, \mathcal A^\mu}^{\vec{p}^{\mathrlap{\prime}}, \vec{p}}
  &= 2G_A \bigl(p^{\prime i}p^\mu+p^ip^{\prime\mu} + m(p^{\prime}+p)^ig^{\mu 0} - g^{i\mu}(m^2+mE^\prime+mE+p^\prime \cdot p)\bigr) \notag \\
  &\quad+ 2G_\tP \frac{q^\mu}{2m} \bigl((m+E^\prime)p^i-(m+E)p^{\prime i}\bigr) \,, \label{eq_gs_trace_A}\\
   B_{P_+^i, \mathcal P}^{\vec{p}^{\mathrlap{\prime}}, \vec{p}}
  &= 2G_P \bigl((m+E^\prime)p^i-(m+E)p^{\prime i}\bigr) \,. \label{eq_gs_trace_P}
\end{align}%
Evaluating these equations for the 4 particular cases depicted in the rows of figure~4, where $\vec p^\prime = \vec 0$ and $\vec p = -\vec q =  (0,0,p)^T$ with $p=\frac{2\pi}{L}$, yields%
\begin{align}
 &\text{row 1:}\quad 4 m (E+m) G_A \,, \label{eq_gs_trace_row1}\\
 &\text{row 2:}\quad (E+m) \bigl(4 m G_A - 2 (E-m) G_\tP \bigr) \,, \label{eq_gs_trace_row2}\\
 &\text{row 3:}\quad p\bigl(4 m G_A - 2 (E-m) G_\tP \bigr) \,, \label{eq_gs_trace_row3}\\
 &\text{row 4:}\quad 4 m p G_P \,. \label{eq_gs_trace_row4}
\end{align}%
For the remaining traces that are needed for the determination of the parametrizations given in section~\ref{sec_final_param} one gets%
\begin{align}
  \operatorname{Tr}\bigl\{P_+^i(\slashed{p}+m)\slashed{r}_+\gamma_5(\slashed{p}+m)\bigr\}
  &= 4\bigl(p^i(mE_\pi+p \cdot r_+)-mr_+^i(m+E)\bigr) \,, \\
  \operatorname{Tr}\bigl\{P_+^i(\slashed{p}^\prime+m)\slashed{r}_-\gamma_5(\slashed{p}^\prime+m)\bigr\}
  &= 4\bigl(p^{\prime i}(mE_\pi+p^\prime \cdot r_-)-mr_-^i(m+E^\prime)\bigr) \,, \\
  \operatorname{Tr}\bigl\{P_+^i\gamma_5(\slashed{p}+m)\bigr\} &= +2p^i \,, \\
  \operatorname{Tr}\bigl\{P_+^i(\slashed{p}^\prime+m)\gamma_5\bigr\} &= -2p^{\prime i} \,.
\end{align}%
\section{Fit ansatz for the subtracted currents\label{app_ansatz_sub}}
For the subtracted correlation functions defined in ref.~\cite{Bali:2018qus} one inserts
\begin{align}
\mathcal A_\perp^\mu &= \left( g^{\mu\nu} - \frac{\bar p^\mu \bar p^\nu}{\bar p^2} \right) \mathcal  A_\nu\,, &
\mathcal P_\perp &= \mathcal  P - \frac{1}{2i\,m_\ell} \frac{\bar p^\mu \bar p^\nu}{\bar p^2} \partial_\mu \mathcal A_\nu \,,
\end{align}%
instead of the usual currents. Here $\bar p = (p^{\mathrlap{\prime}} + p)/2$. By construction, this does not change the ground state contribution at all. In contrast, the excited state contributions are affected very strongly. Therefore, the fit ansatz given in eqs.~\eqref{eq_res_A_noChPT} and~\eqref{eq_res_P_noChPT} has to be adapted to this case. Following the same steps as discussed in detail for the standard currents in section~\ref{sec_eft}, we find
\begin{align} 
\begin{split}
C_{{\rm 3pt},P_+^i}^{\vec{p}^{\mathrlap{\prime}}, \vec{p}, \mathcal A^\mu_\perp} &=\frac{\sqrt{Z^\prime}\sqrt{Z}}{2E^\prime\,2E} e^{-E^\prime(t-\tau)} e^{-E\tau} \\
 &\quad\times \biggl[ \begin{aligned}[t] & B_{P_+^i, \mathcal A^\mu}^{\vec{p}^{\mathrlap{\prime}}, \vec{p}} \biggl( 1 + B_{10} e^{-\Delta E^\prime (t-\tau)} +  B_{01} e^{-\Delta E \tau} + B_{11} e^{-\Delta E^\prime (t-\tau)}  e^{-\Delta E \tau} \biggr) \\
  &+e^{-\Delta E_{N\pi}^\prime (t-\tau)}\frac{E^\prime}{E_\pi} \biggl( r_+^\mu - \bar p^\mu \, \frac{\bar p \cdot r_+}{\bar p^2} \biggr) \biggl(c^\prime p^i + d^\prime q^i\biggr)\\
&+e^{-\Delta E_{N\pi} \tau}\frac{E}{E_\pi}  \biggl( r_-^\mu - \bar p^\mu \, \frac{\bar p \cdot r_-}{\bar p^2} \biggr)\biggl(c \, p^{\prime i} + d \, q^i\biggr) \biggr] \,, \taghere \end{aligned}  \label{eq_res_Aperp_noChPT} \end{split} \\
\begin{split}
C_{{\rm 3pt},P_+^i}^{\vec{p}^{\mathrlap{\prime}}, \vec{p}, \mathcal P_\perp} &=\frac{\sqrt{Z^\prime}\sqrt{Z}}{2E^\prime\,2E} e^{-E^\prime(t-\tau)} e^{-E\tau} \\
 &\quad\times \biggl[ \begin{aligned}[t] & B_{P_+^i, \mathcal P}^{\vec{p}^{\mathrlap{\prime}}, \vec{p}} \biggl( 1 + B_{10} e^{-\Delta E^\prime (t-\tau)} +  B_{01} e^{-\Delta E \tau} + B_{11} e^{-\Delta E^\prime (t-\tau)}  e^{-\Delta E \tau} \biggr) \\
&+e^{-\Delta E_{N\pi}^\prime (t-\tau)}\frac{E^\prime}{E_\pi} \frac{1}{2m_\ell}  \biggl( m_\pi^2 - \frac{(\bar p \cdot r_+)^2}{\bar p^2} \biggr) \biggl(c^\prime p^i + d^\prime q^i\biggr)\\
&-e^{-\Delta E_{N\pi} \tau}\frac{E}{E_\pi}  \frac{1}{2m_\ell}  \biggl( m_\pi^2 - \frac{(\bar p \cdot r_-)^2}{\bar p^2} \biggr) \biggl(c \, p^{\prime i} + d \, q^i\biggr) \biggr] \,.  \end{aligned}  \label{eq_res_Pperp_noChPT} \end{split}
\end{align}
Similar to the situation with unsubtracted correlation functions, the parametrization simplifies for the particular kinematics we are using in our numerical analysis ($\vec p^\prime = \vec 0$ such that $\vec q = - \vec p$).

\input{axFF_cls.bbl}
\end{document}

%% file: axFF_cls.bbl
\providecommand{\href}[2]{#2}\begingroup\raggedright\endgroup